\documentclass[twocolumn]{aa}

\usepackage{graphicx}
\usepackage{amsmath,amsfonts,amssymb}
\usepackage{txfonts}
\usepackage{color}
\usepackage{natbib}
\usepackage{float}
\usepackage{dblfloatfix}
\usepackage{afterpage}
\usepackage{ifthen}
\usepackage[morefloats=12]{morefloats}
\usepackage{placeins}
\usepackage{multicol}
\bibpunct{(}{)}{;}{a}{}{,}
\usepackage[switch]{lineno}
\definecolor{linkcolor}{rgb}{0.6,0,0}
\definecolor{citecolor}{rgb}{0,0,0.75}
\definecolor{urlcolor}{rgb}{0.12,0.46,0.7}
\usepackage[breaklinks, colorlinks, urlcolor=urlcolor,
    linkcolor=linkcolor,citecolor=citecolor,pdfencoding=auto]{hyperref}
\hypersetup{linktocpage}
\usepackage{bold-extra}
\usepackage{import}

\def\setsymbol#1#2{\expandafter\def\csname #1\endcsname{#2}}
\def\getsymbol#1{\csname #1\endcsname}

\def\Planck{\textit{Planck}}

\newbox\tablebox    \newdimen\tablewidth
\def\leaderfil{\leaders\hbox to 5pt{\hss.\hss}\hfil}
\def\endPlancktablewide{\tablewidth=\textwidth 
    $$\hss\copy\tablebox\hss$$
    \vskip-\lastskip\vskip -2pt}
\def\tablenote#1 #2\par{\begingroup \parindent=0.8em
    \abovedisplayshortskip=0pt\belowdisplayshortskip=0pt
    \noindent
    $$\hss\vbox{\hsize\tablewidth \hangindent=\parindent \hangafter=1 \noindent
    \hbox to \parindent{$^#1$\hss}\strut#2\strut\par}\hss$$
    \endgroup}
\def\doubleline{\vskip 3pt\hrule \vskip 1.5pt \hrule \vskip 5pt}

\def\L2{\ifmmode L_2\else $L_2$\fi}

\def\DeltaT{\ifmmode \Delta T\else $\Delta T$\fi}
\def\deltat{\ifmmode \Delta t\else $\Delta t$\fi}
\def\fknee{\ifmmode f_{\rm knee}\else $f_{\rm knee}$\fi}
\def\Fmax{\ifmmode F_{\rm max}\else $F_{\rm max}$\fi}
\def\solar{\ifmmode{\rm M}_{\mathord\odot}\else${\rm M}_{\mathord\odot}$\fi}
\def\Msolar{\ifmmode{\rm M}_{\mathord\odot}\else${\rm M}_{\mathord\odot}$\fi}
\def\Lsolar{\ifmmode{\rm L}_{\mathord\odot}\else${\rm L}_{\mathord\odot}$\fi}
\def\inv{\ifmmode^{-1}\else$^{-1}$\fi}
\def\mo{\ifmmode^{-1}\else$^{-1}$\fi}
\def\sup#1{\ifmmode ^{\rm #1}\else $^{\rm #1}$\fi}
\def\expo#1{\ifmmode \times 10^{#1}\else $\times 10^{#1}$\fi}
\def\,{\thinspace}
\def\lsim{\mathrel{\raise .4ex\hbox{\rlap{$<$}\lower 1.2ex\hbox{$\sim$}}}}
\def\gsim{\mathrel{\raise .4ex\hbox{\rlap{$>$}\lower 1.2ex\hbox{$\sim$}}}}

\def\simprop{\mathrel{\raise .4ex\hbox{\rlap{$\propto$}\lower 1.2ex\hbox{$\sim$}}}}
\def\deg{\ifmmode^\circ\else$^\circ$\fi}
\def\pdeg{\ifmmode $\setbox0=\hbox{$^{\circ}$}\rlap{\hskip.11\wd0 .}$^{\circ}
          \else \setbox0=\hbox{$^{\circ}$}\rlap{\hskip.11\wd0 .}$^{\circ}$\fi}
\def\arcs{\ifmmode {^{\scriptstyle\prime\prime}}
          \else $^{\scriptstyle\prime\prime}$\fi}
\def\arcm{\ifmmode {^{\scriptstyle\prime}}
          \else $^{\scriptstyle\prime}$\fi}
\newdimen\sa  \newdimen\sb
\def\parcs{\sa=.07em \sb=.03em
     \ifmmode \hbox{\rlap{.}}^{\scriptstyle\prime\kern -\sb\prime}\hbox{\kern -\sa}
     \else \rlap{.}$^{\scriptstyle\prime\kern -\sb\prime}$\kern -\sa\fi}
\def\parcm{\sa=.08em \sb=.03em
     \ifmmode \hbox{\rlap{.}\kern\sa}^{\scriptstyle\prime}\hbox{\kern-\sb}
     \else \rlap{.}\kern\sa$^{\scriptstyle\prime}$\kern-\sb\fi}
\def\ra[#1 #2 #3.#4]{#1\sup{h}#2\sup{m}#3\sup{s}\llap.#4}
\def\dec[#1 #2 #3.#4]{#1\deg#2\arcm#3\arcs\llap.#4}
\def\deco[#1 #2 #3]{#1\deg#2\arcm#3\arcs}
\def\rra[#1 #2]{#1\sup{h}#2\sup{m}}

\def\dots{\relax\ifmmode \ldots\else $\ldots$\fi}
\def\WHzsr{\ifmmode $W\,Hz\mo\,sr\mo$\else W\,Hz\mo\,sr\mo\fi}
\def\mHz{\ifmmode $\,mHz$\else \,mHz\fi}
\def\GHz{\ifmmode $\,GHz$\else \,GHz\fi}
\def\mKs{\ifmmode $\,mK\,s$^{1/2}\else \,mK\,s$^{1/2}$\fi}
\def\muKs{\ifmmode \,\mu$K\,s$^{1/2}\else \,$\mu$K\,s$^{1/2}$\fi}
\def\muKRJs{\ifmmode \,\mu$K$_{\rm RJ}$\,s$^{1/2}\else \,$\mu$K$_{\rm RJ}$\,s$^{1/2}$\fi}
\def\muKHz{\ifmmode \,\mu$K\,Hz$^{-1/2}\else \,$\mu$K\,Hz$^{-1/2}$\fi}
\def\MJysr{\ifmmode \,$MJy\,sr\mo$\else \,MJy\,sr\mo\fi}
\def\MJysrmK{\ifmmode \,$MJy\,sr\mo$\,mK$_{\rm CMB}\mo\else \,MJy\,sr\mo\,mK$_{\rm CMB}\mo$\fi}
\def\microns{\ifmmode \,\mu$m$\else \,$\mu$m\fi}

\def\muK{\ifmmode \,\mu$K$\else \,$\mu$\hbox{K}\fi}
\def\microK{\ifmmode \,\mu$K$\else \,$\mu$\hbox{K}\fi}
\def\muW{\ifmmode \,\mu$W$\else \,$\mu$\hbox{W}\fi}
\def\kms{\ifmmode $\,km\,s$^{-1}\else \,km\,s$^{-1}$\fi}
\def\kmsMpc{\ifmmode $\,\kms\,Mpc\mo$\else \,\kms\,Mpc\mo\fi}
\providecommand{\sorthelp}[1]{}

\def\cosmoglobe{\textsc{Cosmoglobe}}
\def\WMAP{\textit{WMAP}}
\def\COBE{\textit{COBE}}
\def\LCDM{$\Lambda$CDM}

\def\healpix{\texttt{HEALPix}}
\def\commander{\texttt{Commander}}

\def\nilc{\texttt{NILC}}

\def\sevem{\texttt{SEVEM}}
\def\smica{\texttt{SMICA}}

\renewcommand{\d}[0]{\vec{d}}

\newcommand{\n}[0]{\vec{n}}

\newcommand{\s}[0]{\vec{s}}
\renewcommand{\a}[0]{\vec{a}}
\newcommand{\m}[0]{\vec{m}}
\newcommand{\x}[0]{\vec{x}}
\newcommand{\f}[0]{\vec{f}}

\newcommand{\B}[0]{\tens{B}}

\newcommand{\Cp}[0]{\tens{C}}
\renewcommand{\L}[0]{\tens{L}}
\newcommand{\g}[0]{\vec{g}}
\newcommand{\N}[0]{\tens{N}}
\newcommand{\M}[0]{\tens{M}}

\renewcommand{\S}[0]{\tens{S}}
\renewcommand{\r}[0]{\vec{r}}

\renewcommand{\P}[0]{\tens{P}}

\newcommand{\Te}[0]{T_{\rm e}}

\newcommand{\BP}{\textsc{BeyondPlanck}}
\newcommand{\lfi}[0]{LFI}

\newcommand{\npipe}[0]{\texttt{NPIPE}}
\newcommand{\Dbp}[0]{\Delta_{\mathrm{bp}}}

\def\Acmb{\ifmmode a_\mathrm{CMB}\else $a_{\mathrm{CMB}}$\fi}
\def\Aquad{\ifmmode a_\mathrm{quad}\else $a_{\mathrm{quad}}$\fi}
\def\Asynch{\ifmmode a_\mathrm{s}\else $a_{\mathrm{s}}$\fi}
\def\Asrc{\ifmmode a_\mathrm{src}\else $a_{\mathrm{src}}$\fi}
\def\Adust{\ifmmode a_\mathrm{d}\else $a_{\mathrm{d}}$\fi}
\def\Asdust{\ifmmode a_\mathrm{sd}\else $a_{\mathrm{sd}}$\fi}
\def\Aame{\ifmmode a_\mathrm{ame}\else $a_{\mathrm{ame}}$\fi}
\def\Aco{\ifmmode a_\mathrm{CO}\else $a_{\mathrm{CO}}$\fi}
\def\AcoOne{\ifmmode a_\mathrm{CO10}\else $a_{\mathrm{CO10}}$\fi}
\def\AcoTwo{\ifmmode a_\mathrm{CO21}\else $a_{\mathrm{CO21}}$\fi}
\def\AcoThree{\ifmmode a_\mathrm{CO32}\else $a_{\mathrm{CO32}}$\fi}
\def\Aff{\ifmmode a_\mathrm{ff}\else $a_{\mathrm{ff}}$\fi}
\def\Tcmb{\ifmmode T_\mathrm{CMB}\else $T_{\mathrm{CMB}}$\fi}
\def\Tdust{\ifmmode T_\mathrm{d}\else $T_{\mathrm{d}}$\fi}
\def\scmb{\ifmmode s_\mathrm{CMB}\else $s_{\mathrm{CMB}}$\fi}
\def\squad{\ifmmode s_\mathrm{quad}\else $s_{\mathrm{quad}}$\fi}
\def\ssynch{\ifmmode s_\mathrm{s}\else $s_\mathrm{s}$\fi}
\def\sdust{\ifmmode s_\mathrm{d}\else $s_{\mathrm{d}}$\fi}
\def\ssdust{\ifmmode s_\mathrm{sd}\else $s_{\mathrm{sd}}$\fi}
\def\same{\ifmmode s_\mathrm{ame}\else $s_{\mathrm{ame}}$\fi}
\def\ssrc{\ifmmode s_\mathrm{src}\else $s_{\mathrm{src}}$\fi}
\def\sco{\ifmmode s_\mathrm{CO}\else $s_{\mathrm{CO}}$\fi}
\def\sff{\ifmmode s_\mathrm{ff}\else $s_{\mathrm{ff}}$\fi}
\def\gff{\ifmmode g_\mathrm{ff}\else $g_{\mathrm{ff}}$\fi}
\def\fsynch{\ifmmode f_\mathrm{s}\else $f_{\mathrm{s}}$\fi}
\def\fsd{\ifmmode f_\mathrm{sd}\else $f_{\mathrm{sd}}$\fi}
\def\fame{\ifmmode f_\mathrm{ame}\else $f_{\mathrm{ame}}$\fi}
\def\alphasrc{\ifmmode \alpha_\mathrm{src}\else $\alpha_{\mathrm{src}}$\fi}
\def\bdust{\ifmmode \beta_\mathrm{d}\else $\beta_{\mathrm{d}}$\fi}
\def\bsynch{\ifmmode \beta_\mathrm{s}\else $\beta_{\mathrm{s}}$\fi}
\def\bsun{\ifmmode \beta_\mathrm{sun}\else $\beta_{\mathrm{sun}}$\fi}
\def\nuzeros{\ifmmode \nu_{0,\mathrm{s}}\else $\nu_{0,\mathrm{s}}$\fi}
\def\nuzeroff{\ifmmode \nu_{0,\mathrm{ff}}\else $\nu_{0,\mathrm{ff}}$\fi}
\def\nuzerod{\ifmmode \nu_{0,\mathrm{d}}\else $\nu_{0,\mathrm{d}}$\fi}
\def\nuzeroame{\ifmmode \nu_{0,\mathrm{ame}}\else $\nu_{0,\mathrm{ame}}$\fi}
\def\nuzerosd{\ifmmode \nu_{0,\mathrm{}}\else $\nu_{0,\mathrm{sd}}$\fi}
\def\nuzerosrc{\ifmmode \nu_{0,\mathrm{src}}\else $\nu_{0,\mathrm{src}}$\fi}
\def\nup{\ifmmode \nu_{\mathrm{p}}\else $\nu_{\mathrm{p}}$\fi}
\def\alphasd{\ifmmode \alpha_{\mathrm{sd}}\else $\alpha_{\mathrm{sd}}$\fi}
\def\Te{\ifmmode T_{\mathrm{e}}\else $T_{\mathrm{e}}$\fi}
\def\lmax{\ifmmode \ell_{\mathrm{max}}\else $\ell_{\mathrm{max}}$\fi}
\def\NHI{\ifmmode N_{\mathrm{H\,\textsc i}}\else $N_{\mathrm{H\,\textsc i}}$\fi}
\def\chisq{\ifmmode \chi^2\else $\chi^2$\fi}
\def\kB{\ifmmode k_\mathrm{B}\else $k_{\mathrm{B}}$\fi}

\def\inv{^{-1}}

\begin{document}

\title{\bfseries{\scshape{BeyondPlanck}} XI. Bayesian CMB analysis \\with sample-based end-to-end error propagation}
\newcommand{\oslo}[0]{2}
\newcommand{\milanoA}[0]{1}
\newcommand{\milanoB}[0]{4}
\newcommand{\milanoC}[0]{3}
\newcommand{\triesteB}[0]{5}
\newcommand{\planetek}[0]{6}
\newcommand{\princeton}[0]{7}
\newcommand{\jpl}[0]{8}
\newcommand{\helsinkiA}[0]{9}
\newcommand{\helsinkiB}[0]{10}
\newcommand{\nersc}[0]{11}
\newcommand{\haverford}[0]{12}
\newcommand{\mpa}[0]{13}
\newcommand{\triesteA}[0]{14}
\author{\small
L.~P.~L.~Colombo\inst{\milanoA}\thanks{Corresponding author: L.~P.~L.~Colombo; \url{loris.colombo@unimi.it}}
\and
J.~R.~Eskilt\inst{\oslo}
\and
S.~Paradiso\inst{\milanoA, \milanoC}
\and
H.~Thommesen\inst{\oslo}
\and
K.~J.~Andersen\inst{\oslo}
\and
\textcolor{black}{R.~Aurlien}\inst{\oslo}
\and
\textcolor{black}{R.~Banerji}\inst{\oslo}
\and
M.~Bersanelli\inst{\milanoA, \milanoB, \milanoC}
\and
S.~Bertocco\inst{\triesteB}
\and
M.~Brilenkov\inst{\oslo}
\and
M.~Carbone\inst{\planetek}
\and
H.~K.~Eriksen\inst{\oslo}
\and
\textcolor{black}{M.~K.~Foss}\inst{\oslo}
\and
C.~Franceschet\inst{\milanoC}
\and
\textcolor{black}{U.~Fuskeland}\inst{\oslo}
\and
S.~Galeotta\inst{\triesteB}
\and
M.~Galloway\inst{\oslo}
\and
S.~Gerakakis\inst{\planetek}
\and
E.~Gjerl{\o}w\inst{\oslo}
\and
\textcolor{black}{B.~Hensley}\inst{\princeton}
\and
\textcolor{black}{D.~Herman}\inst{\oslo}
\and
M.~Iacobellis\inst{\planetek}
\and
M.~Ieronymaki\inst{\planetek}
\and
\textcolor{black}{H.~T.~Ihle}\inst{\oslo}
\and
J.~B.~Jewell\inst{\oslo}
\and
\textcolor{black}{A.~Karakci}\inst{\oslo}
\and
E.~Keih\"{a}nen\inst{\helsinkiA, \helsinkiB}
\and
R.~Keskitalo\inst{\nersc}
\and
G.~Maggio\inst{\triesteB}
\and
D.~Maino\inst{\milanoA, \milanoB, \milanoC}
\and
M.~Maris\inst{\triesteB}
\and
B.~Partridge\inst{\haverford}
\and
M.~Reinecke\inst{\mpa}
\and
A.-S.~Suur-Uski\inst{\helsinkiA, \helsinkiB}
\and
T.~L.~Svalheim\inst{\oslo}
\and
D.~Tavagnacco\inst{\triesteB, \triesteA}
\and
D.~J.~Watts\inst{\oslo}
\and
I.~K.~Wehus\inst{\oslo}
\and
A.~Zacchei\inst{\triesteB}
}
\institute{\small
Dipartimento di Fisica, Universit\`{a} degli Studi di Milano, Via Celoria, 16, Milano, Italy\goodbreak
\and
Institute of Theoretical Astrophysics, University of Oslo, Blindern, Oslo, Norway\goodbreak
\and
INFN, Sezione di Milano, Via Celoria 16, Milano, Italy\goodbreak
\and
INAF-IASF Milano, Via E. Bassini 15, Milano, Italy\goodbreak
\and
INAF - Osservatorio Astronomico di Trieste, Via G.B. Tiepolo 11, Trieste, Italy\goodbreak
\and
Planetek Hellas, Leoforos Kifisias 44, Marousi 151 25, Greece\goodbreak
\and
Department of Astrophysical Sciences, Princeton University, Princeton, NJ 08544,
U.S.A.\goodbreak
\and
Jet Propulsion Laboratory, California Institute of Technology, 4800 Oak Grove Drive, Pasadena, California, U.S.A.\goodbreak
\and
Department of Physics, Gustaf H\"{a}llstr\"{o}min katu 2, University of Helsinki, Helsinki, Finland\goodbreak
\and
Helsinki Institute of Physics, Gustaf H\"{a}llstr\"{o}min katu 2, University of Helsinki, Helsinki, Finland\goodbreak
\and
Computational Cosmology Center, Lawrence Berkeley National Laboratory, Berkeley, California, U.S.A.\goodbreak
\and
Haverford College Astronomy Department, 370 Lancaster Avenue,
Haverford, Pennsylvania, U.S.A.\goodbreak
\and
Max-Planck-Institut f\"{u}r Astrophysik, Karl-Schwarzschild-Str. 1, 85741 Garching, Germany\goodbreak
\and
Dipartimento di Fisica, Universit\`{a} degli Studi di Trieste, via A. Valerio 2, Trieste, Italy\goodbreak
}

\authorrunning{Colombo et al.}
\titlerunning{CMB analysis with end-to-end error propagation}

\abstract{We present posterior sample-based cosmic microwave
  background (CMB) constraints from \Planck\ LFI and
  \WMAP\ observations as derived through global end-to-end Bayesian
  processing within the \BP\ framework. We first use these samples to
  study correlations between CMB, foreground, and instrumental
  parameters, and we identify a particularly strong degeneracy between
  CMB temperature fluctuations and free-free emission on intermediate
  angular scales ($400\lesssim\ell\lesssim600$), which is mitigated
  through model reduction, masking, and resampling. We compare our
  posterior-based CMB results with previous \Planck\ products, and
  find generally good agreement, although with notably higher noise
  due to our exclusion of HFI data. We find a best-fit CMB dipole
  amplitude of $3362.7\pm1.4\muK$, in excellent agreement with
  previous \Planck\ results. The quoted dipole
  uncertainty is derived directly from the sampled posterior
  distribution, and does not involve any ad hoc contributions
  for \Planck\ instrumental systematic effects. Similarly, we
  find a temperature quadrupole amplitude of $\sigma^{TT}_2=229\pm97\muK^2$,
  which is in good agreement with previous results in terms of the
  amplitude, but the uncertainty is an order of magnitude larger than
  the naive diagonal Fisher uncertainty. Relatedly, we find lower
  evidence for a possible alignment between the quadrupole and
  octopole than previously reported due to a much larger scatter in
  the individual quadrupole coefficients, caused both by marginalizing
  over a more complete set of systematic effects, but also by our more
  conservative analysis mask required to mitigate the free-free
  degeneracy. For higher multipoles, we find that the angular
  temperature power spectrum is generally in good agreement with both
  \Planck\ and \WMAP. At the same time, we note that this is the first
  time the sample-based asymptotically exact Blackwell-Rao estimator
  has been successfully established for multipoles up to $\ell\le600$,
  and it now accounts for the majority of the cosmologically important
  information. Overall, this analysis demonstrates the unique
  capabilities of the Bayesian approach with respect to end-to-end
  systematic uncertainty propagation, and we believe it can and should
  play an important role in the analysis of future CMB
  experiments. Cosmological parameter constraints are presented in a
  companion paper~\citep{bp12}.}


\maketitle

\tableofcontents

\section{Introduction}
\label{sec:introduction}

Detailed measurements of the cosmic microwave background (CMB) have
revolutionized modern cosmology during the last three
decades. Offering a unique and crystal clear view of the baby Universe
only 380\,000 years after the Big Bang
\citep[e.g.,][]{bennett2012,planck2016-l01}, its tiny temperature
fluctuations allow scientists to measure a range of cosmological
parameters with sub-percent accuracy, and this work has culminated in
a tremendously successful standard model of cosmology called
$\Lambda$CDM \citep[e.g.,][]{hinshaw2012,planck2016-l06}. According to this
model, the Universe began with a hot Big Bang some 13.8\,billion
years ago; it was filled with Gaussian random density fluctuations
during a cataclysmic quantum mechanical process called inflation
taking place only some $10^{-34}$\,seconds after the beginning, during
which its size grew exponentially; and it is today populated by about
65\,\% dark energy ($\Lambda$), 30\,\% cold dark matter (CDM), and
only 5\,\% ordinary baryonic matter.

While this model is extremely successful in terms of predicting
cosmological observations quantitatively, it leaves unanswered foundational questions, such as ``What is dark matter?''
and ``What is dark energy?'' Perhaps the biggest of them all is
simply, ``What exactly did happen during the very first moments of the Big Bang?'' As of 2022, cosmic
inflation \citep[e.g.,][]{kamionkowski:2016} represents a basic
paradigm for this process that is widely accepted by the community,
simply because it is able to heuristically explain a range of
important observations in modern cosmology, including cosmological
isotropy, flatness, and the absence of topological defects, and its
predictions are largely consistent with CMB measurements, such as
Gaussianity and a nearly scale-invariant, but slightly tilted,
spectrum of initial perturbations. At the same time, inflation as a
general concept is both heavily criticized for being overly flexible
\citep[e.g.,][]{penrose:1989,ijjas:2014}, to the extent that one might
question whether it has any predictive power, and for lacking a robust
theoretical foundation, which may require a proper theory of quantum
gravity.

To make further progress more data are desperately needed. And the
most promising path to such is through deep measurements of
large-scale CMB polarization \citep[e.g.,][]{kamionkowski:2016}. A
firm prediction of the inflationary paradigm is that there should
exist a background of primordial gravitational waves that were excited
during the period of exponential expansion. If so, these super-horizon
gravitational waves should also make an imprint on the CMB field in
the form of so-called $B$-mode polarization. The amplitude of this
signal is typically measured in terms of the tensor-to-scalar ratio,
$r$, and different inflationary models (corresponding to different
inflationary potentials) predict different values for $r$, with
typical values varying between $10^{-4}$ and 0.1 for large model
spaces. The strongest upper limit today is $r<0.032$\footnote{Evaluated at a pivot scale of $0.05 \, \mathrm{Mpc}^{-1}$.} at 95\,\%
confidence, as measured by the combination of \textsc{Bicep2}/\textit{Keck} and
\Planck\ \citep{tristram:2021}. A robust positive detection of $r>0$
would rank among the greatest discoveries in cosmology, providing a
unique signature of ultra-high energy physics almost at the Planck
energy scale. As a reflection of the fundamental importance of such a detection, billions of dollars, euros and yen are currently
being invested in detecting this signal \citep{bp05}.

However, the technical challenges involved in making such a discovery
are massive. For a typical value of $r\sim10^{-3}$, the amplitude of
the $B$-mode polarization signal will not be more than a few tens of
nanokelvins on large angular scales. All sources of systematic errors
must therefore be controlled to unprecedented levels, whether they are
of instrumental or astrophysical origin, and the corresponding
uncertainties must be accurately propagated throughout the entire
analysis process. Underestimating the integrated uncertainty on $r$
by, say, a factor of two could turn an innocent $2.5\,\sigma$ fluke into
a fatal $5\,\sigma$ false claim of new physics.

Most pre-\Planck\ and early \Planck\ CMB analysis pipelines have
effectively relied on systematic errors being relatively small
compared both with the target signal and the noise level of the given
experiment \citep[e.g.,][]{bennett2012,planck2013-p01}. In many cases
it has been an acceptable approximation to account primarily for
(correlated and white) noise uncertainties on the instrument side and
sample and cosmic variance on the CMB side. The impact of
astrophysical foregrounds, whether caused by Milky Way or
extra-galactic sources, has typically been minor, and could often be
accounted for through simple template fitting or internal linear
combination methods
\citep[e.g.,][]{bennett2003b,planck2013-p06}. However, as the
signal-to-noise ratio of a given dataset increases, the relative
importance of systematic errors increases, to the point that these
eventually totally dominate the error budget. A key example of this is
the strong coupling between calibration and astrophysical foregrounds;
since high-sensitivity CMB experiments, such as \Planck, directly
exploit the CMB dipole to estimate their gain, it is key to establish
a robust model of any Galactic foreground that obscure this signal. At
the same time, such a foreground model can only be derived from the same
high-sensitivity dataset, leading to a highly non-linear analysis
problem. For \Planck, this insight eventually led to the development
of highly integrated analysis pipelines \citep{delouis:2019,npipe}
that jointly fit instrumental and astrophysical parameters as part of
the mapmaking process. It is safe to assume that similar integrated
approaches will be even more important for next-generation
inflationary $B$-mode experiments, due to their extreme precision
requirements.

To understand how error propagation may be improved for
next-generation experiments, it is worth noting that two fundamentally
different modes of operations have seen widespread use in the CMB
field until today, corresponding either to the use of simulations or
Bayesian statistics, respectively. In the simulation approach, one
assumes to precisely know the cosmological model, the astrophysical foregrounds, and the instrument, and one derives as realistic time-ordered data
(TOD) simulations as possible of the dataset in question
\citep[e.g.,][]{planck2014-a14}. Each simulation is then processed
with exactly the same algorithms as the real data, and the scatter in
the final quantity is taken as the uncertainty of the point estimate
derived from the data. This mode of operation has traditionally
dominated all lower-level aspects of CMB data processing, including
calibration, mapmaking, and component separation
\citep[e.g,][]{planck2014-a08,planck2016-l04}.

In contrast, the key elements in the Bayesian approach are explicit
models for the data and likelihood in question, and the analysis
process simply amounts to mapping out the corresponding posterior
distribution. In practice, this is typically done using modern Markov
Chain Monte Carlo (MCMC) methods, due to the high dimensionality of
the data model. This approach is typically preferred for the
high-level aspects of the analysis, and in particular for cosmological
parameter estimation \citep[e.g.,][]{cosmomc}. A main reason for this
is that it allows more naturally and efficiently for exploration of
degeneracies between parameters. For \Planck, the Bayesian approach
was used for the final cosmological parameter stage, integrating a
limited instrumental and astrophysical model directly into the
corresponding likelihood \citep{planck2016-l05}, allowing joint
exploration of a few hundred free parameters
\citep{planck2016-l06}. However, most instrumental uncertainties were
still estimated using the low-level simulation approach.

In this paper, we consider error propagation within the context of a
novel end-to-end Bayesian analysis framework called
\BP\ \citep{bp01}. This pipeline is in principle equivalent to the
Bayesian cosmological parameter approach described above, but with one
critical difference: In \BP, the \emph{entire pipeline} is integrated
into the core Monte Carlo sampler \citep{bp01}. As such, the number of
free model parameters is not hundreds, but billions, and there is no
separation between low-level and high-level analysis. Two key
advantages of this global integrated approach are, firstly, joint
exploration of \emph{all} free parameters and, secondly, seamless
end-to-end error propagation. In short, it is the ultimate logical
extension of \Planck's approach of adding a handful of critical
instrumental and astrophysical parameters to the CMB likelihood. It is
also interesting to note that this approach was in fact first
suggested almost 20 years ago by \citet{jewell2004} and
\citet{wandelt2004}, and it took almost two decades of algorithmic and
computer developments before it could be realized in practice.

In this paper, we present CMB results derived from within the
\BP\ pipeline, while a series of companion papers describes individual
instrumental \citep{bp25,bp06,bp07,bp08,bp09} and astrophysical
components \citep{bp13,bp14,bp15}. An important common feature in all
of these papers, however, is the fact that each free parameter is
quantified in the form of a set of \emph{samples} drawn from the joint
posterior distribution. At first glance, these look very similar to
the simulations produced in the traditional low-level approach -- but
they have a fundamentally different statistical interpretation: While
a simulation represents one possible instrument configuration in a
random universe, unconstrained by the actual measurements, a posterior
sample represents one possible instrument configuration in \emph{our}
universe, as constrained by the actual measurements. 

An important consequence of this difference is that the two approaches
have different aspects of the full analysis problem in which they
excel. For questions that may be formulated in terms of numerical
parameter estimates that requires a robust error assessment, for instance
``what is the best-fit value of $r$'', the Bayesian approach is
ideal. For questions that may be formulated in terms of statistical
agreement with a general paradigm, such as ``how likely is the CMB
Cold Spot to appear in a Gaussian and isotropic universe?'', the
simulation-based approach is ideal. That is not to say that either of
the two methods cannot address questions in the other category -- but
they are complementary, and overall better suited to answer different
questions. Going forward, we consider it very likely that most
next-generation experiments will want to implement both pipeline
types, and cross-validate results between them.

In this paper we demonstrate the use of these novel posterior samples
for several classic CMB analysis applications, including CMB dipole
estimation, power spectrum estimation, and low-$\ell$ anomaly studies,
with special attention paid to robust error propagation. However, we
stress that the current \BP\ processing primarily focuses on
\Planck\ LFI data, and in particular does not include
\Planck\ HFI observations in the $100-217$GHz range \citep{bp01}. The results are therefore
significantly less sensitive than the main \Planck\ results in most
respects. In general, the main purpose of the current paper is to
demonstrate the sample-based CMB analysis from an
algorithmic point-of-view, while leaving full integration of
additional state-of-the-art datasets to future work.

The rest of the paper is organized as follows. In Sect.~\ref{sec:bp}
we briefly review the \BP\ data model, and show how CMB samples are
derived within this framework. In Sect.~\ref{sec:degeneracies} we
inspect the raw outputs from the algorithm in the form of posterior
samples, quantify correlations among the various parameters, and
identify one particularly strong degeneracy with respect to free-free
emission. In Sect.~\ref{sec:maps}, we consider posterior mean maps and
power spectra, and compare their properties with those presented by
earlier analyses. In Sect.~\ref{sec:dipole} we present the first fully
Bayesian estimate of the CMB Solar dipole from \Planck\ data, before
we revisit selected low-$\ell$ anomalies in
Sect.~\ref{sec:anomalies}. We conclude in Sect.~\ref{sec:summary}.

\section{\BP\ and end-to-end CMB analysis}
\label{sec:bp}

\subsection{General model and Gibbs sampling scheme}

The starting point of the LFI-oriented Bayesian \BP\ analysis framework is an
explicit parametric model of the time-ordered data of the following
form \citep{bp01},
\begin{align}
    d_{j,t} &= g_{j,t} \P_{tp,j}\left[\B^{\mathrm{symm}}_{pp',j} s^{\mathrm{sky}}_{p',j}  + s^{\mathrm{orb}}_{j,t}  
      + s^{\mathrm{fsl}}_{j,t}\right] 
    + s^{\mathrm{1\,Hz}}_{j,t} + n^{\mathrm{corr}}_{j,t} +
    n^{\mathrm{w}}_{j,t}\\
    &\equiv s^{\mathrm{tot}} + n^{\mathrm{w}}_{j,t}
  \label{eq:todmodel}
\end{align}
In this expression, $j$ is a detector index, $t$ is a time index, $p$
is a sky pixel index. Further, $g$ represents the time-variable
instrumental gain; \P\ is a matrix that describes the satellite
pointing; $\B^{\mathrm{symm}}$ denotes a (assumed azimuthally
symmetric) beam convolution operator; $s^{\mathrm{sky}}$ represents
the total astrophysical sky signal; $s^{\mathrm{orb}}$ is the orbital
CMB dipole; $s^{\mathrm{fsl}}$ are the far sidelobe corrections;
$s^{\mathrm{1\,Hz}}$ represents electronic 1\,Hz spike corrections;
$n^{\mathrm{corr}}$ is the correlated noise; and $n^{\mathrm{w}}$
represents white Gaussian noise. For later notational convenience, the
last line defines all time-ordered data components except the white
noise as $s^{\mathrm{tot}}$.

The total sky signal may be decomposed into individual astrophysical
emission mechanisms, and we assume the following expression in the
current analysis,
\begin{align}
  s^{\mathrm{sky}} &= \left(a^{\mathrm{CMB}}+a^{\mathrm{quad}}(\nu)\right) \frac{x^2 e^x}{(e^x -1)^2}+\label{eq:cmb_astsky}\\
  &+ a^{\mathrm{s}} \left(\frac{\nu}{\nuzeros}\right)^{\bsynch} + \label{eq:synch_astsky}\\
  &+ a^{\mathrm{ff}} \left(\frac{\nuzeroff}{\nu}\right)^2 \frac{g_{\mathrm{ff}}(\nu;\Te) }{g_{\mathrm{ff}}(\nuzeroff;\Te)} +\label{eq:ff_astsky}\\
  &+ a^{\mathrm{ame}} \left(\frac{\nuzeroame}{\nu}\right)^2 \frac{f_{\mathrm{ame}} \left(\nu\cdot \frac{30.0\,\mathrm{GHz}}{\nup}\right)}{f_{\mathrm{ame}} \left(\nuzeroame\cdot \frac{30.0\,\mathrm{GHz}}{\nup}\right)}+ \label{eq:ame_astsky}\\  
  &+ a^{\mathrm{d}} \left(\frac{\nu}{\nuzerod}\right)^{\bdust+1} \frac{e^{h\nuzerod/\kB\Tdust}-1}{e^{h\nu/\kB\Tdust}-1}+ \label{eq:dust_astsky}\\
  &+ U_{\mathrm{mJy}} \sum_{j=1}^{N_{\mathrm{src}}} a_{j,\mathrm{src}} \left(\frac{\nu}{\nuzerosrc}\right)^{\alpha_{j,\mathrm{src}}-2}, \label{eq:ptsrc_astsky}
\end{align}
where $h$ is Planck's constant, $k_\mathrm{B}$ is Boltzmann's
constant, and $x\equiv h\nu/k_{\mathrm{B}}T_{\mathrm{CMB}}$ where
$T_{\mathrm{CMB}}=2.7255\,\mathrm{K}$ is the mean CMB temperature
\citep{fixsen2009}. Each line in this expression represents one
specific astrophysical component, each of which is defined in terms of
an amplitude map, $a$, and a spectral energy density, $f(\nu; \beta)$,
that describe the strength of the component as a function of
frequency, relative to some reference frequency, $\nu_0$, and some set
of free spectral parameters, $\beta$. From top to bottom, the six
lines describe respectively CMB (including a relativistic quadrupole
correction), synchrotron, free-free, AME, and thermal dust emission,
and, finally, a discrete set of point sources. We assume that only
CMB, synchrotron, and thermal dust emission are polarized.  For further
information regarding any of these astrophysical foreground
components, see \citet{bp13} and \citet{bp14}. In practice, each of these terms is
integrated separately with respect to the instrumental bandpass of
each detector, which itself also is associated with a free correction
parameter $\Dbp$, as discussed by \citet{bp09}.

It is convenient to decompose the CMB sky map into spherical
harmonics,
\begin{equation}
a^{\mathrm{CMB}}(\hat{n}) =
\sum_{\ell=0}^{\ell_{\mathrm{max}}}\sum_{m=-\ell}^{\ell} a_{\ell m}
Y_{\ell m}(\hat{n}),
\end{equation}
where $\ell_{\mathrm{max}}$ denotes an harmonic-space bandwidth limit,
and $a_{\ell m}$ are the spherical harmonics coefficients. It is
common to assume that the CMB field is statistically isotropic, in
which case the CMB covariance matrix may be defined as
\begin{equation}
S_{\ell m, \ell' m'}^{\mathrm{CMB}} \equiv \left<a_{\ell m} a_{\ell'
  m'}^* \right> \equiv C_{\ell}\delta_{\ell \ell'} \delta_{mm'},
\label{eq:cmb_cov}
\end{equation}
where the brackets indicate an ensemble average, and $C_{\ell}$ is
called the angular power spectrum. (For simplicity, this notation
applies only to CMB temperature analysis, but the generalization to
polarization is straightforward, and described by
\citealp{zaldarriaga1997}). The angular power spectrum plays a
particularly important role in CMB analysis, as this provides a
computationally efficient path to cosmological parameter estimation
\citep[e.g.,][]{cosmomc}. Estimating the power spectrum distribution
$P(C_{\ell}\mid\d)$, marginalized over all relevant systematic effects,
may in fact be considered the single most important goal of any CMB
analysis pipeline.

Given this parametric signal model, the \BP\ approach to CMB analysis
follows well-established Bayesian methods. That is, let us first
define $\omega \equiv \{a, \beta, g, \Dbp, n^{\mathrm{corr}}, C_{\ell}, \ldots\}$ to be the set of all free parameters in the model;
instrumental, astrophysical and cosmological. By Bayes' theorem, the
joint posterior distribution may then be written as
\begin{equation}
  P(\omega\,\mid\,\d) = \frac{P(\d\,\mid\,\omega)P(\omega)}{P(\d)}
  \propto \mathcal{L}(\omega)P(\omega),
  \label{eq:joint_posterior_full}
\end{equation}
where $\mathcal{L}(\omega) \equiv P(\d\,\mid\,\omega)$ is called the
likelihood, and $P(\omega)$ is a set of user-specified priors. The
likelihood is defined simply by noting that the white noise, which is
equal to $\d-\s^{\mathrm{tot}}$ (Eq. \ref{eq:todmodel}), is assumed to be Gaussian, and 
therefore
\begin{equation}
-2\ln\mathcal{L}(\omega) = \left(\d-\s^{\mathrm{tot}}(\omega)\right)^t\N_{\mathrm{wn}}^{-1}\left(\d-\s^{\mathrm{tot}}(\omega)\right),
\end{equation}
where $\N_{\mathrm{wn}}$ is the white noise covariance matrix. The
prior, $P(\omega)$, is less well defined, and must be specified by the
user. For a summary of the priors adopted in the current analysis, see
\citet{bp01}.

It is important to note that $\omega$ includes a vast number of
parameters with different impact on the final posterior. For instance,
the correlated noise, $n^{\mathrm{corr}}$, contains in principle
billions of degrees of freedom, one for each time sample, but each of
those affect higher-level quantities almost negligibly. Each
astrophysical sky map contains millions of degrees of freedom, each of
which affect the full posterior noticeably. Then there are a handful
of global parameters, for instance the absolute gain and bandpass
corrections, that have a massive impact on almost all other model
parameters. Both the vast number of free parameters and their complex
relationships make it a significant computational challenge to map out
the posterior distribution efficiently. The only computationally
feasible approach suggested to date is Gibbs sampling \citep{geman:1984}, which
allows the user to draw samples from a joint distribution by iterating
over all corresponding conditional distributions. For \BP, this
process may be formally written in terms of the following Gibbs chain,
\begin{alignat}{10}
\g &\,\leftarrow P(\g&\,\mid &\,\d,&\, & &\,\xi_n, &\,\Dbp, &\,\a, &\,\beta, &\,C_{\ell})\label{eq:conditional_gain}\\
\n_{\mathrm{corr}} &\,\leftarrow P(\n_{\mathrm{corr}}&\,\mid &\,\d, &\,\g, &\,&\,\xi_n,
&\,\Dbp, &\,\a, &\,\beta, &\,C_{\ell})\\
\xi_n &\,\leftarrow P(\xi_n&\,\mid &\,\d, &\,\g, &\,\n_{\mathrm{corr}}, &\,
&\,\Dbp, &\,\a, &\,\beta, &\,C_{\ell})\\
\Dbp &\,\leftarrow P(\Dbp&\,\mid &\,\d, &\,\g, &\,\n_{\mathrm{corr}}, &\,\xi_n,
&\,&\,\a, &\,\beta, &\,C_{\ell})\\
\beta &\,\leftarrow P(\beta&\,\mid &\,\d, &\,\g, &\,\n_{\mathrm{corr}}, &\,\xi_n,
&\,\Dbp, & &\,&\,C_{\ell})\label{eq:conditional_beta}\\
\a &\,\leftarrow P(\a&\,\mid &\,\d, &\,\g, &\,\n_{\mathrm{corr}}, &\,\xi_n,
&\,\Dbp, &\,&\,\beta, &\,C_{\ell})\label{eq:conditional_amp}\\
C_{\ell} &\,\leftarrow P(C_{\ell}&\,\mid &\,\d, &\,\g, &\,\n_{\mathrm{corr}}, &\,\xi_n,
	&\,\Dbp, &\,\a, &\,\beta&\,\phantom{C_{\ell}})&,\label{eq:conditional_bp}
\end{alignat}
where $\leftarrow$ indicates sampling from the distribution on the
right hand side. Bayesian CMB analysis as implemented within the
\BP\ framework is nothing but repeated sampling from each of these
distributions, and the main product from this process is a discrete
set of samples drawn from the joint posterior distribution,
$P(\omega\mid\d)$, which naturally and seamlessly allows for detailed
instrumental systematics and astrophysical foreground
marginalization.

CMB sky map and power spectrum estimation are accounted for in the
above Gibbs loop in Eqs.~\eqref{eq:conditional_amp} and
\eqref{eq:conditional_bp}, respectively, and explicit expressions for
these were first derived by \citet{jewell2004} and
\citet{wandelt2004}. All other steps describe either instrumental or
astrophysical effects, and only affect the CMB estimates indirectly.
Sampling algorithms for each of those distributions are described in
detail in \citet{bp01} and references therein, and in the following we
only briefly review the sampling algorithms for
Eqs.~\eqref{eq:conditional_amp} and \eqref{eq:conditional_bp}.

\subsection{CMB sky map sampling, $P(a^{\mathrm{CMB}}\mid\d, g, \beta,
  C_{\ell},  \ldots)$}
\label{sec:amp_sampling}

To derive an appropriate sampling algorithm for
$P(a^{\mathrm{CMB}}\mid\d, g, \beta, \ldots)$, we start with the general
data model defined in Eqs.~\eqref{eq:todmodel}--\eqref{eq:cmb_astsky},
and aim to isolate the component amplitude $\a$ parameter. In
principle, we could even isolate the $\a^{\mathrm{CMB}}$ parameter
alone, but since it generally leads to a shorter Monte Carlo
correlation length to sample partially degenerate components jointly,
we will derive a joint sampling step for both CMB and astrophysical
foreground amplitudes; see \citet{bp13} for further details.

The first step in the algorithm corresponds essentially to
mapmaking. Because all instrumental parameters are conditioned on in
this distribution, we may deterministically form the following residual,
\begin{align}
  r_{j,t} &\equiv \frac{d_{j,t} - (s^{\mathrm{1\,Hz}}_{j,t} +
    n^{\mathrm{corr}}_{j,t})}{g_{j,t}} - \P_{tp,j}\left[s^{\mathrm{orb}}_{j,t} + s^{\mathrm{fsl}}_{j,t}\right] \\
  &= \P_{tp,j}\B_{pp',j} s^{\mathrm{sky}}_{p',j}  +  n^{\mathrm{w}}_{j,t}/g_{j,t},
  \label{eq:resmodel}
\end{align}
which now represents TOD that contain only astrophysical signal signal
convolved with an \citep[implicitly assumed azimuthally symmetric, see][for an in depth discussion]{bp01} beam and
white noise, all in calibrated brightness temperature units. This
residual may be compressed nearly losslessly into a pixelized sky map,
$m_{\nu}$, by solving the so-called mapmaking equation,
\begin{equation}
  \left(\sum_{j}\P_j^t\N^{-1}_{j,\mathrm{wn}}\P_j\right)\m_{\nu} =
  \sum_j \P_{j}^t\N_{j,\mathrm{wn}}^{-1}\r_{j}.
  \label{eq:mapmaking}
\end{equation}
For \Planck, this equation may be solved pixel-by-pixel, and it is
therefore computationally very fast.

The second step in the algorithm corresponds essentially to component
separation. Given the above frequency maps, the data model in
Eqs.~\eqref{eq:todmodel}--\eqref{eq:cmb_astsky} may now be rewritten
compactly in terms of sky maps,
\begin{equation}
  \m_{\nu} = \B^{\mathrm{symm}}_{\nu}\M_{\nu,c}\a + \n^{\mathrm{wn}}_{\nu},
  \label{eq:mapmodel}
\end{equation}
where $\M_{\nu,c}$ is called the mixing matrix, and encodes the
bandpass-integrated SEDs for the various astrophysical components in
each column; when multiplied by the amplitude vector, this matrix
generates the full sky signal at frequency $\nu$ in the appropriate
units for that channel.

It is now straightforward to sample $\a$, again based on the
observation that the white noise component is Gaussian, and therefore
that also $\m_{\nu}-\B^{\mathrm{symm}}_{\nu}\M_{\nu,c}\a$ is Gaussian
with the same covariance. The necessary sampling equation for this
step is therefore structurally identical to the mapmaking equation in
Eq.~\eqref{eq:mapmaking}, except that it has an additional fluctuation
term in order to propagate noise uncertainties,
\begin{equation}
  \left(\sum_{\nu}\B_\nu^t\M_\nu^t\N^{-1}_{\nu,\mathrm{wn}}\M_\nu\B_\nu\right)\a
  = \sum_\nu \B_{\nu}^t\M_{\nu}^t\N_{\nu,\mathrm{wn}}^{-1}\m_{\nu} +
  \sum_{\nu}
  \B_{\nu}^t\M_{\nu}^t\N_{\nu,\mathrm{wn}}^{-\frac{1}{2}}\eta_{\nu},
  \label{eq:compsep}
\end{equation}
where $\eta_{\nu}$ is a random vector of $N(0,1)$ stochastic variates;
for a full derivation of this equation, see Appendix~A in
\citet{bp01}. A computationally efficient Conjugate Gradient (CG)
solver for this equation was presented by \citet{seljebotn:2019}.

Equation~\eqref{eq:compsep} does not account for priors on $\a$. We
support Gaussian priors in our analyses, as defined in terms of some
mean map, $\mu$, and a corresponding prior covariance matrix,
$\S$. The purpose of this prior is two-fold; firstly, for the CMB
component it directly defines the connection to the angular power
spectrum and cosmological parameters, as described by the CMB
covariance matrix in Eq.~\eqref{eq:cmb_cov}. Secondly, for
astrophysical foregrounds it both allows us to introduce useful
information in the form of prior knowledge from other datasets to
break particularly difficult degeneracies, and it allows us to impose
smoothness on small angular scales. With such a Gaussian prior in
place, the full sampling equation for $\a$ reads
\begin{equation}
  \begin{split}
  \left(\S^{-1} + \sum_{\nu}\B_\nu^t\M_\nu^t\N^{-1}_{\nu,\mathrm{wn}}\M_\nu\B_\nu\right)&\a
  = \sum_\nu \B_{\nu}^t\M_{\nu}^t\N_{\nu,\mathrm{wn}}^{-1}\m_{\nu} +\\
  +\S^{-1}\mu &+
  \sum_{\nu}
  \B_{\nu}^t\M_{\nu}^t\N_{\nu,\mathrm{wn}}^{-\frac{1}{2}}\eta_{\nu} +
  \S^{-\frac{1}{2}}\eta_0.
  \end{split}
  \label{eq:compsep_with_prior}
\end{equation}
For a derivation of this expression, see Appendix~A in \citet{bp01},
and for a detailed discussion of foreground priors in \BP, see
\citet{bp13}.

It is worth noting that all of the above equations are general in
terms of basis sets, and apply equally well to objects defined in
terms of pixels or spherical harmonics or any other complete basis on
the sphere. In practice, our current codes model all diffuse
components in terms of spherical harmonics up to some band limit
$\ell_{\mathrm{max}}$. The main reason for this choice is simply that
harmonics are more efficient in terms of the number of free parameters
than pixels; for a HEALPix\footnote{\url{https://healpix.jpl.nasa.gov}} map \citep{gorski2005} with a given $N_{\mathrm{side}}$
resolution parameter, there are $12\,N_{\mathrm{side}}^2$ pixels,
while for a typical maximum band limit of
$\ell_{\mathrm{max}}=3\,N_{\mathrm{side}}$, there are only
$(\ell_{\mathrm{max}}+1)^2 \approx 9\,N_{\mathrm{side}}^2$ spherical
harmonic coefficients. In addition, it is easier to impose additional
smoothness on a given foreground component in harmonic space, simply
by reducing $\ell_{\mathrm{max}}$ for that component. We emphasize,
however, that this is only a practical choice, not a fundamental one;
the algorithm works equally well with both bases.

\subsection{Angular power spectrum sampling, $P(C_{\ell}\mid\d, \a, \ldots)$}
\label{sec:cl_sampling}

Next, we need to derive a sampling algorithm for the power spectrum
distribution, $P(C_{\ell}\mid\d, \a, \ldots)$. This was first
presented by \citet{wandelt2004}, and we will only briefly review the
main steps in the following.

Firstly, we make the trivial observation that ${P(C_{\ell}\mid\d, \a,
\ldots) = P(C_{\ell}\mid\a^{\mathrm{CMB}})}$; if we already know the
CMB map, $\a^{\mathrm{CMB}}$, with infinite precision, no further
instrumental or astrophysical knowledge can possibly provide more
information regarding the CMB power spectrum. Secondly, in the
previous sampling step we assumed only that the CMB SED is defined by
a blackbody spectrum; in this step we additionally assume that the CMB
is statistically isotropic, i.e., that its harmonic space covariance
matrix is diagonal and given by $C_{\ell}$, and that it is Gaussian
distributed. With these additional assumptions, the relevant
distribution may be written as follows,
\begin{align}
  P(C_{\ell}\mid\a) &\propto
  \frac{e^{-\frac{1}{2}\a^t\S^{-1}\a}}{\sqrt{|\S|}}\\
  &=\frac{e^{-\frac{1}{2}\sum_{\ell} \frac{|a_{\ell
          m}|^2}{C_{\ell}}}}{\prod_{\ell}
    C_{\ell}^{\frac{2\ell+1}{2}}}\\
&= \prod_{\ell}
  \frac{e^{-\frac{2\ell+1}{2}\frac{\sigma_{\ell}}{C_{\ell}}}}{C_{\ell}^{\frac{2\ell+1}{2}}},
  \label{eq:invgamma}
\end{align}
where $\sigma_{\ell} = \frac{1}{2\ell+1}|a_{\ell m}|^2$ is the
observed power spectrum of our specific universe.

The distribution inside the product in Eq.~\eqref{eq:invgamma} is
called an inverse Gamma distribution with $2\ell+1$ degrees of
freedom. Its multivariate generalization of this, needed for
polarization analysis, is called the inverse Wishart distribution
\citep{larson:2006}. Sampling from an inverse Gamma distribution is
trivial; simply draw $2\ell-1$ independent Gaussian random variates,
$\eta_i$, and set $C_{\ell} = \sigma_{\ell}/\sum_{i}\eta_{i}^2$
\citep{wandelt2004}.

Unfortunately, as discussed by \citet{eriksen:2004}, this strict Gibbs
sampling algorithm for $\a$ and $C_{\ell}$ has a significant drawback
in the low signal-to-noise regime, namely that the Monte Carlo step
size between two consecutive $C_{\ell}$ samples is determined by
cosmic variance alone, while the full posterior width is defined by
both cosmic variance, sample variance (i.e., masking), and
instrumental noise. In practice, this algorithm therefore has a
prohibitively long correlation length at high multipoles. This problem
was addressed and solved by \citet{jewell:2009} and
\citet{racine:2016}, who proposed a joint sampling step for
$\{\a,C_{\ell}\}$ that moves quickly in the low signal-to-noise
regime. Unfortunately, this step has not yet been fully implemented in
the latest version of the \commander\ code \citep{bp03}, and it is
therefore not used in the \BP\ processing. This work is, however,
on-going, and will be available in the near future for other
projects. An immediate result of this, however, is that, in the following,  we will only present a \BP\ temperature power spectrum up to
$\ell_{\mathrm{max}}=600$, while higher multipoles will, when needed, be
taken from the official \Planck\ processing \citep{planck2016-l05}. 

\subsection{\BP\ data selection}

As described by \citet{bp01}, the \BP\ program has two main goals. The
first goal is to implement and demonstrate the world's first
end-to-end Bayesian sampling algorithm for CMB observations. The
second goal is to resolve a number of outstanding questions regarding
the \Planck\ LFI data that remained after the conclusion of the
official \Planck\ consortium. For both of these reasons, the
\BP\ processing includes significantly less data than if the primary
goal had been to establish a new state-of-the-art sky model and CMB
sky map. Explicitly, we only include \Planck\ LFI 30, 44, and 70\,GHz
data in the time-domain, which are the main target of the entire analysis;
\WMAP\ \textit{Ka}, \textit Q, and \textit V data to constrain low-frequency foregrounds and
poorly observed \Planck\ modes; Haslam 408\,MHz measurements to
constrain synchrotron emission; and \Planck\ PR4 measurements at 857\,GHz (in
temperature) and 353\,GHz (in polarization) to constrain thermal dust
emission.

Critically, the CMB dominated HFI and the \WMAP\ \textit K-band data are
\emph{not} included: Even though they would clearly result in a better
and less degenerate sky model, they would also obscure the impact of
the new algorithm because of their high signal-to-noise ratios, and
they could also potentially introduce unmodelled systematic errors
into the LFI results. Instead, gradual integration of these datasets
falls within the scope of the
\cosmoglobe\footnote{\url{https://cosmoglobe.uio.no}} framework
\citep{bp05,bp17}, which aims to apply these methods to a broad range
of state-of-the-art datasets in the field.

\subsection{Masking, degeneracies, and resampling}
\label{sec:resampling}

\begin{figure}[t]
  \center
  \includegraphics[width=\linewidth]{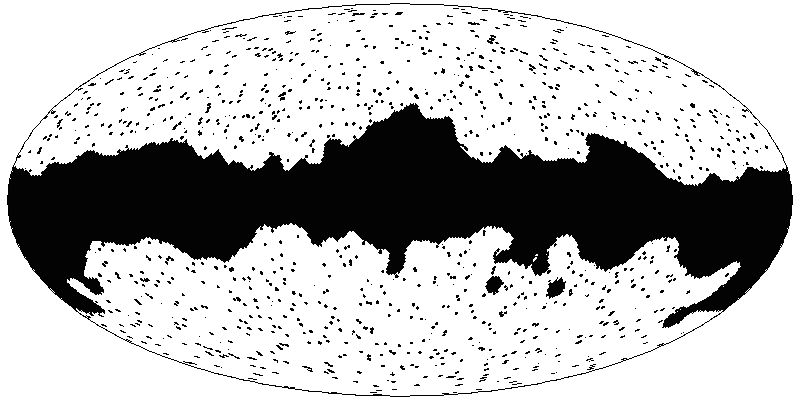}\\
  \includegraphics[width=\linewidth]{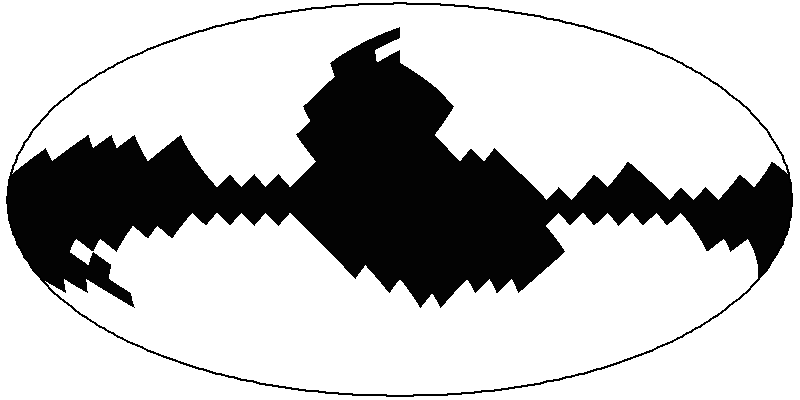}
  \caption{Temperature (\emph{top}) and polarization (\emph{bottom}) confidence
    masks used for \BP\ CMB analysis. The mask allow, respectively, for 
    a sky fraction of 69\,\% and 68\,\%.}
  \label{fig:confmasks}
\end{figure}

For an ideal dataset and a well constrained model, the above algorithm
could in principle be applied without additional
modifications. However, for real-world data there are several
challenges that must be addressed. The first of these is masking:
Despite the notable complexity of the astrophysical data model
described by Eqs.~\eqref{eq:cmb_astsky}--\eqref{eq:ptsrc_astsky}, this
is by no means adequate to model the actual sky to the statistical
precision of the \Planck\ data. As a result, we have to remove parts
of the sky, in particular the Galactic plane and bright point sources,
before actually estimating the CMB power spectrum.

\begin{figure}[t]
  \center
  \includegraphics[width=\linewidth]{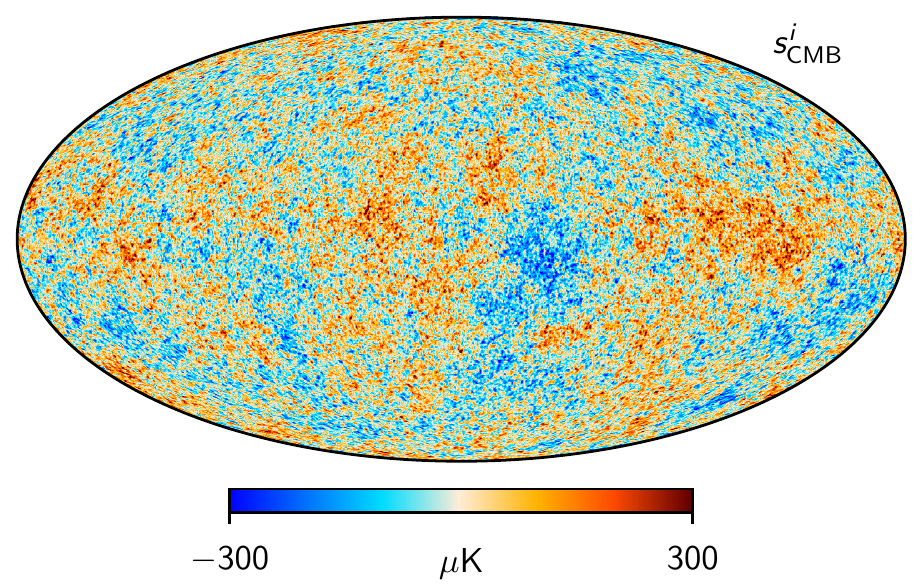}\\
  \includegraphics[width=\linewidth]{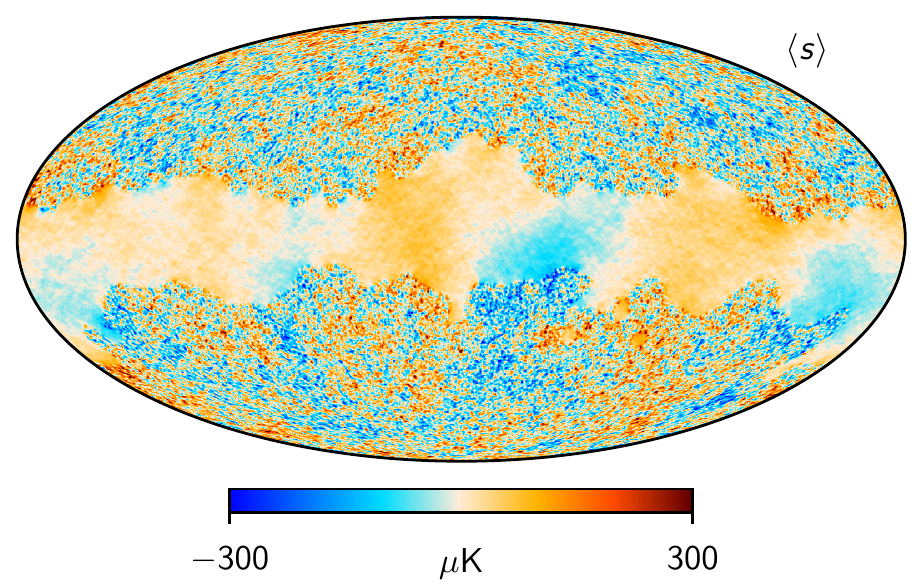}\\
  \includegraphics[width=\linewidth]{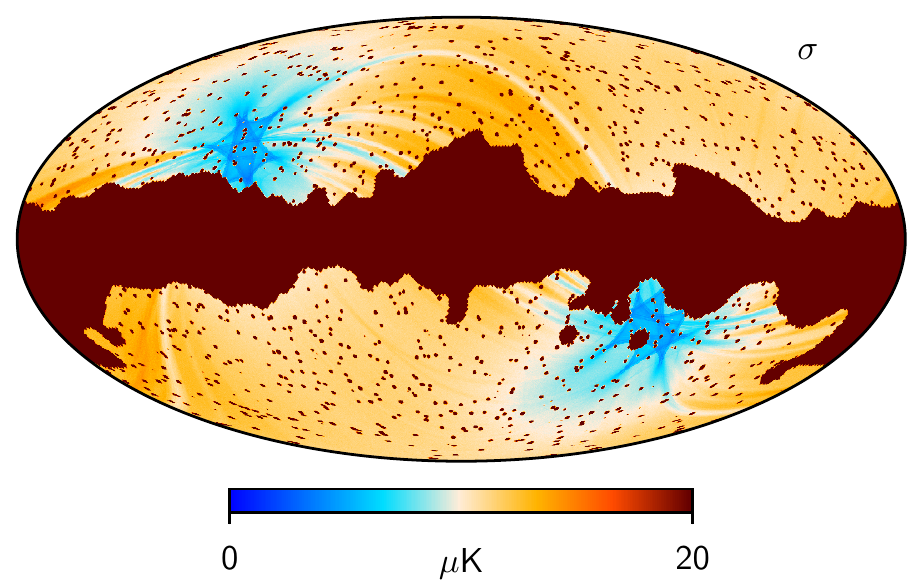}
  \caption{Full-resolution CMB temperature constrained realization
    maps. (\emph{Top}:) Single constrained realization, $\s^i$, drawn
    from $P(\s\mid \d, C_{\ell}, \ldots)$. (\emph{Middle}:) Posterior mean
    map, $\left<\s\right>$, as evaluated from the ensemble of
    constrained CMB realizations; note that the small-scale signal
    amplitude inside the mask decreases smoothly to zero with
    increasing distance from the edge of the mask. (\emph{Bottom}:) CMB posterior
    standard deviation map, as evaluated pixel-by-pixel from the
    ensemble of constrained CMB realizations. This map is dominated by
    instrumental noise outside the mask, and by random fluctuations
    informed by the assumptions of isotropy and Gaussianity inside the
    mask. The CMB Solar dipole has been removed from the top two panels.}\label{fig:cmbmaps_highl}
\end{figure}

The CMB confidence mask used for the current \BP\ processing is shown
in Fig.~\ref{fig:confmasks}, and is generated in a two-step
process. First, we compute data-minus-signal residual maps for each
CMB-dominated frequency. These are smoothed to $1^{\circ}$ angular
resolution, and thresholded in absolute amplitude. These maps serve a
similar purpose as absolute goodness-of-fit tracers as the total
$\chi^2$ map that was used to define the \commander\ confidence mask
in for instance the \Planck\ 2018 analysis. However, the total $\chi^2$ does not provide information on the quality of the individual components, but only on the capability of the model to describe the full set of frequencies. By considering only residual maps for the CMB-dominated frequencies, we instead exclude potential modeling issues that affect only foreground reconstruction but are irrelevant for CMB estimation. These partial single-frequency masks
are multiplied together to form an overall confidence mask. 

\begin{figure*}
  \center	
  \includegraphics[width=0.49\linewidth]{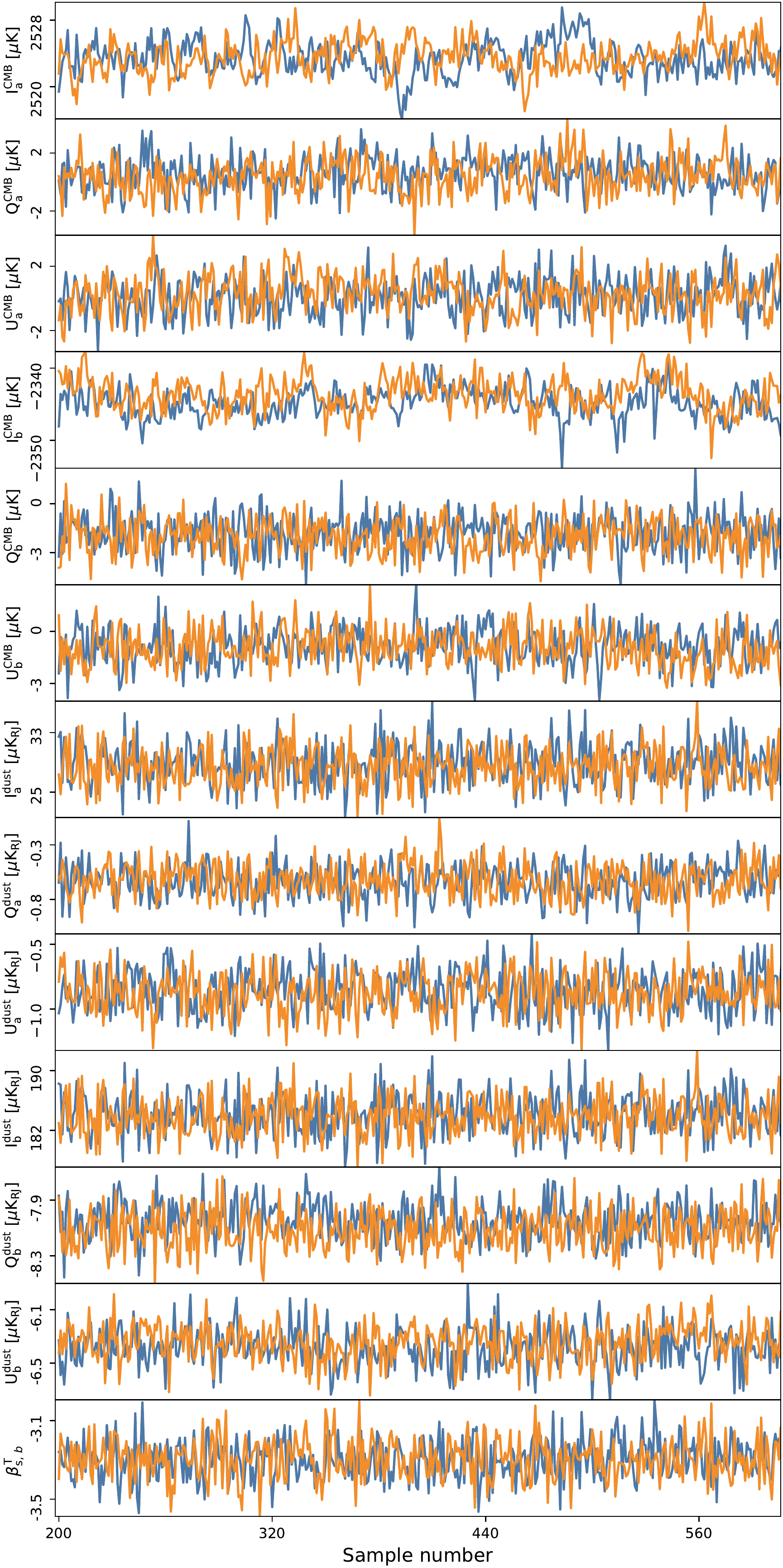}
  \includegraphics[width=0.49\linewidth]{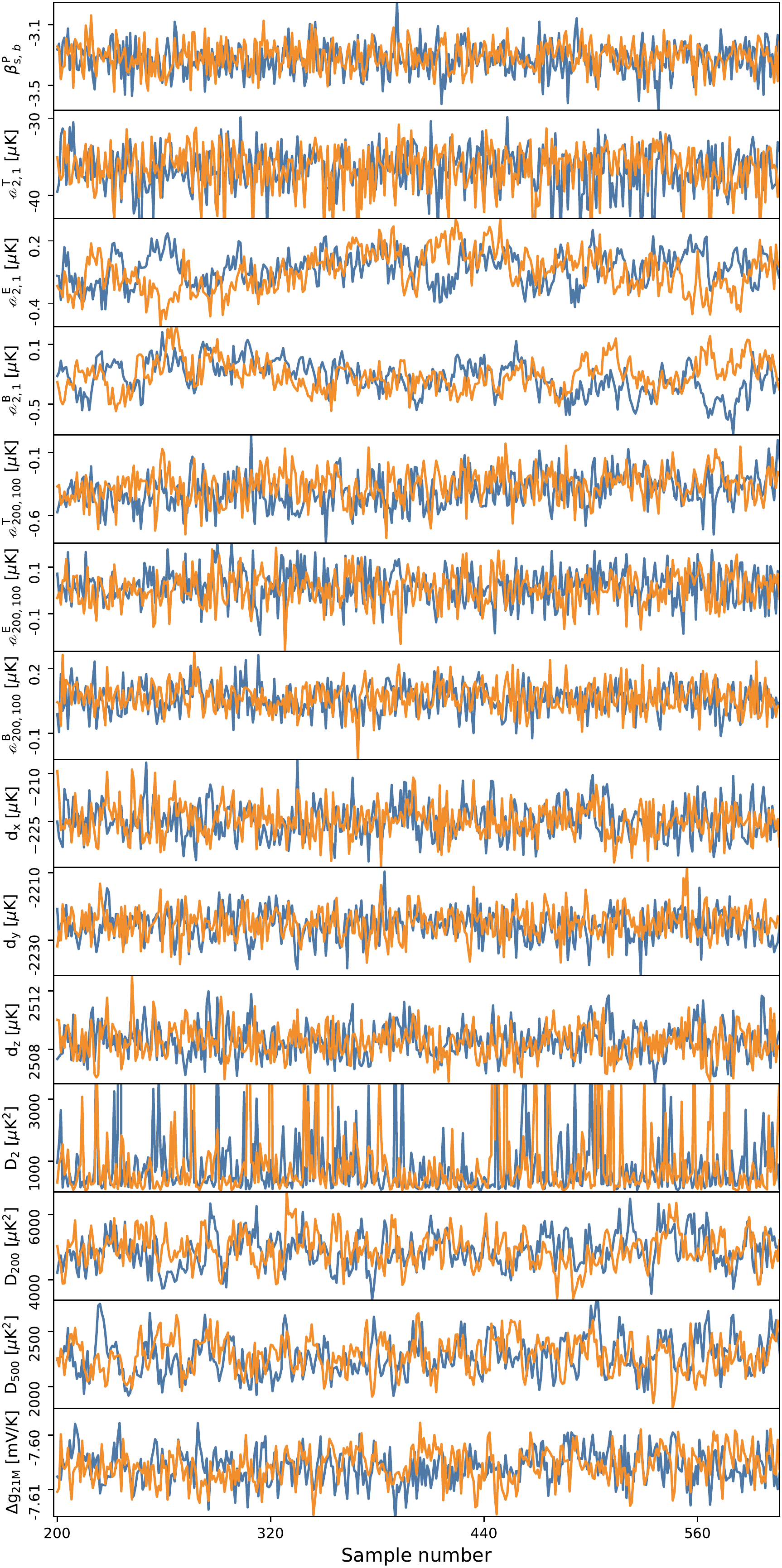}  
  \caption{Trace plots of a set of selected CMB, component separation,
    and instrument parameters; see main text for full definitions. The
    different colors indicate independent Gibbs chains, and 'a' and 'b'
    subscripts indicate HEALPix pixel numbers '340' and '1960' at resolution
    $N_{\mathrm{side}}=16$ in ring ordering, respectively. 
    }
  \label{fig:param_traceplot}
\end{figure*}

The second mask generation step accounts for the resampling procedure
after excluding free-free emission from the model. As discussed in greater detail in Sec.~\ref{sec:degeneracies} below, sample fluctuations in CMB maps are correlated with those in the free-free component at intermediate scales. Intuitively, we do 
not trust any pixel for which the CMB map is significantly different
depending on whether free-free emission is modelled or not. This idea
is implemented in practice by generating a single resampled
constrained realization; computing the difference between this
constrained realization (\emph{without} free-free emission in the
model) and the corresponding CMB map from the main Gibbs analysis
(\emph{with} free-free emission in the model); computing the absolute
value, and smoothing to $4^{\circ}$ FWHM; and exclude all pixels above
a $10\:\mu\rm{K}$ threshold, corresponding to a $\sim 3\sigma$ fluctuation for the difference map defined above. The resulting mask is median filtered
with a $4\;\rm{deg}$ radius to exclude isolated ``islands'' inside the
Galactic plane, and finally we exclude point sources using the
\Planck\ \lfi\ template point source mask. For polarization, we adopt
the same \Planck\ \lfi\ set of masks described in
\citet{planck2016-l05}, and adopt the R1.8 (with
$f_{\mathrm{sky}}=0.68$) for the polarization cosmological analysis;
see \cite{bp12} for further discussion.

Formally speaking, applying a confidence mask in the sampling
algorithms described in Sects.~\ref{sec:amp_sampling} and
\ref{sec:cl_sampling} is trivial; one simply sets the masked pixels in
the inverse frequency covariance matrix, $\N_{\nu}^{-1}$ to zero, and
thereby assign the removed pixels infinite noise. In practice,
however, this also carries a high computational cost for solving
Eq.~\eqref{eq:compsep} by CG, as it massively increases the condition
number of the coefficient matrix on the left-hand side
\citep{seljebotn:2019}. At the same time, the Galactic plane region is
critically important to estimate other parameters in the full data
model, for instance the bandpass corrections \citep{bp09}, and simply
excluding these regions entirely from the analysis is therefore both
computationally expensive and wasteful in terms of throwing away
useful information.

A second complication regards degeneracies between the various
astrophysical components on small angular scales. As discussed by
\citet{bp13}, the \BP\ dataset (comprising \Planck\ LFI, \WMAP, two
HFI channels, and Haslam 408\,MHz observations) simply is not able to
robustly constrain the astrophysical model on its own on multipoles
above $\ell\gtrsim 300$; on these scales, the LFI 30\,GHz and
\WMAP\ \textit{Ka}-band beams start to drop off exponentially, and their
effective signal-to-noise ratio falls quickly. Leaving only
intermediate frequency to constrain the model, one observes a very
strong degeneracy between CMB, AME, and free-free emission. To solve
this problem, \citet{bp13} introduce informative Gaussian priors on
the free-free and AME components, effectively using information from
\Planck\ HFI to constrain the spatial morphology of these components
on small angular scales. The impact of these priors on the CMB
component are explored in Sect.~\ref{sec:degeneracies} in this paper.

To simultaneously mitigate both the masking-induced computational
expense and the degeneracy challenges, we introduce two small
but important additions to the Gibbs chain described in
Eqs.~\eqref{eq:conditional_gain}--\eqref{eq:conditional_bp} that we
refer to as ``resampling''. The first step of this process is to run
the algorithm as described in the previous sections, but without
imposing either a confidence mask or the Gaussian prior on the CMB
component. The outputs from this process are thus full-sky CMB and
astrophysical component maps, together with a full characterization of
the various instrumental parameters. These preliminary CMB maps are,
however, not suitable for high-precision temperature-based power
spectrum and cosmological parameter analysis because of unmasked
foreground residuals and the free-free degeneracy on intermediate
scales discussed above; they can, however, be used for large-angle polarization
analysis, as free-free emission is not expected to be significantly polarized.

To establish CMB intensity maps that actually can be used for
cosmological analysis, we \emph{resample} the original chain. That is,
for each sample in the main chain, we resample the CMB component while
\emph{conditioning} on the instrumental and non-linear astrophysical
parameters derived in the first main sampling phase. During this
process, we make two important changes to the data model: We first
apply the confidence mask, as defined above, to suppress the majority
of the residual foreground contamination. Secondly, we remove the
free-free component in its entirety from the model, leaving only
synchrotron, AME, thermal dust, and point sources to account for any
non-masked signal at the unmasked high Galactic latitudes. Since
free-free emission is generally more localized on the sky than
synchrotron or thermal dust emission \citep{planck2014-a10,bp13}, it
is possible to eliminate most of this signal by masking. On the other
hand, the confidence mask does have to be considerably larger than if
free-free emission had been explicitly modelled, and this is the main
reason that our temperature confidence mask, as defined above, has a
relatively low accepted sky fraction of only
$f_{\mathrm{sky}}=0.64$. To account for possible unmasked residual
free-free emission at high latitudes, we also resample the AME
component amplitude jointly with the CMB component, such that the
resulting AME component at this stage in reality becomes an
``AME-plus-free-free'' component. This is conceptually similar to the
single ``low-frequency foreground'' component used in the
\Planck\ 2018 \commander\ analysis, except that in that case also
synchrotron emission was included. We note, however, that this
``AME-plus-free-free'' component is never used in any further
analysis, but is only a pure phenomenological nuisance parameter as
far as the CMB component is concerned. A single resampled CMB
constrained realization sample is shown in the top panel of
Fig.~\ref{fig:cmbmaps_highl}. The middle panel shows the corresponding
Wiener filter solution alone, in which structures within the Galactic
plane mask may be partially reconstructed due to the assumptions
of statistical isotropy and Gaussianity. The bottom panel shows the
posterior standard deviation in each pixel.

\begin{figure*}
  \center	
  \includegraphics[width=0.9\linewidth]{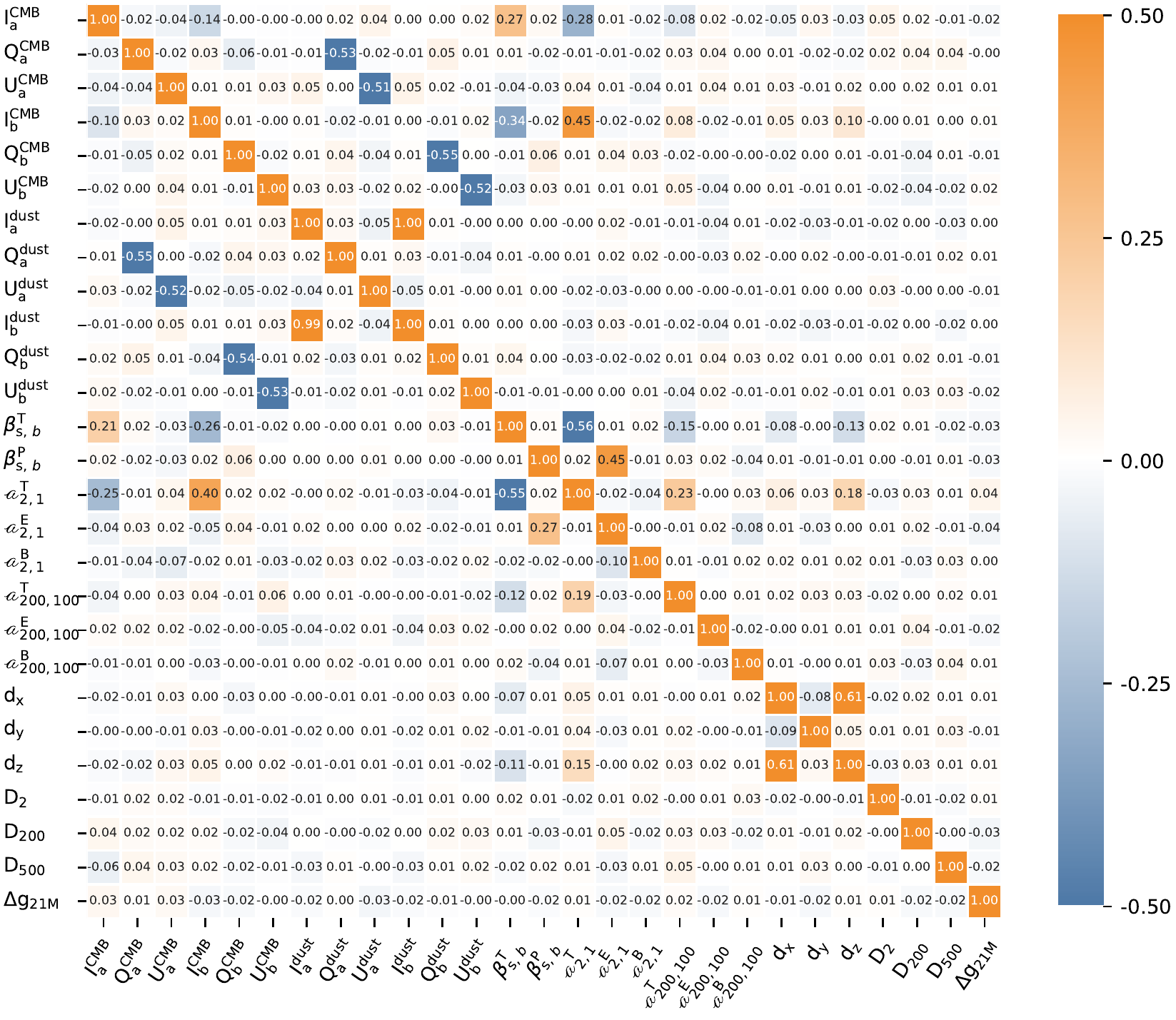}
  \caption{Correlation coefficients between the same parameters as
    shown in Fig.~\ref{fig:param_traceplot}. The lower triangle shows
    raw correlations, while the upper triangle shows correlations
    after high-pass filtering with a running mean with a 10-sample
    window. For further explanation of and motivation for this
    filtering, see \citet{bp13}.  }
  \label{fig:param_corr_local}
\end{figure*}

\begin{figure}[t]
  \center
  \includegraphics[width=\linewidth]{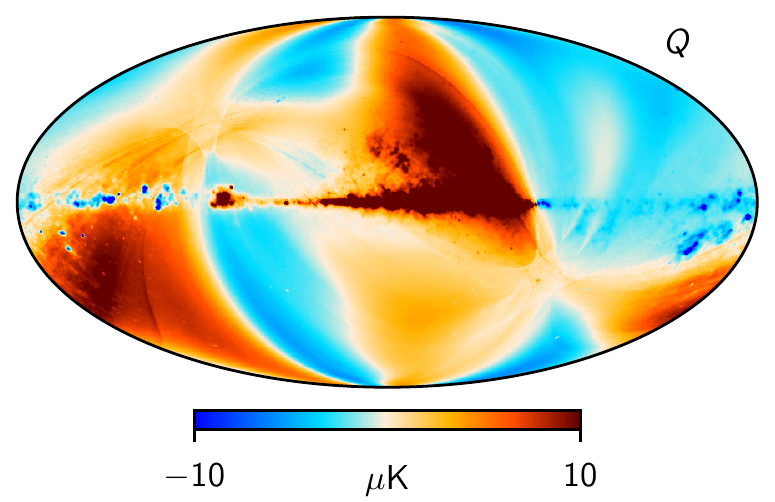}\\
  \includegraphics[width=\linewidth]{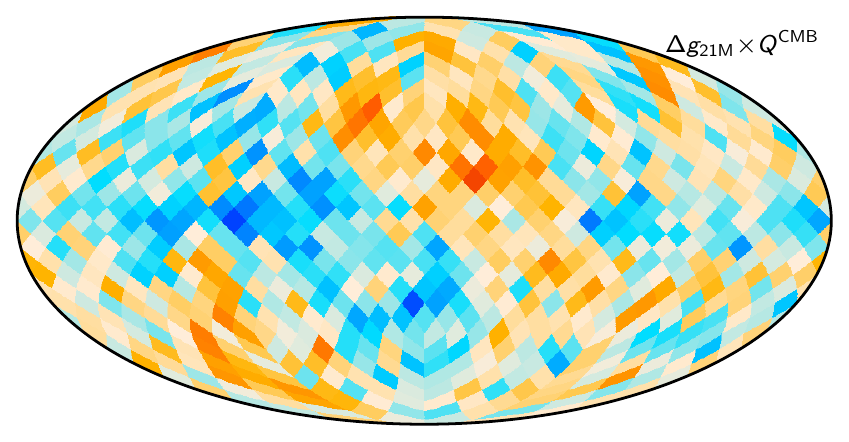}\\
  \includegraphics[width=\linewidth]{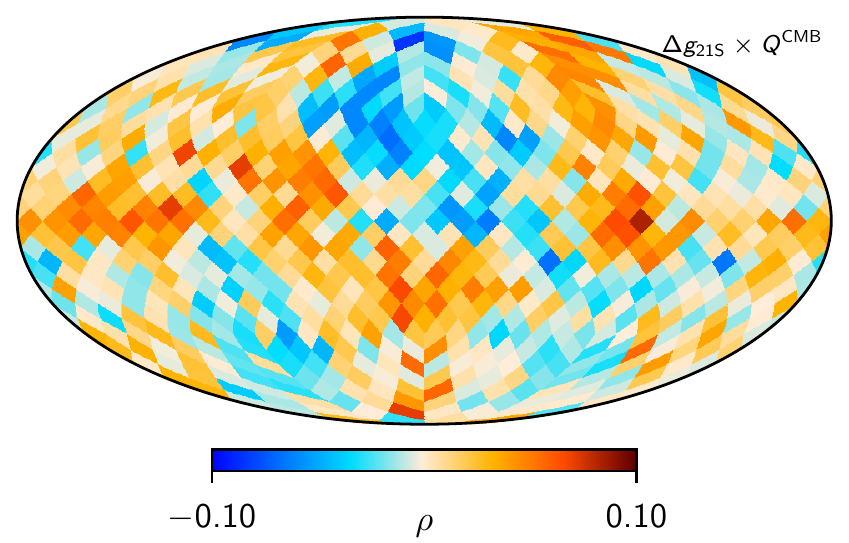}
  \caption{(\emph{Top panel:}) LFI DPC 30\,GHz Stokes $Q$ gain residual template \citep{planck2016-l02}. (\emph{Lower two panels:})  Pixel-by-pixel cross-correlation coefficients between CMB
    Stokes $Q$ and the time-independent absolute gain fluctuation of two
    70\,GHz radiometers, $\Delta g_{21\mathrm{M}}$ (\emph{middle panel}) and $\Delta g_{21\mathrm{S}}$ (\emph{bottom panel}).  }
  \label{fig:crosscorr_map}
\end{figure}

Finally, we also perform an extra resampling step for the CMB
polarization analysis. In this case, we once again condition on the
instrumental and astrophysical parameters from each sample in the main
Gibbs chain, but in this case we perform $N=50$ additional amplitude
sampling steps for each main sample, as defined by
Eq.~\eqref{eq:conditional_amp}, each of which is computationally much
cheaper than a full sample. This resampling step thus involves no
fundamental modifications of the either the algorithm or data model as
such, but is just a computationally convenient way of marginalizing
over white noise, and thereby converging faster at a modest additional
computational cost.


\begin{figure*}
  \includegraphics[width=0.49\linewidth]{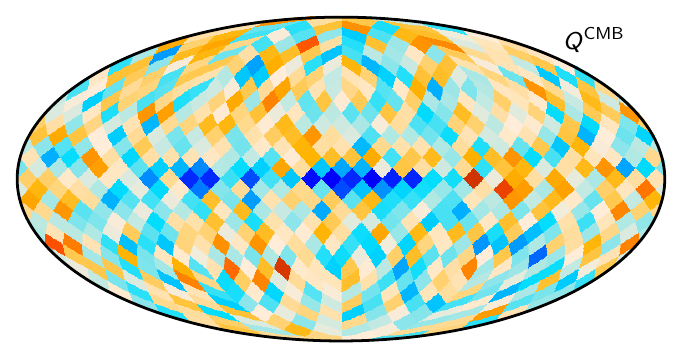}
  \includegraphics[width=0.49\linewidth]{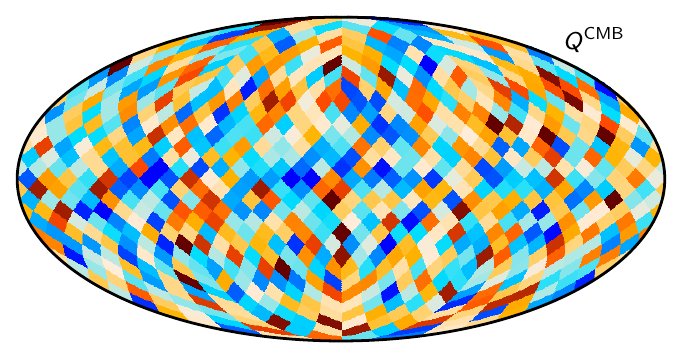}\\
  \includegraphics[width=0.49\linewidth]{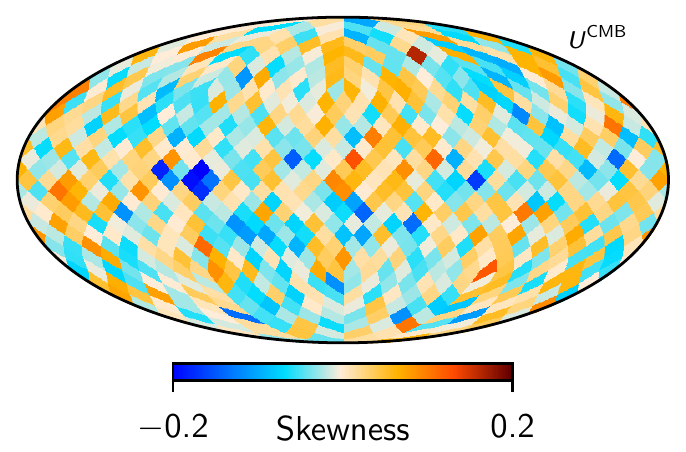}
  \includegraphics[width=0.49\linewidth]{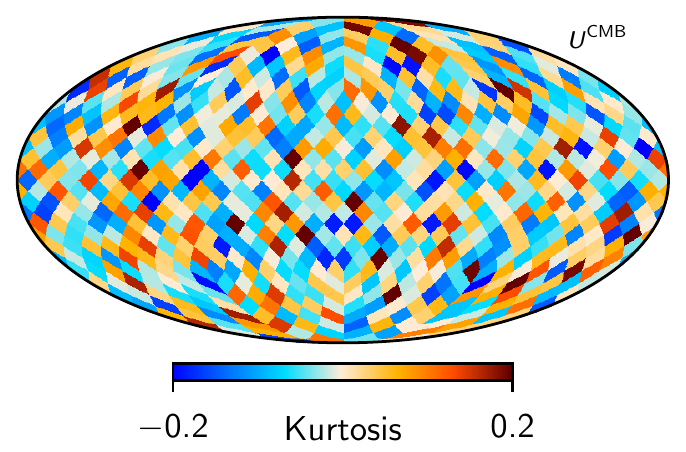}
  \caption{Skewness (\emph{left column}) and kurtosis (\emph{right
      column}) of the CMB Stokes $Q$ (\emph{top row}) and $U$
    (\emph{bottom row}) posterior distributions, evaluated
    pixel-by-pixel at a HEALPix resolution of $N_{\mathrm{side}}=8$.}
  \label{fig:nongauss}  
\end{figure*}

\section{Markov chains and correlations}
\label{sec:degeneracies}

The full \BP\ Gibbs sampler and data configuration are summarized in
\citet{bp01}. The main products from this process are a set of 4000
end-to-end samples evenly distributed over four chains. The first 200 samples
in each chain are conservatively rejected as burn-in, although we have
not identified strong evidence for non-stationary behaviour after the
first few tens of samples. A total of 3200 main Gibbs samples are
retained for science exploitation, and we produce one resampled
high-$\ell$ temperature sample and 50 low-$\ell$ polarization samples
per main Gibbs sample. The total computational cost of the full
analysis is about 800\,kCPU-h \citep{bp03}. The full sample set is
made publicly available through the
\cosmoglobe\footnote{\url{https://cosmoglobe.uio.no}} web page.

Figure~\ref{fig:param_traceplot} shows a collection of trace plots,
i.e., parameter values plotted as a function of chain iteration, for
various CMB and selected ancillary parameters. The quantities marked
with subscripts '$a$' and '$b$' represent sky map pixel values for 
pixel number 340 and 1960, respectively, in maps downgraded by straight averaging to a HEALPix resolution of $N_{\mathrm{side}}=16$ with \texttt{ring}
ordering. Pixel 340 is located in the top right quadrant at high
Galactic latitudes, while pixel 1960 is located near the Southern
center of the Galactic mask edge. From top to
bottom and left to right, the plotted quantities are the three Stokes parameters for CMB
pixels 340 and 1960; the same for thermal dust emission; the
synchrotron spectral index for pixel 1960 in temperature and
polarization; the CMB quadrupole spherical harmonic coefficient
$a_{21}$ for $T$, $E$, and $B$; the same for $a_{200,100}$; the three
components of the CMB dipole in Cartesian coordinates; the CMB angular
temperature power spectrum, $D_{\ell}$, for $\ell=2$, 200, and 500;
and the time-independent radiometer gain fluctuation for the 70\,GHz
21M radiometer, $\Delta g_{21\mathrm{M}}$. Of course, these represent
only 26 parameters out of billions, but they still convey some useful
intuition regarding the overall behaviour of the Gibbs chain as far as
the CMB component is concerned.

The first immediate conclusion that can be drawn from these plots at a
visual level is that the overall correlation length is relatively
short, and the Markov chain mixing is reasonable. Furthermore, all
chains appear stationary, suggesting that the burn-in samples have
been successfully removed. Going into slightly deeper details, we see
that while $a_{21}^{T}$ appears significantly non-Gaussian, with a
pronounced negative tail, $I_{\mathrm{CMB}}$ looks more Gaussian and
symmetric, although with a longer correlation length. As a result,
uncertainties and covariances at low multipoles are generally easier
to summarize in pixel space than in harmonic space. Regarding the
power spectrum coefficients, $D_{\ell}$, we note that these are not
Gaussian distributed at all, but rather follow an inverse gamma (or
inverse Wishart) distribution, which has a very heavy tail toward
positive values at low multipoles. This behaviour is clearly seen for
$D_{2}^{TT}$. 

In Fig.~\ref{fig:param_corr_local}, we show the corresponding matrix of
Pearson's correlation coefficients for each pair of parameters. The
lower triangular part shows raw correlations, while the upper triangular part shows
correlations after high-pass filtering each Markov chain with a boxcar
window of 10 samples; the latter highlights white noise correlation
structures, while the former includes also long trends.

Overall, most correlations are relatively weak, and typically smaller
than 5\,\%, while three are very strong. The first is a 60\,\%
correlation between the $x$- and $z$-components of the CMB Solar
dipole. This is caused by the relative orientation of the Galactic
plane mask, which directly aligns with the $z$-component, and the
diffuse foregrounds at high latitudes, which are anti-symmetric with
respect to Galactic longitude $l=0^{\circ}$, and therefore couples to
the $x$-direction. In contrast, the Galactic plane is symmetric with
respect to Galactic longitudes $l=90^{\circ}$ and $270^{\circ}$, and
therefore it couples weakly to the $y$-dipole.

A second strong correlation is between the CMB and dust Stokes $Q$ and $U$ parameters within a single pixel, which reflects the internal degeneracies of our sky model given \BP\ data selection. While we show here only CMB and thermal dust correlations, similar level of correlations affects also the other sky model components, see \cite{bp10} and \cite{bp13} for further discussion. 
  
The third important strong (anti-)correlation seen in
Fig.~\ref{fig:param_corr_local} is between the large-scale CMB
harmonic $a_{2,1}^T$ (as well as individual temperature pixel values)
and the synchrotron spectral index, $\beta_{\mathrm{s}}$. While
diffuse foregrounds play only a limited part in CMB temperature
reconstruction as measured relative to CMB cosmic variance, and even
relatively simple foreground cleaning methods therefore perform very
well \citep[e.g.,][]{bennett2012,planck2016-l04}, the same foregrounds
still play a very important role as measured relative to the \emph{noise
level} of the experiment, and that is what is probed by these
correlations. For noise-dominated applications, such as CMB $B$-mode
reconstruction, properly accounting for these foreground uncertainties
is therefore key.

The remaining correlations are, as already mentioned, modest, although
not negligible. For instance, there is a 10\,\% correlation between
the two CMB intensity pixels, despite the fact that they are separated
by almost $90^{\circ}$ on the sky, and not located on the same
\Planck\ scanning ring. This correlation is therefore due to general
global parameters, for instance the overall instrument calibration and
gains, the CMB dipole parameters, and the bandpass corrections.

Moving on from individual pixel values to full sky maps, the two
bottom panels in Fig.~\ref{fig:crosscorr_map} show pixel-by-pixel
cross-correlations between the CMB Stokes $Q$ parameter and the
time independent part of the 70\,\GHz\ 21M and 21S radiometer gain variations,
$\Delta g_{\mathrm{21M}}$ and $\Delta g_{\mathrm{21S}}$. (We note that
there is nothing special about the 21M and 21S radiometers in this
respect, beyond the fact that they are 70\,GHz detectors, and the
\BP\ CMB map is strongly dominated by this frequency channel.) In this
case, we see coherent large-scale wave patterns at the 5\,\% level,
with a wave direction that is loosely aligned with the CMB Solar
dipole direction. This pattern is already known, and explained in
terms of correlations between inter-detector gain variations and the
Solar dipole. An example of this is the LFI gain residual template
\citep{planck2016-l02} shown in the top panel; the morphology of this
template is qualitatively very similar to the correlation structures
seen in the \BP\ CMB--gain cross-correlations. We also note that the
21M and 21S correlations are anti-correlated, as expected by the fact
that the polarization angles of these two detectors are rotated
internally by $90^{\circ}$. 

\begin{figure}
  \center	
  \includegraphics[width=\linewidth]{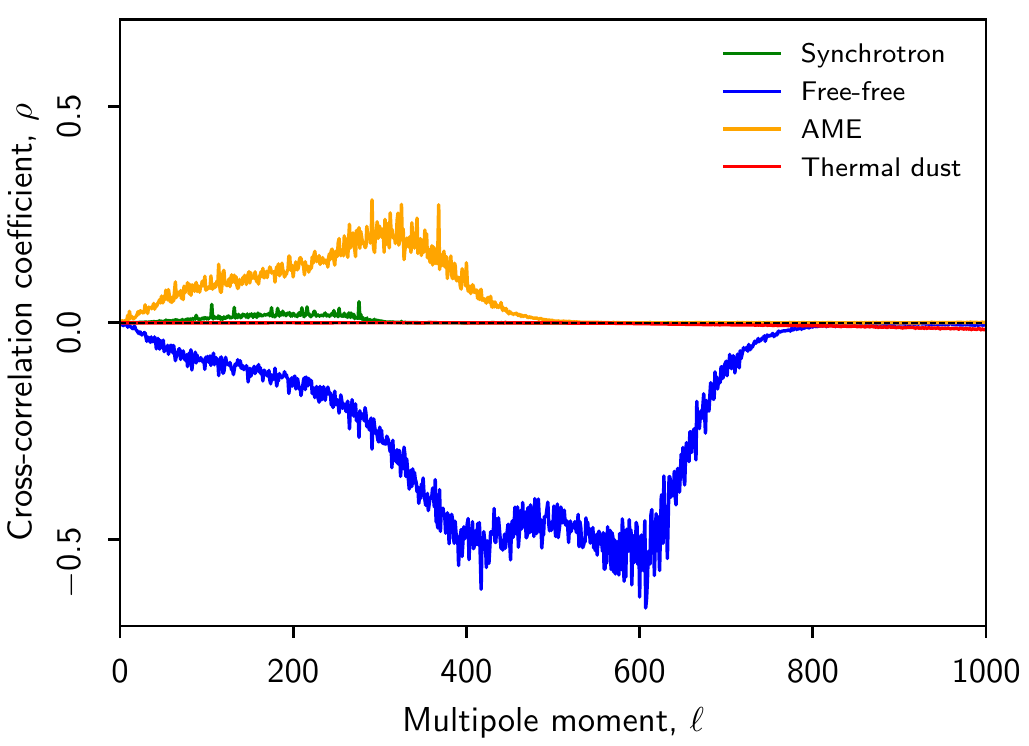}
  \caption{Cross-correlation spectra between CMB and foreground maps. Colors indicate cross-correlations with synchrotron (\emph{green}), free-free (\emph{blue}), AME (\emph{orange}), and thermal dust emission (\emph{blue}). }
  \label{fig:fg_crosspec}
\end{figure}

In Fig.~\ref{fig:nongauss}, we plot the skewness and kurtosis of the
CMB Stokes $Q$ and $U$ parameters per HEALPix $N_{\mathrm{side}}=8$
pixel, which is the same resolution as used by the \BP\ low-$\ell$
likelihood. It is important to note that these estimates do not
measure non-Gaussianity of the CMB signal itself, but rather of the
\emph{uncertainties} of the CMB map. Except for a few statistically
significant non-Gaussian pixel distributions in the center of the
Galactic plane, the skewness and kurtosis maps appear noise dominated,
small in amplitude, and statistically isotropic at high Galactic
latitudes. We also note that the standard deviation of excess skewness
and kurtosis of a random Gaussian sample with $N_{\mathrm{samp}}\gg 1$
are given by $\sqrt{6/N_{\mathrm{samp}}}$ and
$\sqrt{24/N_{\mathrm{samp}}}$, which translate into standard
deviations of 0.043 and 0.086, respectively, for
$N_\mathrm{samp}=3200$. The observed skewness and kurtosis seen in
Fig.~\ref{fig:nongauss} are thus both consistent with zero, and this
suggests that the CMB posterior distribution may be well approximated
in terms of a multi-variate Gaussian distribution \citep{bp12}.

Finally, we conclude this section by measuring the cross-correlation
power spectra between the CMB intensity map and each of the four
diffuse foregrounds included in the \BP\ data model. These
cross-correlations are defined by
\begin{equation}
  \rho^{XY}_{\ell} = \left\langle \frac{C_{\ell}^{XY}}{\sqrt{C_{\ell}^{XX}C_{\ell}^{YY}}} \right\rangle,
\end{equation}
where $C_{\ell}^{XY}\equiv \sum_m|a_{\ell m}^x (a_{\ell m}^Y)^*|/(2\ell+1)$, $X$ denotes CMB and $Y$ any one of the foreground components, and $\langle\ldots\rangle$ denotes average over the chain samples. Note that the cross-correlations spectra are computed between the residual maps, ${\bf m}_i - \langle {\bf m}_i \rangle$, to highlight the impact of model degeneracies, rather than chance correlations between the components.  The results from these calculations are summarized in
Fig.~\ref{fig:fg_crosspec}.

Considering these functions in order from weak to strong correlations,
we first note that the thermal dust emission (red curve) is for all
practical purposes statistically uncorrelated with the CMB
component. It is important to stress that this does \emph{not} imply
that the thermal dust component does not induce foreground modelling
errors in the CMB mean map. Rather, it just shows that the
thermal dust \emph{uncertainty fluctuations} do not correlate with the
CMB uncertainty fluctuations. The reason for this is that the
\BP\ analysis \citep{bp01} relies on the \Planck\ PR4 857\,GHz sky map
as a dust tracer, for which the CMB component is very low, and the
only free intensity thermal dust parameter in the entire model is a
single full-sky power law spectral index. As such, there is in
practice no feedback from the CMB to the thermal dust component in the
model, and we only propagate the thermal dust
uncertainties as predicted by the 857\,GHz channel to the CMB
component, but do not perform a joint fit. In short, the current
analysis effectively assumes that the \Planck\ PR4 analysis is
accurate as far as thermal dust emission is concerned.

A similar consideration holds true for the synchrotron component
(green curve). In this case, the very low frequency of the Haslam
408\,MHz effectively decorrelates the synchrotron and CMB components,
although not quite as strongly as the 857\,GHz map for thermal dust
emission. The smooth drop around $\ell\approx300$ is caused by the
algorithmic smoothing prior discussed by \citet{bp13}, which
suppresses small-scale synchrotron fluctuations.

Significantly higher correlations are seen for the AME component. In
this case, there is no single frequency map that gives a clear picture
of the component in question, but the spatial structure of AME has to
be estimated from the same maps as the CMB itself. The particular data
selection adopted for \BP, which focuses on \Planck\ LFI and
\WMAP\ measurements between 30 and 70\,GHz, leads to correlations at
the 15--20\,\% range for AME.

It is important to stress that significant correlations, such as those
seen for the AME component, by themselves are no cause for alarm as
far as CMB analysis is concerned, as long as the assumed statistical
model is correct. In this case, the corresponding uncertainties are
fully accounted for in terms of the sample distribution. At the same
time, large correlations are nevertheless undesirable, because they
make the CMB component susceptible to \emph{modelling errors} in the
correlating component. For \BP, this is most clearly seen in the
free-free component, which, as seen in Fig.~\ref{fig:fg_crosspec}, is
anti-correlated with the CMB component at the 50\,\% level between
$\ell\approx400$ and 600. The reason for this are two-fold. First, the
free-free emission scales roughly as $\nu^{-2}$, and therefore falls
much more slowly with frequency than both synchrotron emission and
AME. Secondly, it is spatially much more localized than either of the
other two low-frequency components. The maximum multipole required to
model free-free emission without ringing is therefore relatively high
($\ell_{\mathrm{max}}\approx800$). The only \BP\ frequency channels
that provide useful information at such small angular scales and high
frequencies are, primarily, the LFI 70\,GHz channel, and secondarily
the LFI 44\,GHz and \WMAP\ \textit V-band channel. These are also precisely
the same channels that are used to constrain the CMB component, and
the two are therefore highly correlated.

As discussed by \citet{bp13}, a partial solution to this problem is
the introduction of the HFI-dominated spatial free-free prior from
\citet{planck2014-a12}. While this is effective at breaking the
degeneracy in question, which is necessary for constraining important
instrumental parameters such as calibration and bandpasses, it also
introduces an uncontrolled level of unmodelled systematic errors and
uncertainties, both because the LFI and \WMAP\ data were in fact used
to generate the free-free prior template in the first place, and
because of unmodelled systematic and statistical uncertainties in the
HFI data. While we consider these unmodelled uncertainties acceptable
for instrument modelling, which only depend weakly on the free-free
model, they are not acceptable for the CMB component, which is the
main scientific product from the entire analysis. This is a main
reason for performing the \BP\ analysis in a two-step manner, in which
the prior-constrained free-free component is included during the main
Gibbs analysis, but excluded during the CMB resampling stage, while at
the same time applying a large Galactic mask that excludes all regions
with statistically significant free-free emission. For further
discussion, see Sect.~\ref{sec:highell_spec}, \citet{bp12}, and
\citet{bp13}.

\begin{figure*}[t]
  \center
  \includegraphics[width=0.49\linewidth]{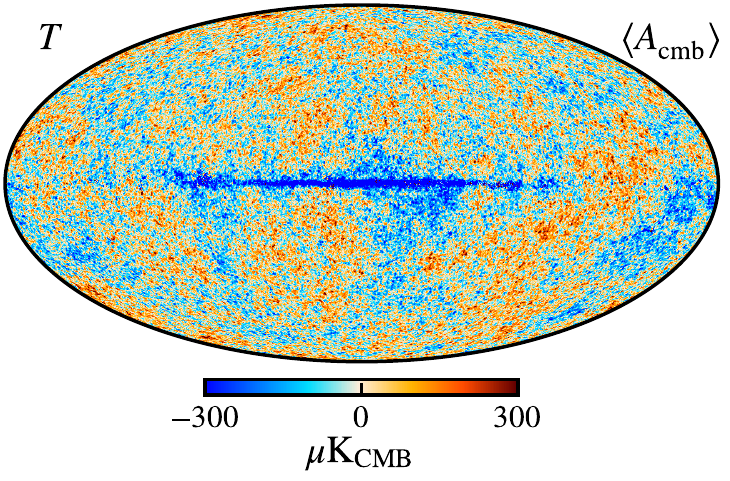}
  \includegraphics[width=0.49\linewidth]{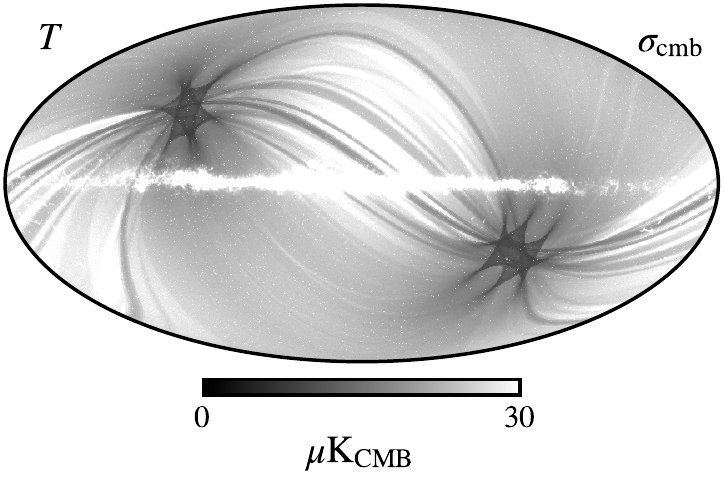}\\  
  \includegraphics[width=0.49\linewidth]{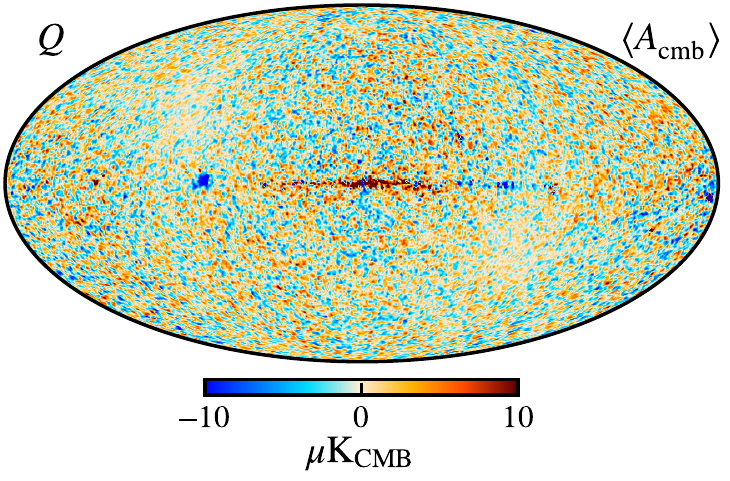}
  \includegraphics[width=0.49\linewidth]{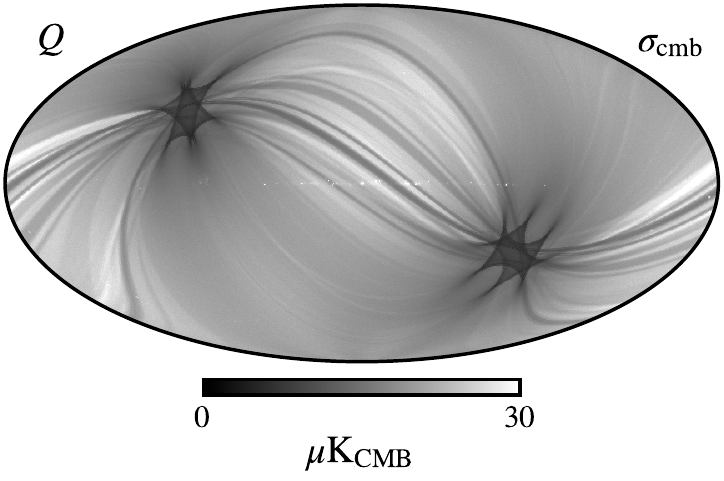}\\  
  \includegraphics[width=0.49\linewidth]{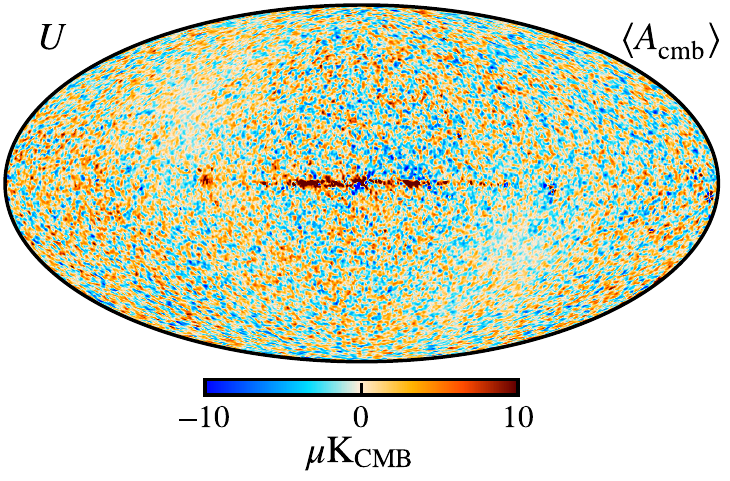}
  \includegraphics[width=0.49\linewidth]{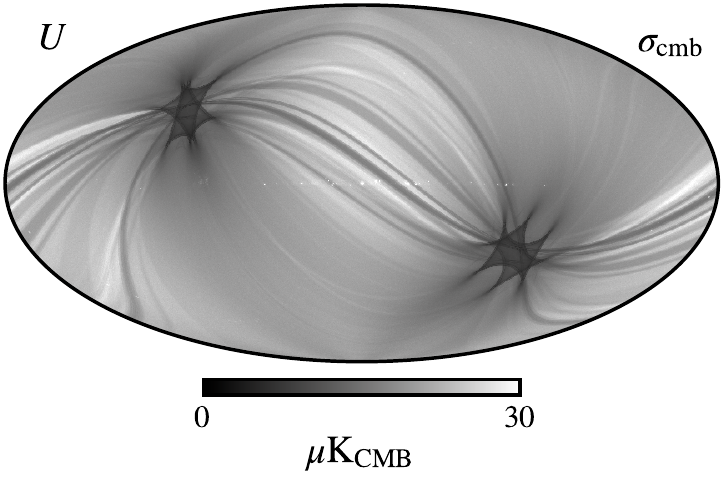}  
  \caption{\BP\ posterior mean (\emph{left column}) and standard
    deviation (\emph{right column}) CMB fluctuation maps. Rows show,
    from top to bottom, temperature and Stokes $Q$ and $U$ parameters,
    respectively. The temperature maps are smoothed to $14\arcm$ FWHM
    resolution, while the polarization maps are smoothed to $1\deg$
    FWHM. }\label{fig:cmb_BP}
\end{figure*}

\begin{figure*}
  \center
  \includegraphics[width=0.47\linewidth]{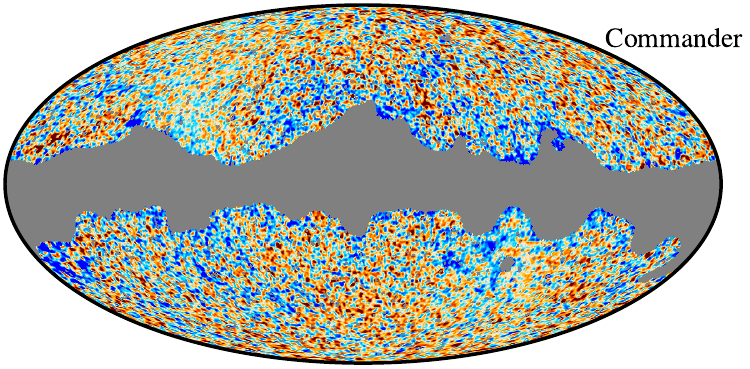}
  \includegraphics[width=0.47\linewidth]{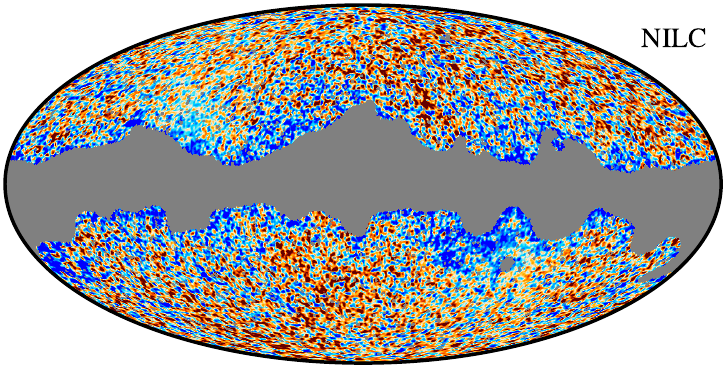}\\
  \includegraphics[width=0.47\linewidth]{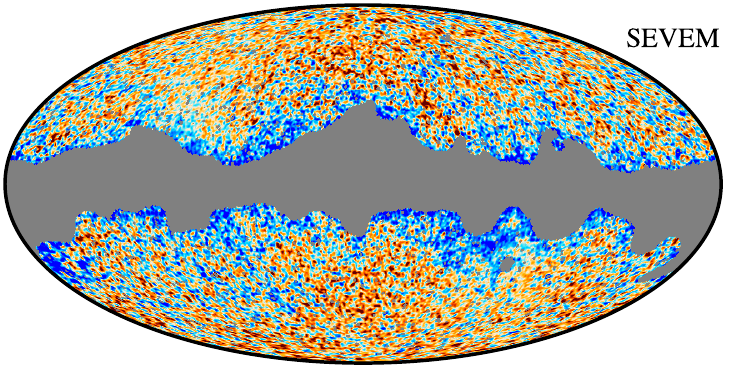}
  \includegraphics[width=0.47\linewidth]{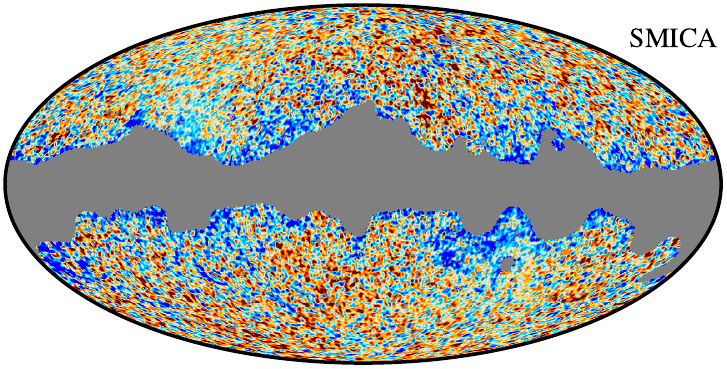}\\
  \includegraphics[width=0.34\linewidth]{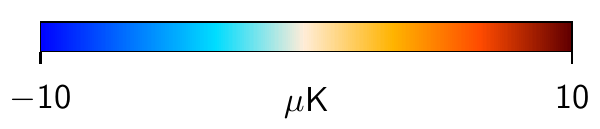}
  \caption{Difference maps between the \BP\ CMB temperature map and
    those derived from the full \Planck\ 2018 data set
    \citep{planck2016-l04}. From left to right and from top to bottom,
    the various panels show differences with respect to \commander,
    \nilc, \sevem, and \smica. All maps are smoothed to a common
    angular resolution of $1^{\circ}$ FWHM.}\label{fig:cmb_diff}
\end{figure*}

\begin{figure*}[t]
  \center
  \includegraphics[width=0.49\linewidth]{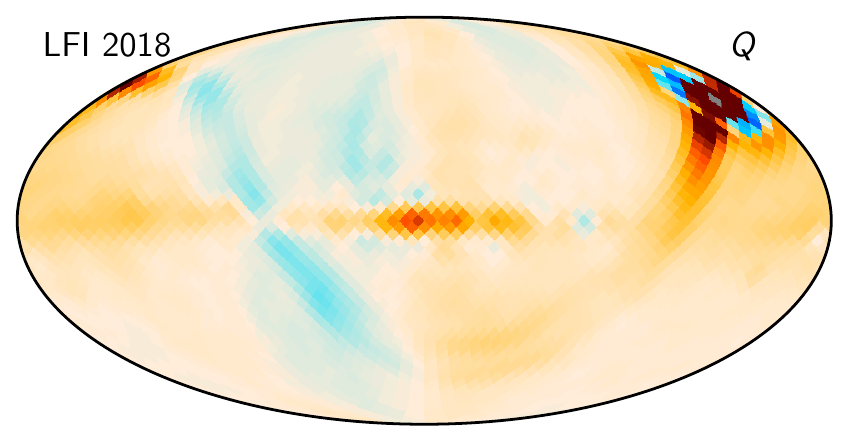}
  \includegraphics[width=0.49\linewidth]{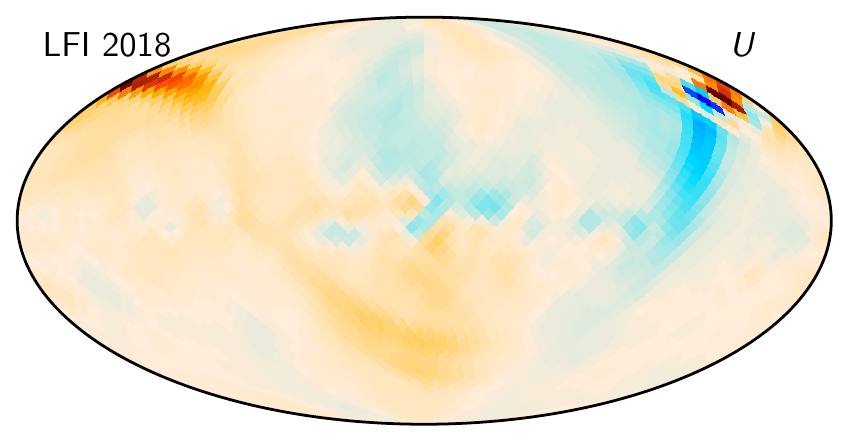}\\
  \includegraphics[width=0.49\linewidth]{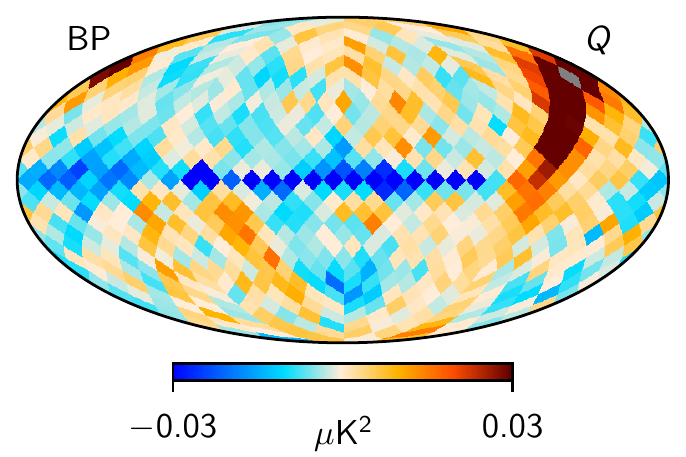}
  \includegraphics[width=0.49\linewidth]{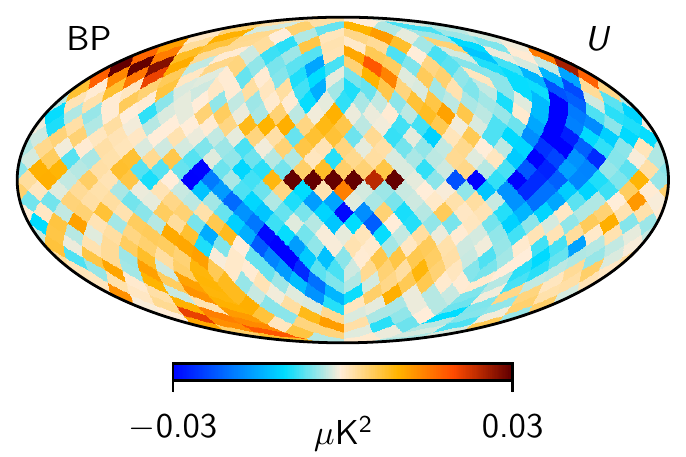}\\
  \caption{Single column of the low-resolution CMB noise covariance
    matrix, as estimated by the LFI DPC (\emph{top row}) and
    \BP\ (\emph{bottom row}). The column corresponds to the Stokes $Q$
    pixel marked in gray, which is located in the top right quadrant
    near the `$Q$' label. Note that the DPC covariance matrix is
    constructed at $N_{\mathrm{side}}=16$ and includes a cosine
    apodization filter, while the \BP\ matrix is constructed at
    $N_{\mathrm{side}}=8$ with no additional filter. }\label{fig:ncov}
\end{figure*}

\begin{figure*}[t]
  \center
  \includegraphics[width=0.49\linewidth]{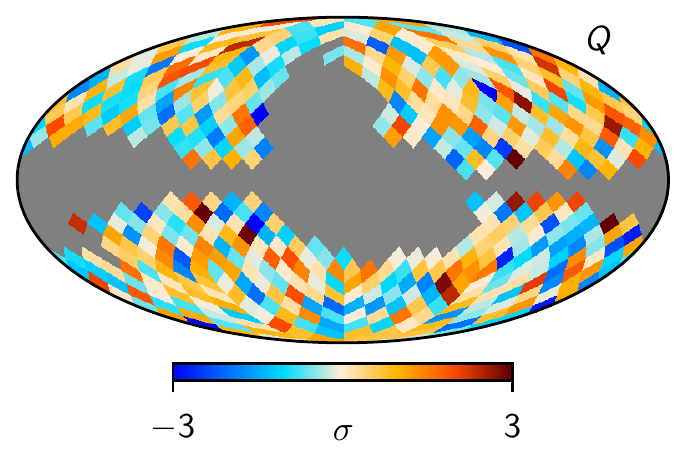}
  \includegraphics[width=0.49\linewidth]{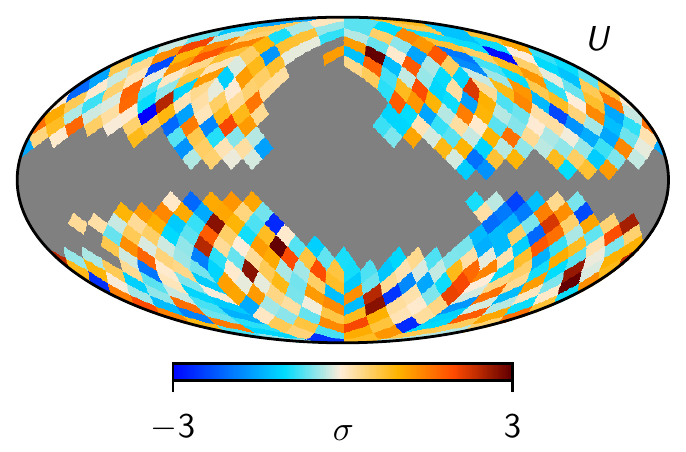}
  \caption{\BP\ low-resolution and ``whitened'' CMB polarization map,
    as defined by $\N_{\mathrm{CMB}}^{-1/2}\s_{\mathrm{CMB}}$ at a HEALPix
    resolution of $N_{\mathrm{side}}=8$ and masked with the
    \BP\ analysis mask. Left and right panel shows Stokes $Q$ and
    $U$ parameters, respectively, and the color scales span
    $\pm3\sigma$.}\label{fig:whitened}
\end{figure*}

\begin{figure}[t]
  \center
  \includegraphics[width=0.93\linewidth]{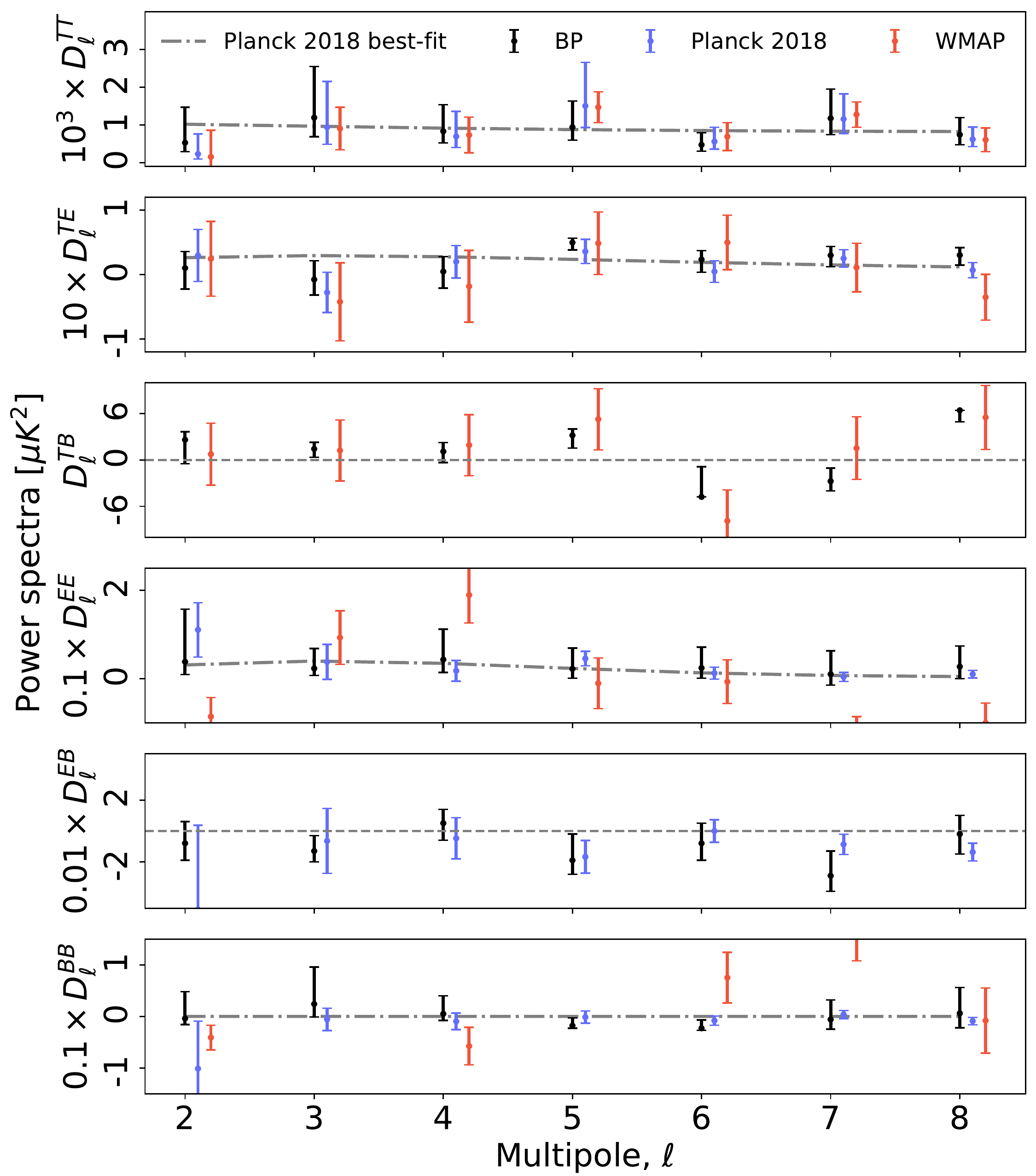}
  \caption{Comparison between low-$\ell$ angular CMB power spectra, as
    derived by the \Planck\ collaboration using both LFI and HFI data
    (blue points; \citealp{planck2016-l05}); by the \WMAP\ team using
    just \WMAP\ data (red points; \citealp{hinshaw2012}); and by
    \BP\ using both LFI and \WMAP\ data (black points;
    this work). Thin black lines indicate the \Planck\ 2018
    best-fit $\Lambda$CDM spectrum \citep{planck2016-l06}. The
    \BP\ data points are evaluated by conditionally slicing the
    posterior distribution $\ell$-by-$\ell$ with respect to the
    best-fit $\Lambda$CDM model, by holding all other multipoles fixed
    at the reference spectrum while mapping out $P(C_{\ell}\mid\d)$,
    to visualize the posterior structure around the peak. For \WMAP,
    the reported $BB$ octopole amplitude is $D_{3}^{BB}=1.12\pm0.03$,
    which is outside the plotted range.}
   \label{fig:powspec_lowl}
\end{figure}

\begin{figure}[t]
  \center
  \includegraphics[width=\linewidth]{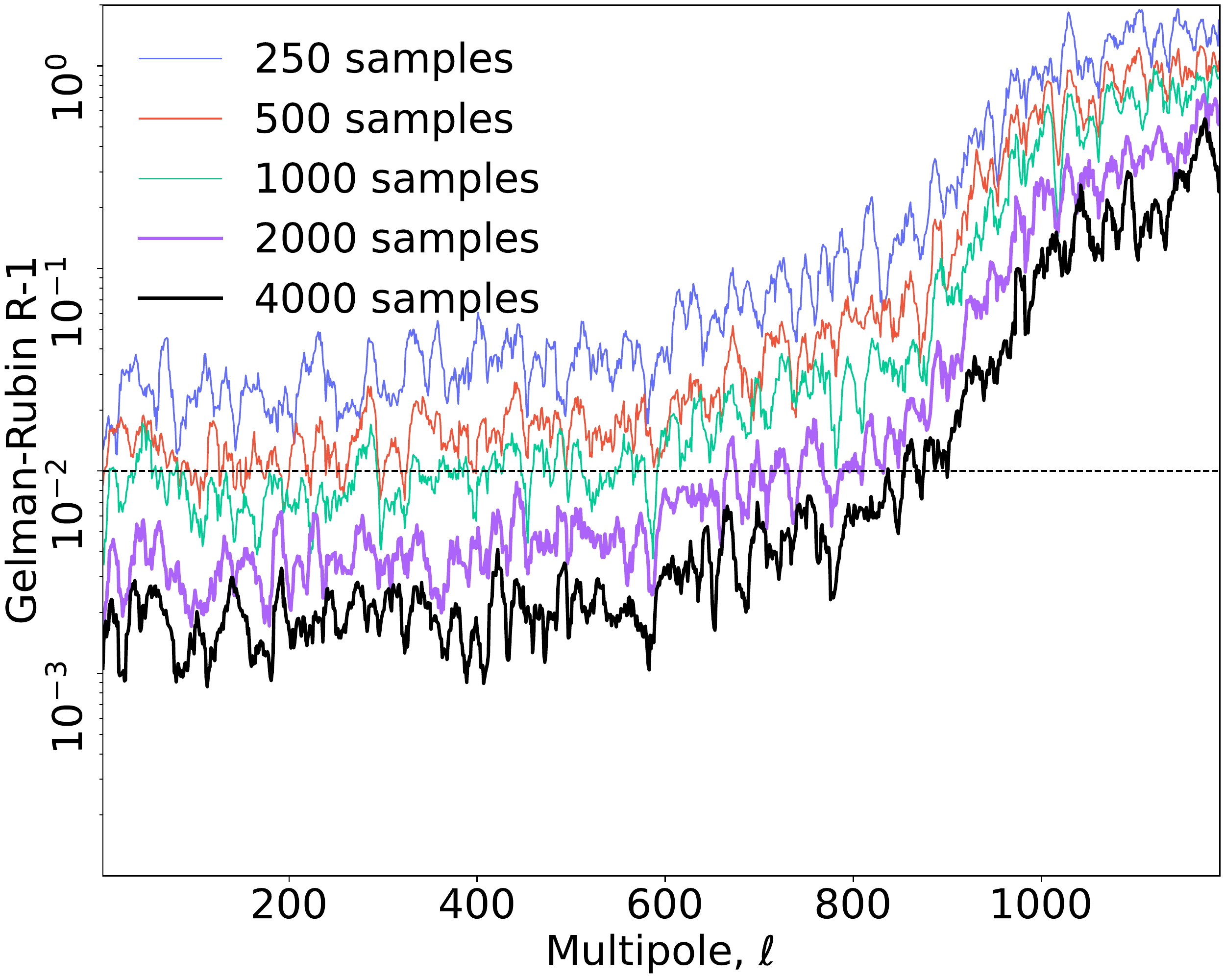}
  \caption{Gelman-Rubin convergence statistic for the \BP\ $TT$
    angular power spectrum, as evaluated from four independent
    $\sigma_{\ell}$ chains. The various curves show results for
    different total number of samples included in the analysis.  A
    value lower than 1.01 (dotted line) typically indicates
    good convergence. Accordingly, \BP\ multipoles above $\ell \sim 800$ should be acceptable for parameter estimation. However, in the current paper, we conservatively include only modes $\ell \le 600$ in the cosmological analysis, see text.}\label{fig:gr_TT}
\end{figure}

\section{CMB maps, covariance matrices, and power spectra}
\label{sec:maps}

\subsection{Posterior mean sky maps}

The individual parameter samples discussed in the previous section
represent the most fundamental products from the current analysis, and
we strongly recommending using the set of such individual Gibbs
samples for any high-level statistical analysis. That ensemble
provides the most convenient approach to fully propagate uncertainties
into any given statistic. In order to do so, one simply analyzes all
available samples individually, as if they were ideal CMB map
estimates, and then reports the full distribution as final results.
Worked examples of this procedure are given in
Sect.~\ref{sec:anomalies} for select previously reported low-$\ell$
anomalies.

\begin{figure*}[t]
  \center
  \includegraphics[width=\linewidth]{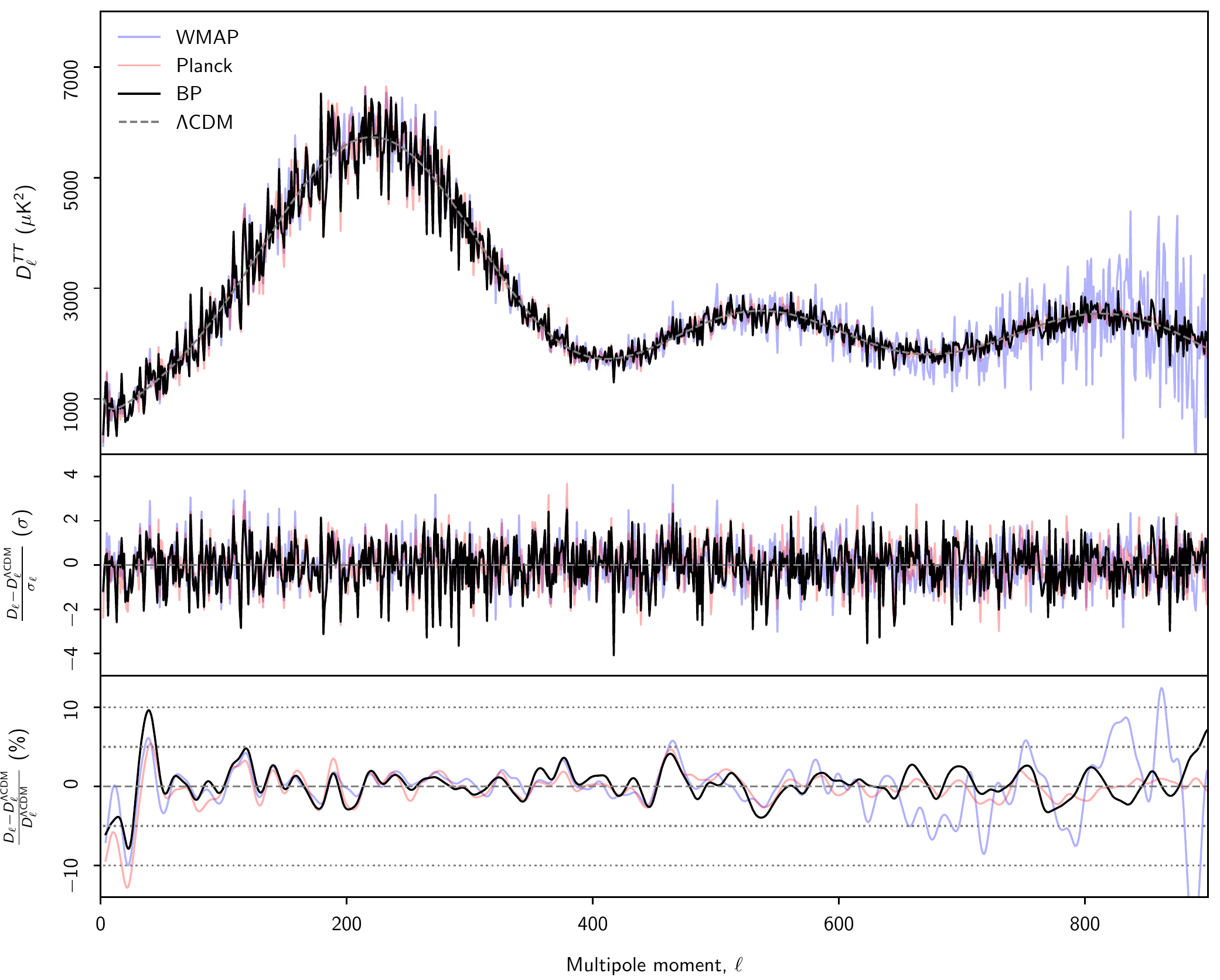}
  \caption{(\emph{Top}:) Angular CMB temperature power spectrum,
    $D_{\ell}^{TT}$, as derived by \BP\ (black), \Planck\ (red), and
    \WMAP\ (blue). The best-fit \Planck\ 2018 $\Lambda$CDM power
    spectrum is shown in dashed gray. (\emph{Middle}:) Residual power
    spectrum relative to $\Lambda$CDM, measured relative to full
    quoted error bars, $r_{\ell} \equiv
    (D_{\ell}-D_{\ell}^{\Lambda\mathrm{CDM}})/\sigma_{\ell}$. For
    pipelines that report asymmetric error bars, $\sigma_{\ell}$ is
    taken to be the average of the upper and lower error
    bar. (\emph{Bottom}:) Fractional difference with respect to the
    \Planck\ $\Lambda$CDM spectrum. In this panel, each curve has been
    boxcar averaged with a window of $\Delta\ell=100$ to suppress
    random fluctuations. }\label{fig:cl_TT}
\end{figure*}

\begin{figure}[t]
  \center
  \includegraphics[width=\linewidth]{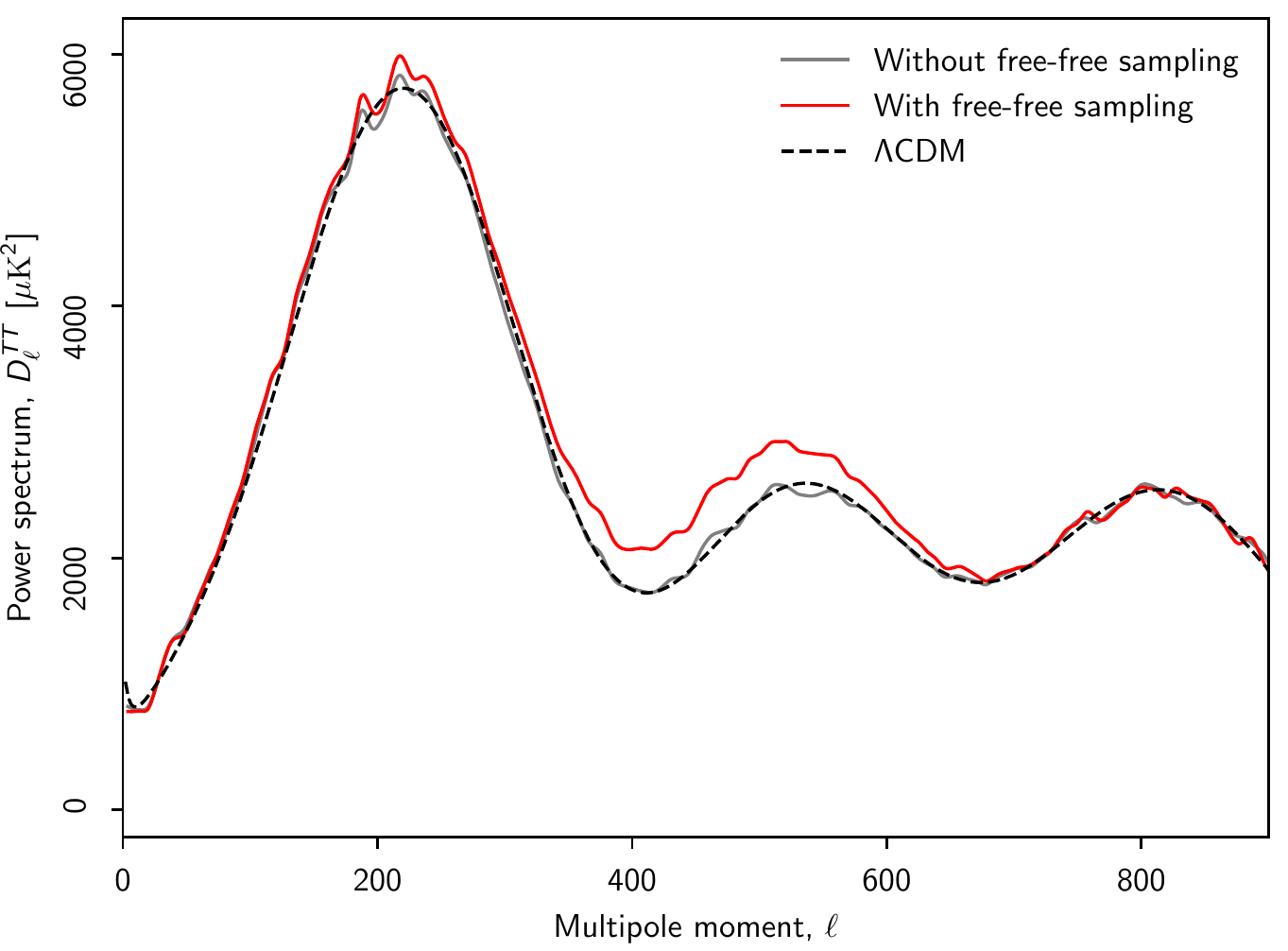}
  \caption{Comparison of power spectra derived with (\emph{red}) and without (\emph{gray}) including prior-constrained free-free emission in the $TT$ resampling procedure.}
  \label{fig:freefree}
\end{figure}

Still, for visualization and comparison purposes it is still
convenient to consider sample averaged mean and standard deviation
maps, which correspond most closely to the best-fit CMB maps derived
with traditional pipelines. These are shown in
Fig.~\ref{fig:cmb_BP}. In this figure, each sample is convolved with a
Gaussian azimuthally symmetric beam of $14\arcm$ FWHM for temperature
and $1^{\circ}$ FWHM for polarization, before projecting into sky
maps. The first 200 samples from each chain are conservatively
discarded as burn-in, leaving a total of 3200 samples for actual
analysis. The mean and standard deviation is then evaluated
pixel-by-pixel from these samples. (Note that the CMB Solar dipole has
been removed from the temperature maps in these plots; this component
is discussed separately in the next section).

Starting with the temperature mean map in the top left panel, we see
that this CMB map is visually similar to the \commander\ CMB map
presented by in the \Planck\ 2015 analysis \citep{planck2014-a12}. At
high Galactic latitudes, the familiar isotropic CMB fluctuations are
visually obvious, while at low Galactic latitudes, there is a clear
negative foreground imprint. This is due to over-subtraction of
thermal dust and free-free emission, and it can be removed through
detailed foreground modelling that also includes \Planck\ HFI
observations; see \citet{planck2016-l04} for a \commander-based
analysis that successfully eliminates this effect. For the current
analysis, which does not include HFI observations and only fit the
thermal dust SED with a single spectral index, $\beta_{\mathrm{d}}$,
across the full sky, this foreground leakage represents the main
limiting effect at low Galactic latitudes, and clearly shows why a
large Galactic mask is needed.

The upper right panel shows the corresponding standard deviation map,
and we see that this is dominated by three main effects. At high
latitudes, the dominant feature is the \Planck\ scanning strategy, and
individual features are associated with the white
noise distribution of the 70\,GHz LFI channel \citep{bp10}. One may
also see a number of bright dots, corresponding to individual point
sources, as described by Eq.~\eqref{eq:ptsrc_astsky}. At low Galactic
latitudes, the uncertainties are dominated by diffuse foregrounds, and
the morphology is visually dominated by free-free and thermal dust
emission \citep{bp13}.

The two bottom rows show the same for the Stokes $Q$ and $U$
polarization maps. In these cases, the mean maps are visually
dominated by white noise over most of the sky, as evidenced by the
fact that one may see the Ecliptic pole regions also in the mean
maps. Of course, this is fully expected, given that the average
standard deviation per pixel is about 20\,\muK, while the expected CMB
signal for an ideal $\Lambda$CDM CMB map smoothed to $1^{\circ}$ FWHM
is $\lesssim3\,\muK$. Thus, the signal-to-noise ratio is less than 0.5
per pixel.

The only obvious visually recognizable features are Galactic plane
residuals with an alternating sign, which is a classic signature of
temperature-to-polarization leakage from bandpass mismatch
\citep{bp09}. This is, however, confined to a narrow region of less
than 1\,\% of the full sky. The polarization CMB confidence mask shown
in the bottom panel Fig.~\ref{fig:confmasks} is more than
sufficient to eliminate these residuals from higher-level analysis.

In Fig.~\ref{fig:cmb_diff} we show difference maps between the
\BP\ posterior mean CMB temperature map and the four
foreground-reduced CMB maps presented by \citet{planck2016-l04},
generated by \commander\ \citep{eriksen2008},
\nilc\ \citep{2012MNRAS.419.1163B, 2013MNRAS.435...18B},
\sevem\ \citep{leach2008,fernandez-cobos2012}, and
\smica\ \citep{cardoso2008}, respectively. The gray region indicates
the \BP\ temperature CMB confidence mask, and a constant offset has
been removed from each map outside this mask. First, we note that the
color range is $\pm10\muK$, which is the same range as was used in
Figs.~6 and 7 of \citet{planck2016-l04} to show differences between
the 2015 and 2018 CMB maps, and internally among the four
\Planck\ component separation algorithms. As such, the \BP\ CMB map
agrees about as well with either of those maps as the \Planck\ maps do
internally. However, a closer comparison of our
Fig.~\ref{fig:cmb_diff} with their Fig.~7 reveals two important
differences, namely a large white noise contribution at high Galactic
latitudes, and a blue edge around the Galactic plane mask. Both of
these effects have fundamentally the same explanation, namely that the
current analysis does not involve the CMB-dominated and
high-sensitivity HFI frequency channels, and the current map therefore
has both higher noise and more free-free and thermal dust
contamination. The latter of these effects dictates our larger
confidence mask for high-level analysis.

\subsection{Low-$\ell$ polarization power spectrum}

We now turn our attention to CMB power spectrum estimation, and we
start with low-$\ell$ polarization. For this task, we adopt the
well-established machinery of multivariate Gaussian likelihood
estimation
\citep[e.g.,][]{tegmark1997,page2007,planck2016-l05,gjerlow2015}, and
map out the following distribution with respect to $C_{\ell}$,
\begin{equation}
  \label{eq:gauss_like}
  P(C_\ell\mid\hat{\s}_{\mathrm{CMB}}) \propto \frac{\exp{(-\frac{1}{2}\hat{\s}_{\mathrm{CMB}}^t
  \left( \S(C_\ell)+\N \right)^{-1}\hat{\s}_{\mathrm{CMB}}})}{\sqrt{|\S(C_\ell)+\N|}},
\end{equation}
where $\hat{\s}_{\mathrm{CMB}}$ denotes the posterior mean CMB map,
$\N$ is the corresponding noise covariance matrix, and $\S(C_\ell)$ is
the signal covariance matrix, which is fully defined by the angular
power spectrum.  For a detailed review of the implementation used in
\BP, we refer the interested reader to \citet{bp12}.

The main scientific goal of the entire \BP\ framework is precisely to
derive estimates of $\hat{\s}_{\mathrm{CMB}}$ and $\N$ for which
astrophysical and instrumental systematic effects are fully
marginalized over. Given the Gibbs samples described above, this may
be done very conveniently as follows,
\begin{align}
  \hat{\s}_{\mathrm{CMB}} &= \langle\s^i_{\mathrm{CMB}}\rangle \\
  \N &= \left\langle\left( \s^i_{\mathrm{CMB}}-\s_{\mathrm{CMB}} \right)
  \left( \s^i_{\mathrm{CMB}}-\s_{\mathrm{CMB}} \right)^t\right\rangle
  \label{eq:postmean}
\end{align}
where $i$ indicates sample number, and brackets denote averages over
all available Monte Carlo samples. As described by \citet{bp12}, we
evaluate both these quantities at a HEALPix resolution of
$N_{\mathrm{side}}=8$ after smoothing the temperature component to
$20^{\circ}$ FWHM.

The bottom row of Fig.~\ref{fig:ncov} shows a slice through $\N$,
centered on the Stokes $Q$ pixel marked in gray in the upper right
quadrant. This plot effectively summarizes all the various systematics
corrections described in Sect.~\ref{sec:bp} to the extent that they
are significant for large-scale polarization reconstruction. For
comparison, the top row shows the corresponding CMB covariance matrix
slice computed by the \Planck\ DPC \citet{planck2016-l02}; note that
this was evaluated at $N_{\mathrm{side}}=16$, and also that the DPC
analysis applied an additional cosine smoothing kernel not used by
\BP.

Comparing the \BP\ and DPC covariance matrices is useful for building
intuition regarding these products. First, we note that the
\BP\ covariance appears noisier than the DPC matrix. This is due to
the fact that it is constructed by Monte Carlo sampling as opposed to
analytic calculations. A computational disadvantage of the sampling
approach, relative to the analytic approach, is that any high-level
product derived from the covariance matrix must be accompanied by a
corresponding convergence analysis that verifies that the final result
is robust with respect to the number of samples; for \BP, this is done
explicitly for the optical depth of reionization, $\tau$, by
\citet{bp12}.

However, this minor disadvantage is more than compensated for by the
fact that the sampling approach is able to jointly account for many
more systematic effects than the analytic approach, and this is
clearly seen in Fig.~\ref{fig:ncov}: While the DPC matrix only models
correlated noise (seen as the ring passing through the gray pixel) and
simple template-based foreground corrections, the \BP\ matrix
additionally accounts for absolute detector calibration differences
(seen as large-scale red and blue regions aligned along the Solar CMB
dipole direction; \citealp{bp07,bp10}); time-dependent gain
fluctuations (seen as additional power along the scanning ring passing
through the gray pixel; \citealp{bp07}); and bandpass leakage (seen as
the sharp Galactic plane; \citealp{bp09}). There are also many other
effects that are not as visually obvious, but they still contribute to
the final results, such as spatially varying foreground spectral
indices \citep{bp13,bp14} and time-dependent noise power spectral
density parameters \citep{bp06}. Propagating all these effects
analytically into a final joint covariance matrix can be for all practical
purposes impossible, while with the novel sampling approach introduced
here it is quite straightforward.

Figure~\ref{fig:whitened} shows the corresponding noise-weighted (or
``whitened'') posterior mean map,
$\N_{\mathrm{CMB}}^{-1/2}\s_{\mathrm{CMB}}$; when plotted directly in
terms of $\s_{\mathrm{CMB}}$, the maps are dominated by the
\Planck\ scanning pattern and poorly constrained large-scale modes,
which complicates the visual interpretation of actually statistically
significant features. Note that the color scale ranges over
$\pm3\,\sigma$. Overall, we see that these maps appear noise
dominated, with most pixels having values below $2\sigma$. However,
there are a handful of saturated pixels as well, in particular close
to the Galactic plane and near the Orion complex (lower right
quadrant). Most likely, these are due to unmodelled foreground errors,
and should in principle be removed. However, since they are isolated,
they only contribute with high-$\ell$ power, well above
$\ell\gtrsim10$, and they are therefore of minor concern for the
current low-$\ell$ focused analysis; \citet{bp12} explicitly shows
that all main large-scale polarization results are stable with respect
to mask variations, from $f_{\mathrm{sky}}\approx0.25$ to 0.75. We do
also see some fainter coherent structures on larger angular scales,
but these are all well below $1.5\,\sigma$. Some of those structures
are real CMB signal, and some are just coherent large-scale noise
fluctuations generated by the same effects as are seen in the
covariance matrix slices in Fig.~\ref{fig:ncov}. As reported by
\citet{bp12}, the total signal-plus-noise $\chi^2$ has a
probability-to-exceed of 32\,\% when evaluated for the best-fit
$\Lambda$CDM power spectrum with $\tau=0.066\pm0.013$, which indicates
that the data are fully consistent with the model.

Figure~\ref{fig:powspec_lowl} compares the low-$\ell$ \BP\ power
spectra with corresponding results reported by
\Planck\ \citep{planck2016-l05} and \WMAP\ \citep{hinshaw2012}. The
\BP\ constraints shown here is computed by slicing the full
probability distribution in Eq.~\eqref{eq:gauss_like}
$\ell$-by-$\ell$, while fixing all other coefficients at their
reference values; error bars indicate asymmetric 68\,\% confidence
ranges. Overall, we find good agreement between \BP, \Planck, and
\WMAP. For $TT$, we see that the \BP\ uncertainties are generally
somewhat larger than either of the other two, and that is due to the
larger analysis mask. The most notable multipole in this spectrum is
$\ell=2$, with a peak value of 526\,$\muK^2$, which is substantially
higher than the typical values of about 200\,$\muK^2$ reported
previously. However, the reason for this is algorithmic in nature, and
driven by our conditioning on $\Lambda$CDM $TE$ and $EE$ in this
particular plot; when marginalizing over polarization, we do recover a
quadrupole amplitude of 181\,$\muK^2$, fully consistent with previous
results; for further discussion of this multipole, see
Sect.~\ref{sec:quadrupole}.

For both $TE$ and $EE$, the most notable feature is that our
uncertainties fall between \Planck\ and \WMAP\ in magnitude, which is
expected given that the current analysis include both \WMAP\ and LFI
data, but not HFI. For \BP\, the most significant outlier in the full set of results is at $\ell=8$ in $TB$. \Planck\ has not publicly released $TB$ measuremets for the default HFI cross-spectrum based pipeline, while the LFI pixel-base results show a qualitatively similar outltier at that multipole, but with a lower statistical significance.  We note,
however, that the full probability distribution for this multipole is
highly asymmetric, and a full inspection shows that for \BP\ this is discrepant
with respect to $\Lambda$CDM at the $3\,\sigma$ level, with a PTE of
0.2\,\%. The probability of having one such outlier among 49
measurements by random chance is 9\,\%. This multipole may thus
provide some slight evidence for residual systematics, for instance
associated with the saturated pixels in Fig.~\ref{fig:whitened}, but
the statistical significance is low. In the case of $EB$, \WMAP\ does
not report any results, while both \BP\ and \Planck\ are consistent
with zero, as they are for $BB$. In the latter case, \WMAP\ do report
results, but with large error bars; note that $\ell=3$  and $\ell=5$ fall far
outside the plotted range.

\subsection{High-$\ell$ temperature power spectrum}
\label{sec:highell_spec}

Next, we consider the high-$\ell$ temperature power spectrum, and in
this case we employ the Gaussianized Blackwell-Rao (GBR) estimator
\citep{chu2005,rudjord:2009,planck2016-l05} to map out the posterior
distribution; for specific details on the
\BP\ implementation of this estimator, see \citet{bp12}. In short,
this estimator is defined by averaging the inverse Gamma distribution,
which is the appropriate distribution  for the ideal CMB sky,
\begin{equation}
  P(C_{\ell}\mid\s^{\mathrm{CMB}}) =
  \sum_{i = 1}^{n_{\mathrm{samp}}} \frac{\exp({-\frac{2\ell+1}{2}\frac{\sigma^i_{\ell}}{C_{\ell}}})}{|C_{\ell}|^{\frac{2\ell+1}{2}}},
\end{equation}
over all available Monte Carlo samples, where $\sigma_{\ell}^{i}$ is
the measured full-sky power spectrum of sample $i$. The resulting
marginal distribution is then Gaussianized through a non-linear
mapping, $x_{\ell}(C_{\ell})$, by matching percentiles to a standard
normal distribution, and the final likelihood expression takes the
following form,
\begin{equation}
  \label{eq:GBR}
  P(C_\ell\mid\d)\approx \left(\prod_\ell\frac{\partial C_\ell}{\partial x_\ell}\right)^{-1} e^{-\frac{1}{2}(\x-\mu)^T\Cp^{-1}(\x-\mu)},
\end{equation}
where the first factor denotes the Jacobian resulting from the
change-of-variables.

This expression formed the basis of the default low-$\ell$ temperature
likelihood in both the \Planck\ 2015 and 2018 data releases
\citep{planck2014-a13,planck2016-l05} for $\ell\le30$, and was in the
latter also used as an experimental likelihood up to $\ell\le250$. The
main limitation from extending it to even higher multipoles stemmed
from the fact that the samples that defined the \Planck\ GBR estimator
were computed from foreground-cleaned CMB maps, and those have
effectively smoothed white noise contributions which are difficult to
describe accurately at high multipoles. In contrast, the novel
\BP\ approach generates the samples from \emph{foreground-subtracted
frequency maps}, which do have unsmoothed white noise
contributions. As such, there is no noise modelling limitation
associated with the new implementation, and the GBR estimator can
therefore in principle be used to arbitrary high multipoles.

In practice, however, the effective range of the GBR estimator is
limited by Monte Carlo convergence. This is illustrated in
Fig.~\ref{fig:gr_TT}, which shows the Gelman-Rubin statistic, $R$
\citep{gelman:1992}, for each power spectrum multipole for different
numbers of Monte Carlo samples. This statistic measures the ratio
between the intra-chain and inter-chain variances, and values of
$R<0.01$ typically indicate good convergence. In this figure, we see
that $R$ increases rapidly above $\ell\approx 800$, where the CMB
signal-to-noise ratio of the \BP\ dataset falls below unity. This
behaviour is theoretically well understood
\citep[e.g.,][]{eriksen:2004}, and may be solved by introducing
additional sampling steps \citep{jewell:2009,racine:2016};
implementing this in the latest version of \commander\ is currently
on-going. For now, we conservatively restrict the range for which this
estimator is used to $\ell\le600$.

The final \BP\ temperature power spectrum is shown in
Fig.~\ref{fig:cl_TT} together with \Planck\ 2018 and \WMAP. The top
panel shows the full power spectrum; the middle panel shows the
difference with respect to the best-fit \Planck\ 2018 $\Lambda$CDM
spectrum in units of $\sigma_{\ell}$, and the bottom panel shows the
fractional difference with respect to $\Lambda$CDM in units of
percent. Overall, we see that all three analyses agree very well. For
\BP, the most significant outliers is $\ell=416$, which is anomalous
at the $4\,\sigma$ level; we note that this multipole is also low in
the HFI-dominated \Planck\ 2018 spectrum, although at a slightly lower
significance of about $3\,\sigma$. The probability of having one such
outlier among 599 trials by random chance is about 8\,\%.

Before concluding this section, we return to the issues of strong
free-free correlations and resampling discussed in
Sect.~\ref{sec:resampling}. Specifically, Fig.~\ref{fig:freefree}
compares the angular power spectra (convolved with a Gaussian
smoothing kernel) derived from chains that samples free-free emission
per pixel (red curve) with the baseline approach that excludes this
component. Here we see a highly statistically significant excess
between $\ell=300$ and 600, with a general behaviour that overall
mirrors the CMB-vs-free-free cross-correlation shown in
Fig.~\ref{fig:fg_crosspec}. The explanation for this behaviour is
quite simple: Taking into account the beam sizes and white noise
levels of the data involved in the \BP\ analysis, by far most of the
constraining power for $\ell\gtrsim300$ comes from the LFI 70\,GHz
channel alone, with only slight additional support from the LFI
44\,GHz \WMAP\ \textit Q- and \textit V-bands. This leaves the free-free and CMB
components highly degenerate. At the same time, accurate modelling of
free-free emission on larger scales is key for obtaining a robust
calibration and foreground model. As a temporary solution to this
problem, the current main analysis adopts the (HFI-dominated)
\Planck\ 2015 free-free map as a spatial template prior
\citep{bp13}. While this prior does regularize the foreground fit as
such, it also biases the CMB component at intermediate angular
scales. For this reason, we only include the prior-constrained
free-free component while estimating the instrumental and
astrophysical parameters, but exclude it when estimating the final CMB
parameters. This issue will of course be resolved in a future Bayesian
end-to-end analysis that jointly analyzes both LFI and HFI data from
scratch.

\begin{figure*}[t]
  \center
  \includegraphics[width=0.9\linewidth]{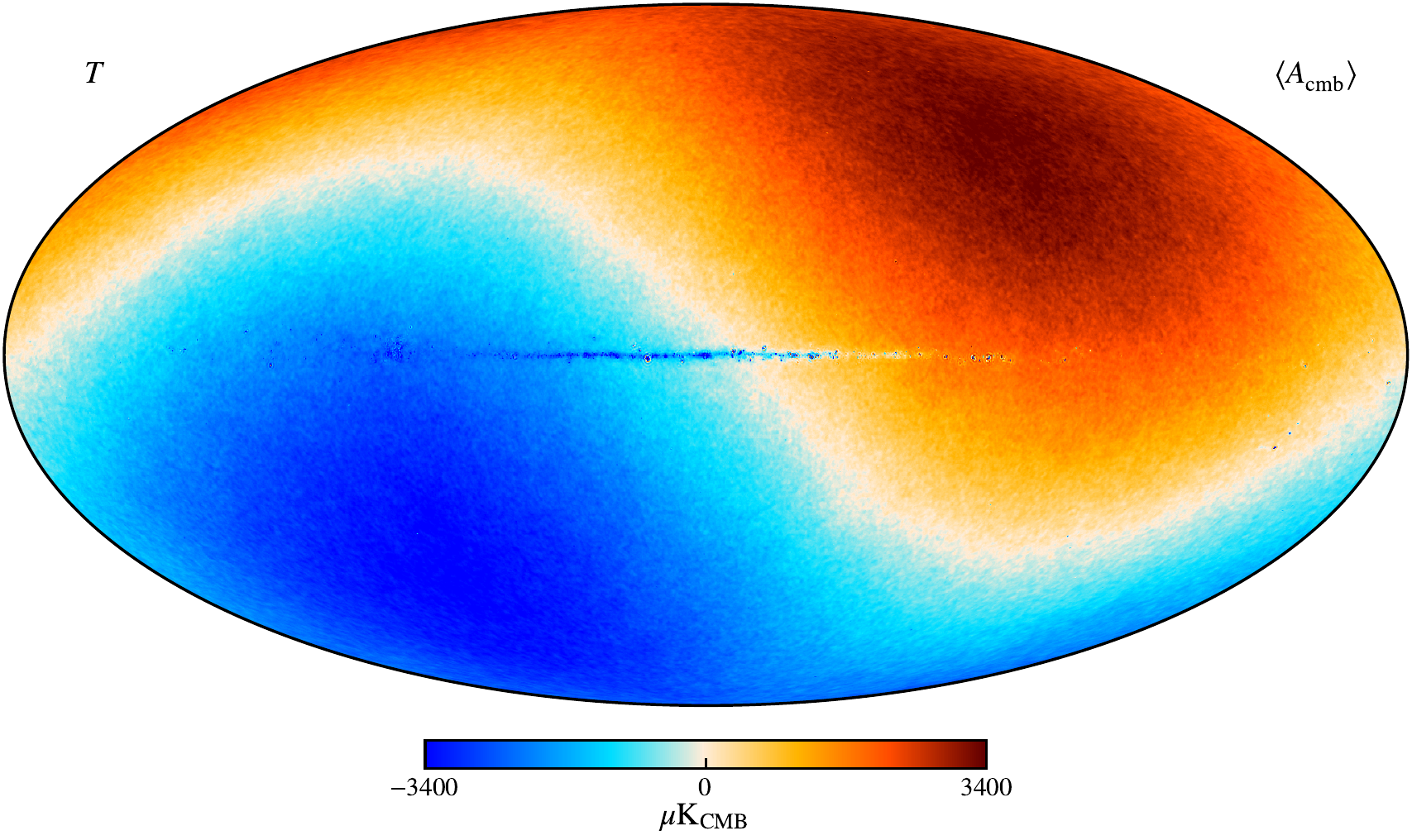}
  \caption{Posterior mean CMB \BP\ temperature map, smoothed to an
    angular resolution of $14\arcm$ FWHM.}\label{fig:cmb_with_dipole}
\end{figure*}

\section{The CMB Solar dipole}
\label{sec:dipole}

We now turn our attention to the CMB Solar dipole. In the
\BP\ framework, this component is in principle estimated on completely
the same footing as any other mode in the CMB sky, and is represented
in terms of three spherical harmonic coefficients in
$\s_{\mathrm{CMB}}$. No special-purpose component separation
algorithms are applied to derive the CMB dipole, nor does any
individual frequency play a more important role than others, except as
dictated by the relative instrumental noise level in each channel.

\begin{figure*}[t]
  \center
  \includegraphics[width=0.33\linewidth]{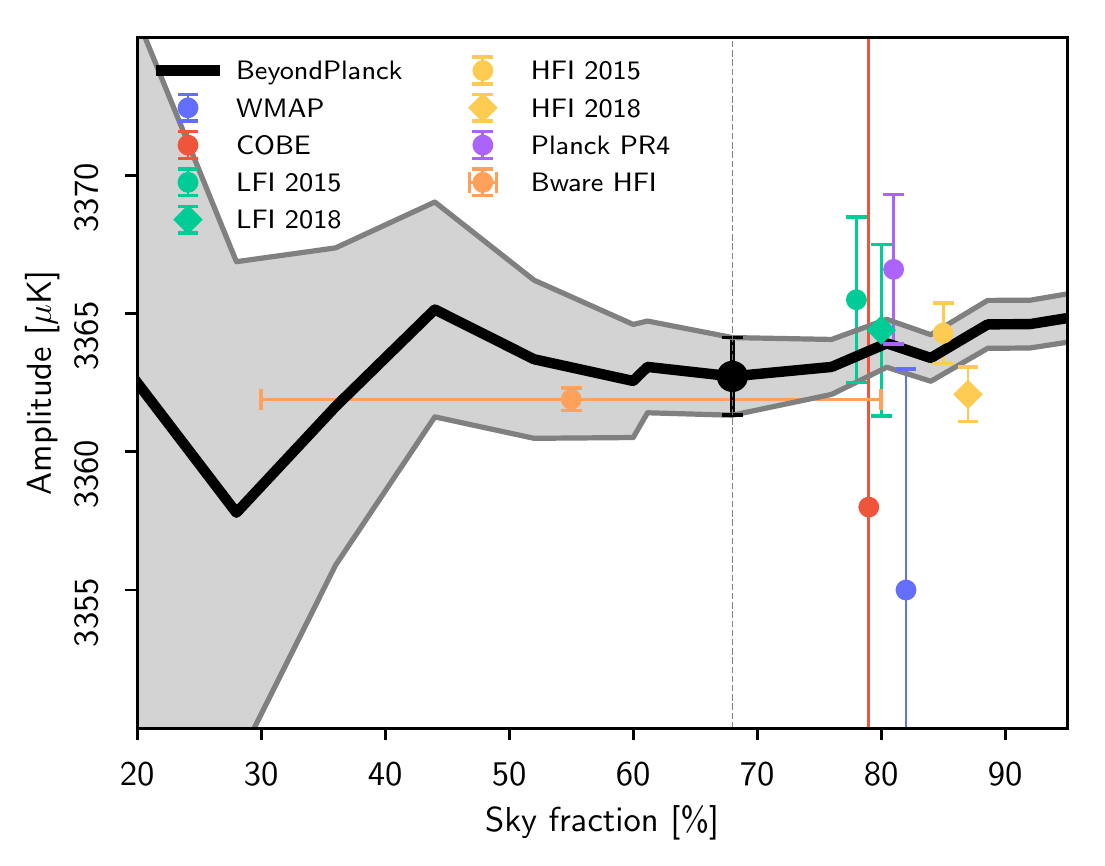}
  \includegraphics[width=0.33\linewidth]{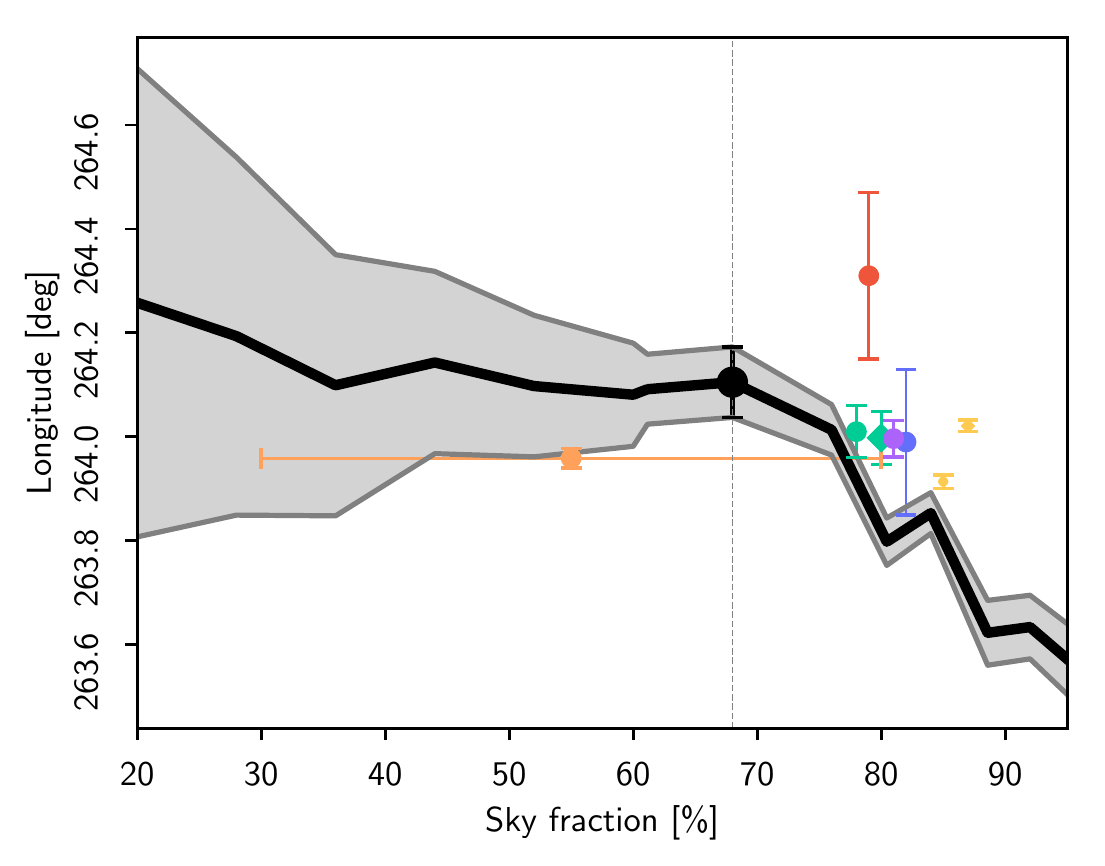}
  \includegraphics[width=0.33\linewidth]{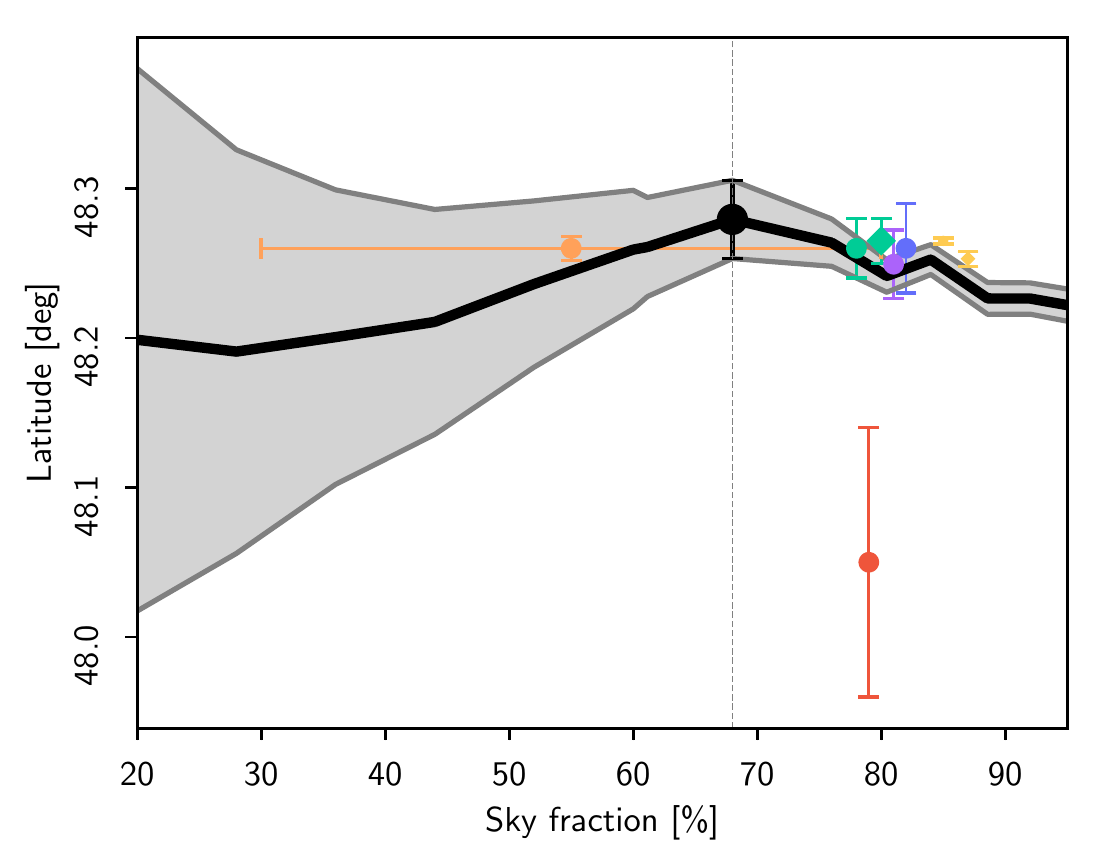}
  \caption{CMB dipole parameters as a function of sky fraction. 
    From left to right, the panels show the dipole amplitude, longitude, and latitude. Gray bands indicate 68\,\% posterior confidence regions.}\label{fig:cmb_dipole}
\end{figure*}

However, as discussed by \citet{bp06}, \citet{bp07}, and \citet{bp10},
this apparent algorithmic simplicity does by no means imply that
robust CMB dipole estimation is \emph{easy} in the
\BP\ procedure. Indeed, the CMB dipole is quite possibly the single
most difficult parameter to estimate in the entire model, simply
because it both affects, and relies on, a wide range of partially
degenerate parameters. The first and foremost of these is the
absolute calibration, $g_0$. This parameter directly scales the
amplitude of the entire CMB map, including the Solar dipole. This
parameter is itself constrained from the orbital dipole, which is both
 weaker in terms of absolute amplitude, and for significant
parts of the mission it is nearly aligned with, and thereby obscured by, the
Galactic plane \citep{bp07}.

Secondly, astrophysical foregrounds have in general both a non-zero
dipole moment, as well as higher-order moments with unknown
parameters, and these must be estimated jointly with the CMB
dipole. Considering that the current data set includes five
astrophysical components, each with a free value in each pixel, and
there are only eight significantly independent frequency channels, the
full system is rather poorly constrained. It is therefore possible to
add a significant dipole to the CMB map and subtract appropriately
scaled dipoles from each of the foreground maps, with only a minimal
penalty in terms of the overall $\chi^2$. In practice, we observe
particularly strong degeneracies between the CMB, AME and free-free
components, when exploring the full system without priors
\citep{bp13}.

Thirdly, correlated noise, $n_{\mathrm{corr}}$, is only weakly
constrained through its $1/f$-style PSD parameters, and this component
is therefore able to account for a wide range of modelling errors,
including calibration errors \citep{bp06,bp17}. In particular,
incorrectly estimated gains leave a spurious dipole-like residual in
the time-ordered data. Since this residual is detector-dependent, it
will typically be interpreted by the algorithm as correlated noise,
and thereby excite a dipolar structure in
$\n_{\mathrm{corr}}$. Coherent large-scale patterns aligned with the
Solar CMB dipole in $\n_{\mathrm{corr}}$ is one of the most typical
signs of overall calibration errors.

Finally, the coupling between the large-scale CMB quadrupole,
foreground, and bandpass corrections all affect the Solar CMB
dipole. While the CMB $E$-mode polarization quadrupole by itself is
predicted by current $\Lambda$CDM models to have a very small
quadrupole, with a variance of typically less than $0.05\,\muK^2$,
there is nothing in the current parametric \BP\ model that explicitly
enforces this. This particular mode therefore opens up a particularly
problematic degeneracy for \Planck\ through coupling with the gain and
bandpass shift as follows: An error in the absolute gain leads to an
apparently wrong orbital dipole. However, this can be countered by
adding a \emph{polarized CMB quadrupole}, which has the same SED and
nearly the same spin harmonics as the orbital dipole, due to the
\Planck\ scanning strategy that observes along nearly perfect great
circles  (see Fig.~1 in \citet{bp07}).\footnote{This particular
  degeneracy does not exist for \WMAP, because of its more complex
  scanning strategy.}  Second-order residuals in the total polarized
sky signal as observed at each frequency can then finally be countered
by adjusting the combination of relative gains, polarized foreground
signals, and bandpass corrections between radiometers, leaving the
total $\chi^2$ nearly unchanged. To break this degeneracy, we actually
do impose a $\Lambda$CDM power spectrum prior on the $E$ $\ell=2$ mode
\emph{during gain estimation alone}, and marginalize over its
amplitude; this prevents the polarization quadrupole from taking on
obviously pathological values. In addition, we note that we include
the large-scale \WMAP\ polarization in the CMB fit, and this also
helps regularizing the large-scale polarization signal. For
comparison, we note that both the \Planck\ LFI DPC and PR4 pipelines
set the entire CMB polarization signal to zero during the gain
estimation process; this is a far more aggressive approach to
resolving this degeneracy, and for \Planck\ PR4 it results in a
non-negligible transfer function on large angular scales (see
\citep{npipe} for full details). For the \Planck\ LFI detectors alone,
the signal-to-noise ratio is too low to make any measurable difference
during gain calibration, resulting in an effectively unbiased
algorithm for \Planck\ LFI, even with this strong prior
\citep{planck2014-a04}.

\begin{table*}
\newdimen\tblskip \tblskip=5pt
\caption{Comparison of Solar dipole measurements from \COBE, \WMAP, and \Planck. }
\label{tab:dipole}
\vskip -4mm
\footnotesize
\setbox\tablebox=\vbox{
 \newdimen\digitwidth
 \setbox0=\hbox{\rm 0}
 \digitwidth=\wd0
 \catcode`*=\active
 \def*{\kern\digitwidth}
  \newdimen\dpwidth
  \setbox0=\hbox{.}
  \dpwidth=\wd0
  \catcode`!=\active
  \def!{\kern\dpwidth}
  \halign{\hbox to 2.5cm{#\leaderfil}\tabskip 2em&
    \hfil$#$\hfil \tabskip 2em&
    \hfil$#$\hfil \tabskip 2em&
    \hfil$#$\hfil \tabskip 2em&
    #\hfil \tabskip 0em\cr
\noalign{\doubleline}
\omit&&\multispan2\hfil\sc Galactic coordinates\hfil\cr
\noalign{\vskip -3pt}
\omit&\omit&\multispan2\hrulefill\cr
\noalign{\vskip 3pt} 
\omit&\omit\hfil\sc Amplitude\hfil&l&b\cr
\omit\hfil\sc Experiment\hfil&[\muK_{\rm
CMB}]&\omit\hfil[deg]\hfil&\omit\hfil[deg]\hfil&\hfil\sc Reference\hfil\cr
\noalign{\vskip 3pt\hrule\vskip 5pt}
\COBE \rlap{$^{\rm a,b}$}&                  3358!**\pm23!**&     264.31*\pm0.16*&
     48.05*\pm0.09*&\citet{lineweaver1996}\cr
\WMAP\ \rlap{$^{\rm c}$}&                  3355!**\pm*8!**&     263.99*\pm0.14*&
     48.26*\pm0.03*&\citet{hinshaw2009}\cr
\noalign{\vskip 3pt}
LFI 2015 \rlap{$^{\rm b}$}&              3365.5*\pm*3.0*&     264.01*\pm0.05*&
     48.26*\pm0.02*&\citet{planck2014-a03}\cr
HFI 2015 \rlap{$^{\rm d}$}&              3364.29\pm*1.1*&     263.914\pm0.013&
     48.265\pm0.002&\citet{planck2014-a09}\cr
\noalign{\vskip 3pt}
LFI 2018 \rlap{$^{\rm b}$}&              3364.4*\pm*3.1*&     263.998\pm0.051&
     48.265\pm0.015&\citet{planck2016-l02}\cr
HFI 2018 \rlap{$^{\rm d}$}&              3362.08\pm*0.99&     264.021\pm0.011&
     48.253\pm0.005&\citet{planck2016-l03}\cr
\noalign{\vskip 3pt}
Bware & 3361.90\pm*0.40 & 263.959\pm0.019 & 48.260\pm0.008  & \citet{delouis:2021}  \cr
\Planck\ PR4\ \rlap{$^{\rm a,c}$}& 3366.6*\pm*2.6*& 263.986\pm0.035&
48.247\pm0.023&\citet{planck2020-LVII}\cr
\noalign{\vskip 3pt}
\bf\BP\ \rlap{$^{\rm e}$} & \bf3362.7*\pm*1.4*& \bf264.11*\pm0.07*&
 \bf48.279\pm0.026&This paper\cr
\noalign{\vskip 5pt\hrule\vskip 5pt}}}
\endPlancktablewide
\tablenote {{\rm a}} Statistical and systematic uncertainty estimates are added in quadrature.\par
\tablenote {{\rm b}} Computed with a naive dipole estimator that does not account for higher-order CMB fluctuations.\par
\tablenote {{\rm c}} Computed with a Wiener-filter estimator that estimates, and marginalizes over, higher-order CMB fluctuations jointly with the dipole.\par
\tablenote {{\rm d}} Higher-order fluctuations as estimated by subtracting a dipole-adjusted CMB-fluctuation map from frequency maps prior to dipole evaluation. \par
\tablenote {{\rm e}} Estimated with a sky fraction of 68\,\%. Error bars include only statistical uncertainties, as defined by the global \BP\ posterior framework, and they thus account for instrumental noise, gain fluctuations, parametric foreground variations etc. 
\par
\end{table*}

During the initial test phase of the \BP\ pipeline, the Markov chain
was allowed to explore all the above degeneracies freely, without any
informative or algorithmic priors. These early runs resulted in a full
marginal uncertainty on the dipole amplitude of more than $40\muK$, as
compared to $3\muK$ reported by \Planck\ LFI for the 70\,GHz channel
alone \citep{planck2016-l02}, or $1\muK$ as reported by HFI
\citep{planck2016-l03}. Although this value by itself could be
considered acceptable, given the limited cosmological importance of
the CMB dipole, it was also strikingly obvious that all component maps
were compromised by the poorly constrained calibration, ultimately
leading to non-physical Galactic component maps with large
dipolar residuals. With the introduction of the spatial free-free and
AME priors discussed by \citet{bp13}, and the $\Lambda$CDM-based
$E$-mode quadrupole prior discussed by \citet{bp07}, these
degeneracies are effectively broken.

Figure~\ref{fig:cmb_with_dipole} shows the marginal CMB temperature
fluctuation posterior mean map as derived in \BP, given both the
data, model and priors described above. This map is massively
dominated by the CMB Solar dipole, with only a small imprint of the
Galactic plane being visible in the very center. At high latitudes,
CMB temperature fluctuations may be seen as tiny ripples superimposed
on the dipole.

Because of the small but non-negligible Galactic plane residuals, we
must impose an analysis mask before estimating final CMB Solar dipole
parameters. For this purpose, we use the Wiener filter estimator
described by \citet{thommesen:2019}, which in-paints the Galactic mask
with a constrained realization prior to parameter estimation; this is
necessary in order to account for, and marginalize over, coupling to
higher-order CMB fluctuations. This method was also adopted for the
dipole estimates presented in \citet{planck2020-LVII}, although we
introduce one significant difference to that analysis: In the current
analysis we estimate the magnitude of systematic uncertainties
directly from the \BP\ Gibbs samples, as opposed to putting in it by
hand. Specifically, instead of producing 9000 constrained realizations
from a single maximum likelihood map, as was done by
\citet{thommesen:2019} and \citet{planck2020-LVII}, we now produce 100
constrained realizations from each of the 3200 available end-to-end
Gibbs samples. Since each of these realizations have different gain,
correlated noise, and foreground residuals, the full ensemble accounts
seamlessly for all relevant systematic uncertainties. The only
additional term we put by hand into to the error budget is a
contribution of 0.7\muK\ from the CMB monopole uncertainty
\citep{fixsen2009}.

Using this methodology, we estimate the CMB dipole parameters over a
series of Galactic masks, ranging in sky fraction from 20 to
95\,\%. The results from these calculations are shown in
Fig.~\ref{fig:cmb_dipole}. Overall, we see that the posterior
distributions are quite stable with respect to sky
fraction. Furthermore, we note that the uncertainties do not decrease
after $f_{\mathrm{sky}}\approx 0.75$, as they would if the full error
budget could be described in terms of white noise and sky
fraction. Rather, the weight of the additional sky coverage is
effectively reduced when marginalizing over the various systematic
contributions, as desired. We conservatively adopt a sky fraction of
$f_{\mathrm{sky}}=0.68$ to define our final dipole estimates,
corresponding to the sky fraction close to that used for the main CMB
temperature analysis. The resulting values are plotted as black points
in Fig.~\ref{fig:cmb_dipole}, and tabulated together with previous
estimates in Table~\ref{tab:dipole}.

Several points are worth noting regarding these results. First, we see
that the reported best-fit \BP\ dipole amplitude is
${3362.7\pm1.4}$\,\muK, which is slightly lower than the latest LFI
2018 estimate of ${3364.4\pm3.1}$\,\muK, which in turn is lower than
the \npipe\ estimate of ${3366.6\pm2.6}$\,\muK. On the other hand, it
is very close to the latest HFI estimate of
${3362.08\pm0.99}$\,\muK, which is derived from an almost completely
independent data set. Overall, the agreement between these various
data sets and methods is excellent.

Regarding the directional parameters, two observations are worth
pointing out. First, we see that the \BP\ uncertainties are larger
than any of the previous \Planck-dominated results. Here it is worth
recalling again that no additional systematic error contributions are
added by hand to the \BP\ directional uncertainties, and the reported
values are thus the direct result of degeneracies within the model
itself. Perhaps the biggest algorithmic difference in this respect is
the fact that the current implementation explicitly marginalizes over the
full foreground and calibration model, while most other approaches
condition on external constraints. The second observation is that the
\BP\ latitude is very slightly higher than any of the previous
results, except \COBE. The statistical significance of this difference
is low, only about 1.5\,$\sigma$, but compared with the remarkable
internal agreement between \Planck\ and \WMAP, it is still
noteworthy. In this respect, we once again recall that we are
currently using the \Planck\ 2015 free-free map as an informative
prior in the current processing, and CMB and free-free emission are
known to be strongly correlated for the current data set; see
\citet{bp13}. Performing a joint analysis of LFI, HFI, and
\WMAP\ without an external free-free prior might be informative
regarding this point.



\section{Low-$\ell$ CMB anomalies}
\label{sec:anomalies}

\begin{figure}
  \includegraphics[width=\linewidth]{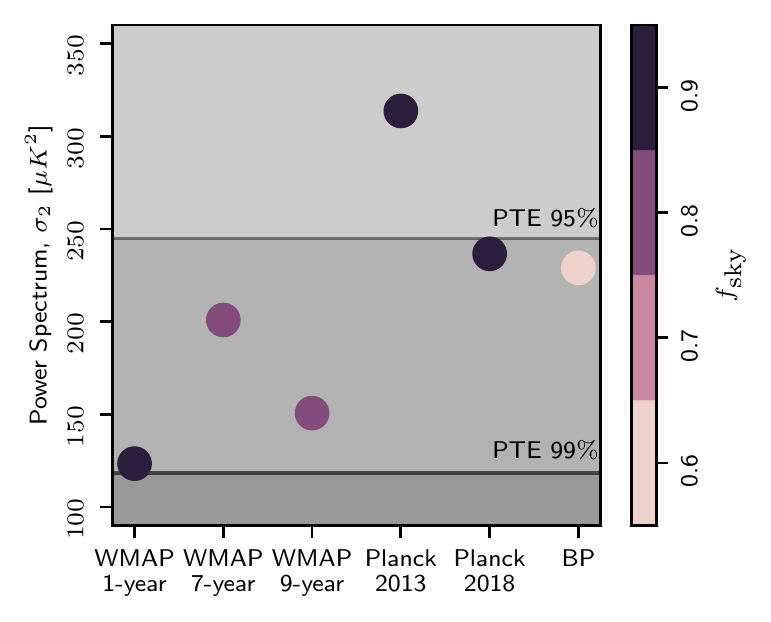}
  \caption{Estimates of the realization-specific quadrupole
    amplitude of our universe, $\sigma_2$, from \WMAP, \Planck, and
    \BP. The gray background indicate PTEs relative to the best-fit
    \Planck\ 2018 \LCDM\ model, and the color of the dots indicate the
    sky fraction used by the respective analysis. The radius of the
    dots correspond to the diagonal Fisher matrix uncertainty reported
    by \citet{hinshaw2012}, and provides a very naive noise-only
    estimate.}
  \label{fig:sigma_2_experiment}
\end{figure}

The posterior CMB sky map samples generated by the Gibbs sampler
discussed in Sect.~\ref{sec:bp} may be used for any scientific
analysis to which standard foreground-reduced CMB maps are
subjected. The main practical difference between these maps and
traditional maximum-likelihood maps is simply that in the Bayesian
case one must analyze an entire ensemble of different CMB maps, rather
than just one, and the resulting answer is typically defined in terms
of a histogram, rather than a single value.

The main advantage of this approach is full propagation of all
modelled systematic effects, some of which are very difficult to
account for with traditional approaches. One important example of this
is time- and detector-dependent gain variations. As already noted,
calibration uncertainties modulate the large CMB Solar dipole, and can
consequently also excite other large-scale modes through coupling from
the satellite scanning strategy, noise weighting, and confidence
mask. This issue is particularly pertinent to the question of
large-scale CMB anomalies, several of which were reported after the
release of the first \WMAP\ sky maps, including a low quadrupole
amplitude \citep{bennett:1992}, lack of large-scale correlations
\citep{spergel:2003}, quadrupole-octopole alignment and octopole
planarity \citep{dOC2004}, hemispherical power asymmetry
\citep{asymmetry2004}, a large non-Gaussian cold spot
\citep{vielva:2004}, and a low low-$\ell$ $TT$ power spectrum
\citep{planck2013-p08}.

Most of these effects are, however, typically only statistically
significant at the $3\,\sigma$ level, and unmodelled systematic errors
could therefore often be relevant in ways that are difficult to
quantify with traditional CMB maps. As such, the new CMB posterior
samples presented in this paper offer a unique opportunity to more
fully assess the significance of these anomalies, and in the following
we consider four examples that are implementationally straightforward
to evaluate, namely 1) the low quadrupole amplitude, 2) the
quadrupole-octopole alignment, 3) the octopole planarity, and 4) the
low low-$\ell$ $TT$ spectrum. We encourage other research groups to
revisit the remaining anomalies using their own tools on the new
posterior products provided here.

\begin{figure}[t]
  \includegraphics[width=\linewidth]{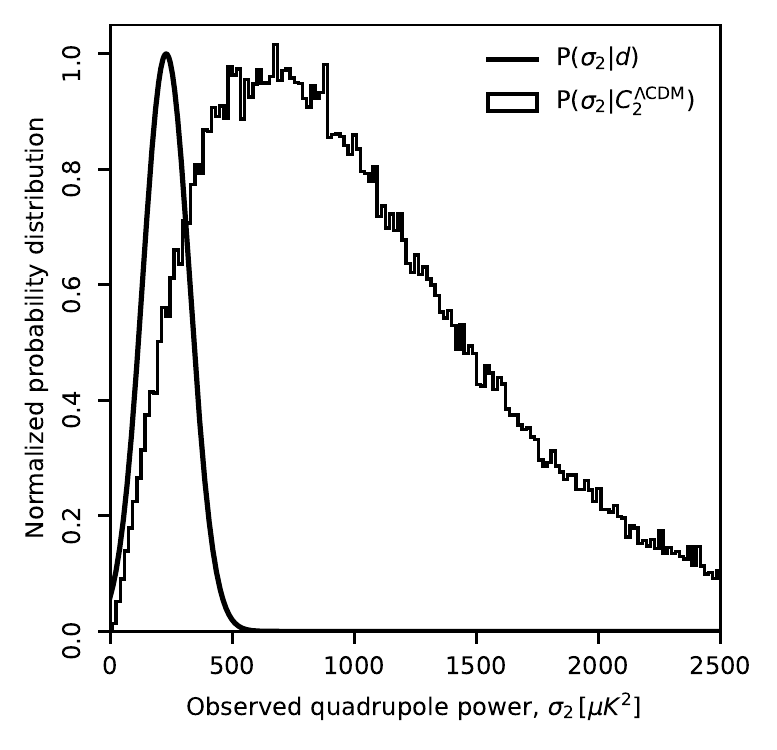}
  \caption{Comparison of the realization-specific quadrupole amplitude
    distribution derived from \BP\ (smooth black curve) and the
    predicted distribution from the \Planck\ 2018 best-fit
    \LCDM\ power spectrum (histogram).}
  \label{fig:P_sig_d_c}
\end{figure}

\subsection{Low quadrupole amplitude}
\label{sec:quadrupole}

As already mentioned, the $TT$ quadrupole amplitude has been measured
to be relatively weak compared to the \LCDM\ predictions ever since
COBE-DMR \citep{bennett:1992}, and this observation has been confirmed
both by \WMAP\ \citep{hinshaw2003b} and
\Planck\ \citep{planck2013-p08}. However, it is interesting to note
that the various experiments and analyses report quite different
values when it comes to the precise value for the quadrupole, as
illustrated in Fig.~\ref{fig:sigma_2_experiment}. Here we show the
reported quadrupole amplitudes,\footnote{Recall that $\sigma_2$ denotes
  the realization-specific quadrupole amplitude of \emph{our}
  universe, while $C_{2}$ (or $D_{2}=C_2 \cdot 6/2\pi$) denotes the ensemble-averaged quadrupole
  amplitude.} $\sigma_2$, for \WMAP, \Planck, and \BP; the gray
background indicates the PTE relative to the best-fit \Planck\ 2018
$\Lambda$CDM model, while the color of the dots indicate sky fraction. For
reference, the \citet{hinshaw2012} reports a diagonal Fisher
uncertainty for this mode of $9\muK^2$, which is comparable to the dot
radius.

All analyses report a generally low amplitude compared to \LCDM, with
all PTEs except one being higher than 0.95. At the same time, it is
also striking to note that even very similar analyses that rely on
highly correlated datasets, use almost identical techniques,
and are performed by the same research group, find results that vary
internally by many sigmas: The 7-year \WMAP\ analysis reports a
best-fit value of $201\,\muK^2$ \citep{larson2010}, while the
corresponding 9-year analysis reports $151\,\muK^2$
\citep{hinshaw2012}, which are formally different by more than
$5\,\sigma$. Furthermore, the confidence sky mask used in these two
analyses are identical, and sample variance does therefore not
contribute at all to this difference. Likewise, \Planck\ 2013 and 2018
reports values of 299 and $226\,\muK^2$, respectively, discrepant at
more than $8\,\sigma$, as measured by naive Fisher uncertainties,
taking into account \Planck's higher signal-to-noise ratio.

\begin{figure}
  \includegraphics[width=\linewidth]{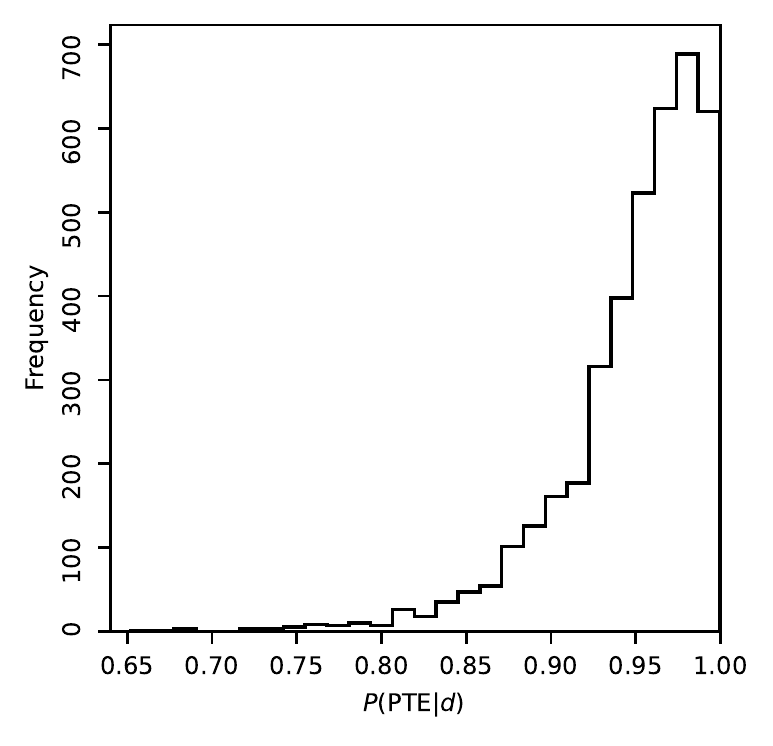}
  \caption{PTE probability distribution, $P(\mathrm{PTE}\mid\d)$, for
    the realization-specific quadrupole amplitude, $\sigma_2$ to
    exceed the \LCDM\ prediction after marginalizing over all modelled
    uncertainties.}
  \label{fig:C_2_PTE}
\end{figure}

What these results clearly show is simply that white noise
uncertainties only account for a small fraction of the total CMB
temperature quadrupole uncertainty. With the \BP\ posterior samples,
we are finally in a position where this statement may be quantified
more precisely in terms of the full marginal posterior distribution
$P(\sigma_2\mid\d)$, which now accounts for important contributions from
calibration and foreground uncertainties, and their correlations. This
distribution is shown in Fig.~\ref{fig:P_sig_d_c} as a solid smooth
curve, while the histogram is derived from $10^5$ ideal realizations
of $\sigma_2$ drawn from the \Planck\ 2018 best-fit \LCDM\ prediction,
$C^{\Lambda\mathrm{CDM}}_2 = 1064.7 \muK^2$. The posterior distribution may
be summarized as a Gaussian with $\sigma_2=229\pm97$, and the full marginal uncertainty,
including contributions from the instrument, astrophysics and
confidence mask, is thus more than 10 times larger than the naive
diagonal Fisher estimate quoted above, despite the fact that more data
are included in the current analysis.

We are now interested in deriving a total significance for the low
quadrupole amplitude. In a classic frequentist simulation-based
analysis this would be done simply by counting how many of the
simulated realizations in the histogram Fig.~\ref{fig:P_sig_d_c} have
a lower value than the observed peak posterior value. However, in our
case the PTE of the peak position carries no particular statistical
significance, and instead the full distribution of possible $\sigma_2$
values must be considered. Accordingly, Fig.~\ref{fig:C_2_PTE} show
the probability distribution of PTEs, $P(\mathrm{PTE}\mid\d)$, and here
we see that the 95\,\% confidence limit on this PTE is 0.85. Thus, the
observed quadrupole value is certainly on the low side compared to the
\LCDM\ prediction, but the effect is not highly significant with the
current dataset. A smaller confidence mask and a better constrained
instrument model are required to shed further light on the effect.

\begin{figure}[t]
  \includegraphics[width=\linewidth]{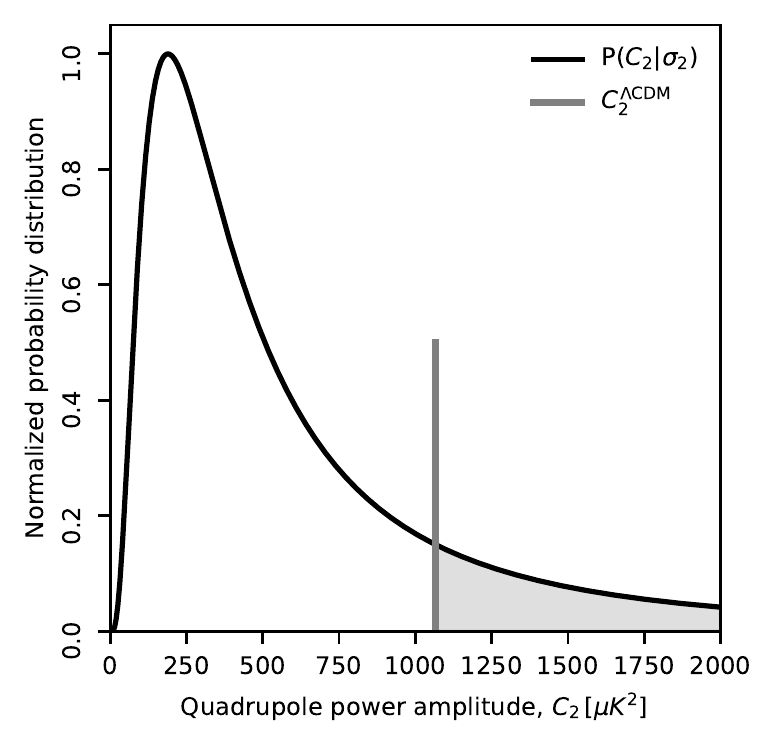}
  \caption{Marginal probability distribution of the ensemble-averaged
    quadrupole power spectrum, $P(C_{\ell}\mid\d)$, estimated in
    \BP\ (solid line). The vertical line at $C_2 = 1064.7$ indicates
    the value predicted by the \Planck\ 2018 best-fit $\Lambda$CDM
    model; 21.7\,\% of the marginal distribution exceeds this value.}
  \label{fig:norm_P_sigma_P_c_2}
\end{figure}

To conclude this section, we also turn the question around, and ask
``what is the probability distribution for $C_2$ given the measured
values of $\sigma_2$''? To answer this, we evaluate the GBR estimator
discussed in Sect.~\ref{sec:highell_spec} as a function of $D_{\ell}$,
as shown in Fig.~\ref{fig:norm_P_sigma_P_c_2}. The PTE for $C_2$
relative to $C_2^{\Lambda \mathrm{CDM}} = 1064.7\,\muK^2$ is 21.7\,\%.

\subsection{Quadrupole-octopole alignment}
\label{sec:alignment}

A second anomaly regarding the very largest angular scales in the CMB
map was first reported by \citet{Tegmark:2003ve} and \citet{dOC2004}, who found
that the quadrupole and octopoles appeared morphologically
aligned. This was quantified by first defining a preferred direction,
$\hat{\mathbf{n}}_\ell$, for each mode separately by maximizing the
angular momentum dispersion of the wave function,
\begin{equation}
  \langle \psi | (\hat{\textbf{n}}_\ell \cdot \textbf{L})^2 |\psi\rangle = \sum_m m^2|a_{\ell m}(\hat{\textbf{n}})|^2,
\end{equation}
and then computing the angular separation between $\hat{\textbf{n}}_2$
for the quadrupole and $\hat{\textbf{n}}_3$.  \cite{dOC2004} found
that this angle was smaller than the isotropic expectation with a PTE
of 0.984 for the first-year \WMAP\ data. This observation was
qualitatively later confirmed by the \WMAP\ team \citep{bennett2012},
who found the angle to be around $3^\circ$ in the nine year data,
corresponding to a probability of 0.14\% for such an alignment or
stronger to occur assuming isotropy. The \WMAP\ team did however note
that the foreground removal procedure was a limiting factor in the
measurement of the misalignment. Likewise, \cite{Planck:2013lks}
reported an alignment in the interval $9^\circ$ and $13^\circ$
depending on the component separation method, corresponding to PTEs in
between 1.2 and 3.4\%, respectively.

\begin{figure}
  \includegraphics[width=\linewidth]{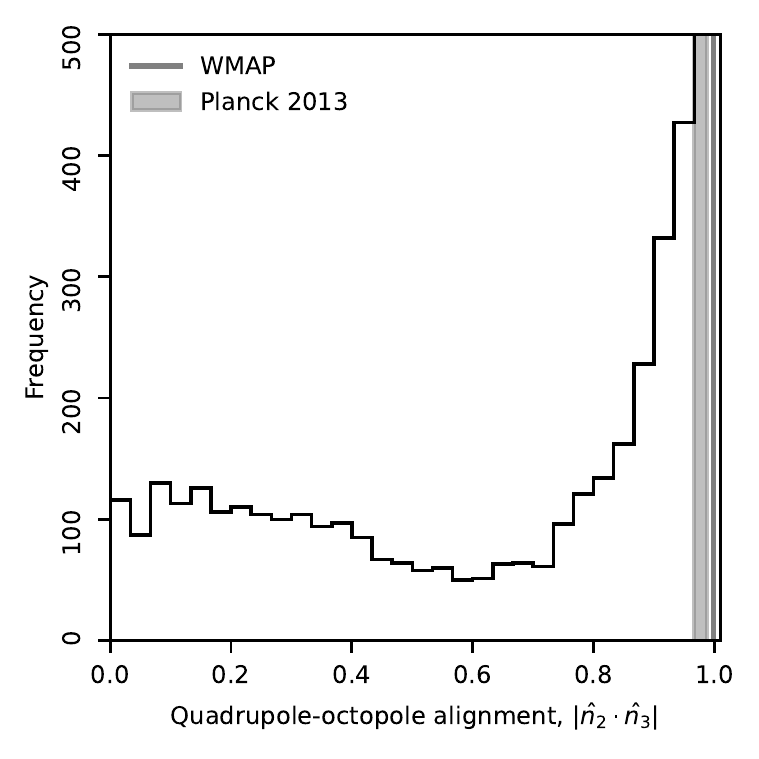}
  \caption{Histogram of the quadrupole-octopole alignment, $|\hat{n}_2\cdot\hat{n}_3|$. The official WMAP nine-year result reported the value $|\hat{n}_2\cdot\hat{n}_3| = 0.9986$ \citep{bennett2012} from a misalignment of 3 degrees, while Planck \citep{Planck:2013lks} reported the interval $[0.9663, 0.9877]$ corresponding to 9 and 13 degrees, respectively, depending on the component separation method.}
  \label{fig:n_2_n_3}
\end{figure}

Given the fact that instrumental systematic uncertainties affect the
absolute quadrupole amplitude by a large factor, as discussed in the
previous section, it is reasonable to assume that also the preferred
quadrupole direction is affected by the same uncertainties. In this
section, we therefore apply the methodology of \cite{dOC2004} to the
same set of constrained CMB realizations discussed above, and derive
the full distribution of alignment PTEs after full systematics
marginalization. This is summarized in the form of a histogram in
Fig.~\ref{fig:n_2_n_3}, with the 9-year \WMAP\ and \Planck\ 2013
results shown as gray vertical bars. The width of the \Planck\ bar
indicates the uncertainty derived among the four \Planck\ component
separation codes.

In this figure we see that the agreement between the \BP\ results and
previous results is excellent in terms of single-point maximum
posterior values. However, we also see that the full \BP\ posterior
distribution is very broad, to the extent that all possible angles are
in fact allowed by the data, from 0 to $90^{\circ}$. Part of this
larger uncertainty does come from the somewhat more conservative
analysis mask with $f_{\mathrm{sky}}=0.64$ employed in the current
analysis, as compared to 0.72 for \citet{Planck:2013lks}. At the same
time, we also note that foreground modelling details appear to have
only a small impact of the final results, as very different methods
reach quite similar conclusions: The \WMAP\ result was derived from an
Internal Linear Combination (ILC) map with low-resolution foreground
eigenmode error propagation, while the \Planck\ results were derived
using four qualitatively different methods coupled with end-to-end
simulations. All these methods agree internally qualitatively very
well, and they also agree with the maximum-posterior \BP\ result.

The fundamental difference between the \BP\ and previous analyses does
not lie in different foreground modelling, but rather in the
instrument modelling and the general statistical treatment and error
propagation. Most importantly, while previous analyses only accounted
for relatively simple foreground and noise uncertainties, the
\BP\ processing additionally accounts for full gain uncertainties and
their coupling to the CMB Solar dipole and foregrounds. When doing so,
the statistical evidence for a quadrupole-octopole alignment
diminishes significantly. Of course, it is also important to emphasize
that a substantial contributor to this additional variance is the
exclusion of the \Planck\ HFI measurements, which would allow both
better CMB constraints (and thereby indirectly also stronger LFI
calibration constraints), as well as a smaller Galactic plane by
properly fitting free-free and thermal dust emission. Future work that
also includes HFI data will therefore need to revisit this question.

\subsection{Planar octopole}
\label{sec:planar}

\begin{figure}
  \includegraphics[width=\linewidth]{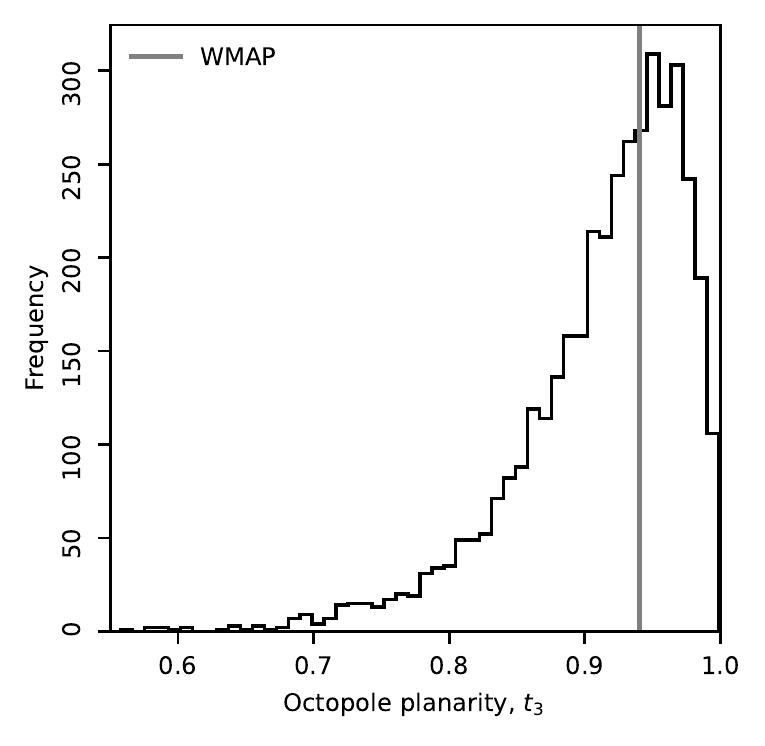}
  \caption{Histogram of the planarity test statistics $t_3$ derived
    from the set of \BP\ constrained CMB realizations. We also mark
    the value of $t_3=94\%$ originally reported by \citet{dOC2004}
    from the first-year \WMAP\ ILC map.}
  \label{fig:t_3}
\end{figure}

\citet{Tegmark:2003ve} and \citet{dOC2004} also noted that not only is the plane
of the temperature octopole closely aligned with the quadrupole, but
the octopole is also intrinsically highly planar. \citet{dOC2004}
quantified this through the test statistic $t_3$,
\begin{equation}
  t_3 \equiv \underset{\hat{\textbf{n}}}{\textrm{max}}  \frac{|a_{3-3}(\hat{\textbf{n}})|^2 + |a_{33}(\hat{\textbf{n}})|^2}{\sum^{3}_{m=-3}|a_{3m}(\hat{\textbf{n}})|^2},
\end{equation}
which measures the ratio of the total octopole power that may be
contributed to $a_{3\pm3}$, maximized over all coordinate
systems. This is shown in terms of a histogram for the \BP\ CMB
samples in Fig.~\ref{fig:t_3} together with the original
\WMAP\ measurement of $t_3=94\%$ by \citet{dOC2004}. Once again, we
see that the agreement is very good in terms of the single-point peak
value -- but we also see that the distribution is quite broad when
marginalizing over the full set of uncertainties. This distribution is
in qualitatively good agreement with the results of
\citet{rassat:2014}, who measured the octopole planarity for six
different foreground-reduced \Planck\ 2013 CMB maps, and found values
ranging between 0.84 and 0.95 (corresponding to PTEs between 7 and
37\,\%) depending on foreground cleaning and mask details. When
additionally marginalizing over instrumental systematic effects in
\BP, we see that the range broadens further.

\subsection{Low-$\ell$ temperature amplitude}
\label{sec:lowell_amp}

\begin{figure}
  \includegraphics[width=\linewidth]{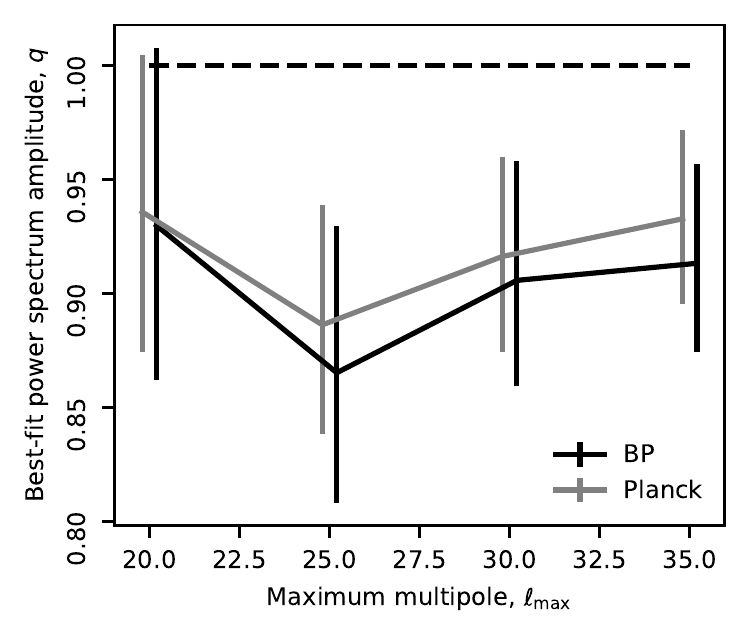}
  \caption{Best-fit amplitude, $q$, as a function of maximum
    multipole, $l_{\mathrm{max}}$ measured relative to the best-fit
    \Planck\ \LCDM\ power spectrum, as derived for both \Planck\ 2015
    (\emph{gray}) and \BP\ (\emph{black}).}
  \label{fig:best_fit_q}
\end{figure}

Finally, we revisit the ``low low-$\ell$ temperature power spectrum
anomaly'', first reported by \citet{planck2013-p08} \citep[an analogous result at the angular correlation function level had been previously observed in \WMAP\ data, albeit at lower significance level, e.g.][]{Bennett_2011}. In this case, the \Planck\ team fitted the low-$\ell$ temperature power spectrum with an
amplitude-scaled model $C_{\ell}(q) = q C^{\Lambda
  \textrm{CDM}}_{\ell}$ between $2\le\ell\le\ell_{\mathrm{max}}$, and
varied $\ell_{\mathrm{max}}$. Doing so, they found best-fit amplitudes
typically ranging between 0.87 and 0.95, which are low with
statistical significances typically at the 1.5--$2.5\,\sigma$ level.

Figure~\ref{fig:best_fit_q} shows corresponding results for both
\BP\ and \Planck\ 2015. Once again, we see that the results generally
agree well. In this case, we see on the one hand that the \BP\ mean
results are in fact slightly lower than the \Planck\ results, possibly
hinting towards a stronger anomaly. On the other hand, the
uncertainties are also larger due to the larger confidence mask and
more complete instrument error marginalization, and the overall
significance of the effect is therefore essentially unchanged.

\section{Discussion and conclusions}
\label{sec:summary}

In this paper, we have presented the CMB results from the Bayesian
\BP\ analysis \citep{bp01}. This represents the first example of an
end-to-end posterior sample-based CMB analysis for which the inputs
are defined in terms of raw time-ordered data and the final outputs
are CMB sky maps and power spectra. This method was first suggested in
a CMB setting by \citet{jewell2004} and \citet{wandelt2004}, and it
took almost twenty years of computer hardware and algorithm
development to realize this in practice.

Two of the most fundamental advantages of integrated end-to-end CMB
analysis are full joint exploration of all free parameters --
instrumental, astrophysical, and cosmological -- and true end-to-end
error propagation. In principle, this algorithm has a similar
statistical foundation as the traditional low-$\ell$ brute-force CMB
likelihood approach used by both \WMAP\ \citep{hinshaw2012} and
\Planck\ \citep{planck2016-l05}, but with a few key differences:
Rather than just accounting for correlated noise and template-based
foreground residuals at low angular resolution, this method can
account for \emph{all} degrees of uncertainty at full angular
resolution. It achieves this through Monte Carlo sampling, as opposed
to analytic construction of dense covariance matrices, and neither
angular resolution nor model complexity therefore carry a similar
prohibitive computational cost as the traditional method.

It is important to note the Bayesian method, whether implemented
analytically or through sampling, is fundamentally different from the
frequentist forward simulation-based method that is commonly used in
the CMB field for error propagation. Intuitively, the main difference
lies in that, while forward simulations describe some \emph{random}
instrument and universe, the Bayesian approach describes \emph{our}
instrument and universe. Because of this difference, the two methods
are naturally geared toward answering different types of statistical
questions. For the Bayesian approach, it is easy to address questions
like ``what are the most likely $\Lambda$CDM parameters for our
universe?'', but difficult to ask ``is our data set consistent with
the $\Lambda$CDM model?''. For the frequentist simulation approach,
the opposite holds true. It is also interesting to note that the
simulation-based approach becomes indistinguishable from the Bayesian
approach if constrained realizations are used to generate the
instrument and sky model, as opposed to statistically independent
realizations, as is typically done. Indeed, the current
\BP\ implementation may in many respects simply be considered as a
constrained realization-based simulator.

In this paper, we have used this sampling framework to address several
classical problems in CMB analysis. We have studied cross-correlations
between the CMB component and instrumental and astrophysical
parameters, and we have identified and mitigated a particularly strong
degeneracy with free-free emission. We have compared the resulting
posterior mean CMB maps and power spectra with previously published
results, and found good agreement. We have also derived a CMB Solar
dipole amplitude of $3362.7\pm1.4\muK$, which is in excellent
agreement with previous results -- but it is important to note that
the quoted uncertainty is derived directly from the global statistical
model, and not associated with any additional \Planck-specific
systematic error.

Given that all of the above are in good agreement with previous
results, one may ask, what is the point of this approach? Does this
not simply show that the traditional method works just as well? The
main answer to this question may be formulated in terms of
signal-to-noise ratio: As long as the statistic or quantity in
question is signal dominated, such as the \Planck\ $TT$ spectrum on
large and intermediate scales, the current method provides little or
no obvious advantage. However, when the statistic in question is
either systematics- or noise-dominated, then these methods become very
powerful through their end-to-end error propagation capabilities. This
was explicitly demonstrated in this paper by revisiting a number of
previously reported large-scale anomalies in the CMB temperature
anisotropies. In many cases, we found that the significances of these
anomalies were significantly reduced after accounting for both
low-level instrumental parameters and the full non-Gaussian shape of
the posterior distribution. Such effects are very difficult to model
accurately by non-sampling methods.

We posit that the same will also hold true for any next-generation
high-sensitivity CMB $B$-mode experiment that aims to detect
primordial gravitational waves. These experiments are looking for a
signal that is five or more orders of magnitude weaker than the CMB
dipole, and at least a few orders of magnitude weaker than the
Galactic diffuse foregrounds. As such, accurate and joint error
propagation of both instrumental and astrophysical uncertainties will
be key to claiming a robust detection. Indeed, developing methods
applicable to this task was the main motivation behind the
\BP\ project in general. The current analysis has shown in practice
that end-to-end Bayesian CMB analysis is both computationally and
implementationally feasible.

\begin{acknowledgements}
  We thank Prof.\ Pedro Ferreira and Dr.\ Charles Lawrence for useful suggestions, comments and 
  discussions. We also thank the entire \Planck\ and \WMAP\ teams for
  invaluable support and discussions, and for their dedicated efforts
  through several decades without which this work would not be
  possible. The current work has received funding from the European
  Union’s Horizon 2020 research and innovation programme under grant
  agreement numbers 776282 (COMPET-4; \BP), 772253 (ERC;
  \textsc{bits2cosmology}), and 819478 (ERC; \textsc{Cosmoglobe}). In
  addition, the collaboration acknowledges support from ESA; ASI and
  INAF (Italy); NASA and DoE (USA); Tekes, Academy of Finland (grant
   no.\ 295113), CSC, and Magnus Ehrnrooth foundation (Finland); RCN
  (Norway; grant nos.\ 263011, 274990); and PRACE (EU).
\end{acknowledgements}

\bibliographystyle{aa}

\bibliography{Planck_bib,BP_bibliography}

\end{document}